\documentclass[manuscript]{aastex}

\shorttitle{Collapsed Cores in Globular Clusters}
\shortauthors{Djorgovski et al.}

\usepackage{color} 
\begin{document}


\title{Rotational properties of Maria asteroid family}


\author{Myung-Jin Kim\altaffilmark{1,2}, Young-Jun Choi\altaffilmark{2}, Hong-Kyu Moon\altaffilmark{2}, Yong-Ik Byun\altaffilmark{1,3},\\ 
Noah Brosch\altaffilmark{4}, Murat Kaplan\altaffilmark{5}, S\"{u}leyman Kaynar\altaffilmark{5}, \"{O}mer Uysal\altaffilmark{5},\\ 
Eda G\"{u}zel\altaffilmark{6}, Raoul Behrend\altaffilmark{7}, Joh-Na Yoon\altaffilmark{8}, Stefano Mottola\altaffilmark{9},\\ 
Stephan Hellmich\altaffilmark{9}, Tobias C. Hinse\altaffilmark{2}, Zeki Eker\altaffilmark{5}, and Jang-Hyun Park\altaffilmark{2}}

\email{skarma@galaxy.yonsei.ac.kr}


\altaffiltext{1}{Department of Astronomy, Yonsei University, 50 Yonsei-ro, Seodaemun-gu, 120-749 Seoul, Korea}
\altaffiltext{2}{Korea Astronomy and Space Science Institute, 776 Daedeokdae-ro, Yuseong-gu, 305-348 Daejeon, Korea}
\altaffiltext{3}{Yonsei University Observatory, 50 Yonsei-ro, Seodaemun-gu, 120-749 Seoul, Korea}
\altaffiltext{4}{Tel Aviv University, P.O. Box 39040, Tel Aviv 69978, Israel}
\altaffiltext{5}{Akdeniz Universitesi, Fen Fakultesi, Dumlupinar Bulvari, Kampus, 07058 Antalya, Turkey}
\altaffiltext{6}{Department of Astronomy and Space Sciences, University of Ege, Bornova, 35100 Izmir, Turkey}
\altaffiltext{7}{Geneva Observatory, Maillettes 51, 1290 Sauverny, Switzerland}
\altaffiltext{8}{Chungbuk National University Observatory, 802-3 Euntan-ri, Jincheon-gun, Chungcheongbuk-do, Korea}
\altaffiltext{9}{German Aerospace Center (DLR), Rutherfordstra$\beta$e 2, 12489 Berlin, Germany}


\begin{abstract}
Maria family is regarded as an old-type ($\sim$3 $\pm$ 1 Gyr) asteroid family which has experienced substantial collisional and dynamical evolution in the Main-belt. It is located nearby the 3:1 Jupter mean motion resonance area that supplies Near-Earth asteroids (NEAs) to the inner Solar System. We carried out observations of Maria family asteroids during 134 nights from 2008 July to 2013 May, and derived synodic rotational periods for 51 objects, including newly obtained periods of 34 asteroids. We found that there is a significant excess of fast and slow rotators in observed rotation rate distribution. The two-sample Kolmogorov-Smirnov test confirms that the spin rate distribution is not consistent with a Maxwellian at a 92\% confidence level. From correlations among rotational periods, amplitudes of lightcurves, and sizes, we conclude that the rotational properties of Maria family asteroids have been changed considerably by non-gravitational forces such as the YORP effect. Using a lightcurve inversion method \citep{kat01,kaa01}, we successfully determined the pole orientations for 13 Maria members, and found an excess of prograde versus retrograde spins with a ratio ($N_p/N_r$) of 3. This implies that the retrograde rotators could have been ejected by the 3:1 resonance into the inner Solar System since the formation of Maria family. We estimate that approximately 37 to 75 Maria family asteroids larger than 1 km have entered the near-Earth space every 100 Myr.
\end{abstract}


\keywords{minor planets, asteroids: general}




\section{Introduction}

An asteroid family is an group of asteroidal objects in the proper orbital element space ({\it a}, {\it e}, and {\it i}), considered to have been produced by a disruption of a large parent body through a catastrophic collision (first identified by Hirayama 1918; see Cellino et al. 2009 and references therein). Family members have usually similar surface properties such as spectral taxonomy types \citep{cel02}, SDSS colors \citep{ive02,par08}, and visible geometric albedo \citep{mas11}. Therefore an asteroid family can be seen as a natural solar system experiment and is regarded as a powerful tool to investigate space weathering \citep{nes05} and non-gravitational phenomena such as the Yarkovsky and YORP (Yarkovsky-O'Keefe-Radzievskii-Paddack) effects. 

The Maria asteroid family has long been known as one of the Hirayama families \citep{hir22}, and is a typical old population ($\sim$3 $\pm$ 1 Gyr) \citep{nes05} that is expected to have experienced significant collisional and dynamical evolution in the history of the inner Solar System. The Maria family is located close to the outer border of the 3:1 mean-motion resonance (MMR) ($\sim$2.5 AU) with Jupiter, thus it might be regarded as a promising source region candidate for a couple of giant S-type near-Earth asteroids, 433 Eros and 1036 Ganymed \citep{zap97}. Our knowledge about the properties of the Maria family, however, are still limited. To date, rotational periods of the family members among 3,230 cataloged objects \citep{nes10} have been known only for 58 of the relatively large asteroids (LCDB\footnote{http://www.minorplanet.info/lightcurvedatabase.html}, March 2013), accounting for less than 2 percent of the family.

The study of the rotational properties of an asteroid family, i.e. rotational period, pole orientation, and overall shape of lightcurve, can offer a unique opportunity to obtain insight both on the collisional breakup process and on the dynamical evolution of asteroids. \citet{pao05} proposed that the statistical properties of the spin period and pole orientation contain information on the collisional transfer of angular momentum of small bodies. Laboratory experiments \citep[and reference therein]{hol02} from catastrophic disruption events show that the distribution of asteroidal spin periods resembles a Maxwellian, which implies that members have reached an approximate equilibrium status after a chain of isotropic collisions. There must be differences between the behavior of laboratory experiments and the catastrophic collisions of asteroids in their natural environment due to various approaches on scaling law \citep[see, e.g.,][]{miz90,hou90,hou91,dav94}. \citet{gib98} have detected that, on average, the smaller asteroids rotate faster from the catastrophic disruption experiments, and it has already been predicted by the semi-empirical model of catastrophic impact processes by \citet{pao96}. In addition, this tendency is also found in spin rate-size distribution from the Asteroid Lightcurve Database (LCDB) \citep{pra02,war09} as well as on a statistical analysis of C and S-type MBAs \citep{car10}.

A typical asteroid family shows a V-shape distribution in the proper semi-major axis vs. absolute magnitude plane \citep{nes04,vok06,mil10}. The dynamical evolution of the family members due to the Yarkovsky effect tends to spread the members out in semi-major axis, over the evolution time. For this reason, the dispersions of the semi-major axes can be used as a clue to estimate the age of an asteroid family \citep{bot01,nes05}.

The lightcurve inversion method \citep{kat01,kaa01} is a powerful tool to obtain rotational properties, especially pole orientation, from time-series photometric data. \citet{han11} derived 80 new asteroid models based on the combination of classical dense-in-time (hereafter dense) and sparse-in-time (hereafter sparse) photometric lightcurve data. They found the distribution of pole latitudes for small asteroids (D $< 30$ km) in the Main-belt to be depleted near the ecliptic plane, an effect attributed to the YORP thermal effect.

There are several observational studies on the rotational properties of asteroid families. \citet{sli02} and \citet{sli03} discovered an alignment of the spin vectors in the Koronis family, which is commonly referred to as the Slivan effect. \citet{kry12} showed that the distribution of rotational periods in the Flora asteroid family is non-Maxwellian. Taking advantage of the analysis of lightcurves obtained from Flora family members, \citet{kry13} recently found the Slivan effect in the Flora family. \citet{alv04} tried to find correlations among rotation periods, lightcurve amplitudes, and sizes of objects for the Themis, Eos, and Maria families, but no specific relationship was found. \citet{ito10} focused on the observation of young asteroid families, which are expected to have experienced little collisional and orbital evolution. In general, we have some difficulty in obtaining high S/N lightcurves, especially for smallest and hence, faintest family members.

The purpose of this paper is to investigate the rotational properties of the Maria family; rotation periods, orientations of the spin axis, and shapes, mainly based on our observations but also with data available in the literature. In Sect. 2 we describe the photometric observations of the Maria family asteroids with an introduction on the several telescopes we used. In Sect. 3 we explain the method of data reduction and acquisition from the available dense and sparse datasets from the AstDyS (Asteroids Dynamic Site\footnote{http://hamilton.dm.unipi.it/astdys}) database. Lightcurve plots for each asteroid are given in Appendix Figs. A. 1 $\sbond$ 61. Using the lightcurve inversion method, we derive the orientations of spin axis for 13 Maria asteroids. The results of this analysis and related discussions about the rotational properties of the Maria family are given in Sect. 4. Finally, in Sect. 5, we summarize our conclusions.

\section{Observations}


Observations of the Maria family asteroids were conducted during 134 nights from 2008 July to 2013 May, using 0.5 m- to 2 m- class telescopes at 7 observatories in the northern hemisphere. We used CCD cameras on the Wise Observatory (WO) 0.46 m telescope in Negev desert, Israel, the Tubitak Ulusal Gozlemevi (TUG) 1.0 m telescope in Bakirlitepe, Turkey, the Bohyunsan Optical Astronomy Observatory (BOAO) 1.8 m telescope on Bohyunsan, Korea, the Sobaeksan Optical Astronomy Observatory (SOAO) 0.6 m telescope on Sobaeksan, Korea, the Korea Astronomy and Space Science Institute (KASI) 0.6 m telescope in Jincheon, Korea, the Lemmonsan Optical Astronomy Observatory (LOAO) 1.0 m telescope on Mt. Lemmon, USA, and the Calar Alto (CA) Astronomical Observatory 1.2 m telescope in Almeria, Spain. The details of the observatories and instruments are shown in Table 1. 
All telescopes except for the BOAO 1.8 m and the CA 1.2 m telescopes were guided at sidereal tracking rate. For sidereal tracking of asteroids, we considered two factors: (1) the apparent motion of the asteroid should be less than the FWHM of the stellar profiles on each observatory; (2) the signal-to-noise ratio of the asteroid measurement to be at least 50. Accordingly, the maximum exposure time did not exceed 300 seconds while tracking at sidereal rate during the observations. Because several asteroids were relatively faint, some images taken with the BOAO 1.8 m telescope were acquired through a non-sidereal tracking mode that corresponded to the predicted motion of the objects. On the other hand, the CA 1.2 m telescope was tracked with a tracking vector halfway between sidereal rate and that of asteroids. In this way both asteroids and the background stars were trailed by the small amount. We made time-series observations mainly with Johnson R filter in order to characterize the rotational states of the asteroid, since the R-band filter with an optical imager is the most sensitive to small bodies in the Solar System.

\subsection{Target selection}

The Maria asteroid family consists of 3,230 known members with identification based on Hierarchical Clustering Method \citep[HCM;][]{nes10}. The latest family classification is available at the AstDys web site\footnote{http://hamilton.dm.unipi.it/$\sim$astdys2/propsynth/numb.members}. As of November 2013, the database lists 2,085 Maria members based on synthetic proper orbital elements of numbered asteroids. We checked the membership as listed in the AstDys, and found that most of our target asteroids were confirmed as Maria members except for asteroids 652, 4122, 5977, 13679, 19184, 19333, 32116, 43174, 50511, 109792, and 114819. However, memberships of an asteroid family can be subject to some unavoidable uncertainties, since they depend on the datasets of proper orbital elements available at the epoch of family search, on the adopted identification technique and membership assignment criteria.

We investigated the cumulative distribution N ($< H$) of absolute magnitudes H for all the members of the Maria family. We used a power-law approximation of N ($< H$) $\varpropto$ 10$^{\gamma{H}}$ in the magnitude range between 12 and 14.5, and obtained $\gamma \sim$ 0.54 (see Fig. 1). The value of $\gamma$ obtained is close to the steady-state mass distribution of collisional fragmentation for which \citet{doh69} derived $\gamma$ = 0.5, but discordant with a considerably steeper distribution constructed by Pareto power laws \citep[e.g.][]{tan99}. This power law index implies that Maria family has undergone a significant collisional and dynamical evolution and has currently reached an equilibrium state \citep[e.g.][]{doh69,obr03,bot05}. The slope depends highly on the number of objects brighter than 12th absolute magnitude, while the slope is thought to have been contaminated by observational bias for objects fainter than H = 14.5 magnitude. We chose our targets in the H range between 12 and 14.5; the total number of targets in this magnitude range is 981. The catalogued absolute magnitudes in the standard (H, G) system have large uncertainties due to the transformation from an observed magnitude system to the Johnson V system, and to the phase correction \citep[see][for more details]{cel09}. It might either overestimate or underestimate the size distribution of a family.

We generated ephemeris for each object from the 981 target asteroids using the JPL Horizons service\footnote{http://ssd.jpl.nasa.gov/?horizons}, then produced target lists of observable asteroids during any given night. Because the rotational periods of most targets are unknown, we decided to observe one or two target asteroids per night at least, assuming typical rotational periods between 2 and 24 hours. Since a preliminary rotational lightcurve of an asteroid is determined only after a prompt data reduction, it is better to allocate follow-up observation so as to cover its full-phased lightcurve. For the sake of increasing the observational efficiency, we developed an observation scheduler to carry out asteroid follow-up observation in a timely manner. When we input observational parameters such as initial rotational period, observed time, and observatory code, the scheduling program suggests to the user the next proper observing time to cover gaps in the lightcurve.

\subsection{Coordinating the observations}

Our observations were focused on asteroids in the Maria family that lack a known rotational period. To increase the number of observable asteroids as much as possible, we observed two objects alternately on a single night with the BOAO 1.8 m, TUG 1.0 m, and CA 1.2 m telescopes. Moreover, further observations during the years in 2012 - 2013 were also performed for 21 targets at the Wise telescope in 2008, and spin axis information was obtained from published lightcurves (see Sect. 3.2. for more details).

It is difficult to cover entire rotational phase of asteroids with rotational periods longer than 8 hours during a single night. In addition, if the rotational period of an asteroid aliases with 24 hours $\sbond$ the rotational period of the Earth, it is essentially impossible to have a fully covered lightcurve from a single observatory. A network of follow-up telescopes can be used to solve these problems. In order to maximize phase coverage of an asteroid, taking advantage of a network observation at different time zones, we organized observation campaigns with three observatories in Korea (UT + 9 h), Turkey (UT + 2 h), and USA (UT $\sbond$ 8 h). These network observations were carried out in 2012 with TUG and LOAO on June 28, 30; BOAO and TUG on Oct. 12, 14, 19, and SOAO and TUG on Oct. 12, 14, 17. In order to calibrate all the data gathered from various observatories, the same CCD fields obtained during the previous observations were taken in the next runs.

The detailed observational circumstances of each asteroid are listed in Table 2: UT date corresponding to the mid time of the observation, the topocentric equatorial coordinates (RA and Dec, J2000), the heliocentric (r) and the topocentric distances ($\Delta$), the solar phase (sun-asteroid-observer's) angles ($\alpha$), the apparent predicted magnitude (V), and the telescopes used.

\section{Data reduction and analysis}

All observational data reduction procedures were performed using the Image Reduction and Analysis Facility (IRAF) software package. Individual images were calibrated using the standard processing routine of the IRAF task {\it noao.imred.ccdred.ccdproc}. Bias and dark frames with relatively high standard deviations were not used for analysis. Twilight sky flats were acquired before sunrise and after sunset, and combined to produce a master flat image for each night. Instrumental magnitudes of asteroids were obtained using the IRAF {\it apphot} package; aperture radii were set to be equal to the FWHM of the typical stellar profile on each frame in order to maximize the S/N ratio \citep{how89}. The lightcurves of most asteroids were constructed based on relative magnitudes, that is the difference between the instrumental magnitude of an asteroid and the average magnitude of each comparison star.

In order to choose a set of comparison stars, we inspected whether there exists any short-term variability in brightness of a star, according to the following method. We overlapped each frame that was taken during the same night in order to calculate differences in magnitude of each star and the median value of an ensemble of the remaining stars. In such a way, we were able to determine standard deviations of each star in the same frame, and hence checked if they showed variability during the course of a single night’s observation. Likewise, we repeatedly applied the same procedure to the other CCD fields, and successfully selected stars with minimum standard deviation that show the least amount of light variation in each field. Finally we selected 3 to 5 comparison stars with a typical scatter of 0.01 to 0.02 mag. The observation time (UT) was corrected for light travel time.

\subsection{Rotational period and lightcurve}

Out of 134 nights observations in total, we obtained 218 individual lightcurves for 74 Maria family members and derived synodic rotational periods for 51 objects including obtained periods for 34 asteroids for the first time. In order to find the periodicity of the lightcurve, the Fast Chi-Squared (F$\chi^2$) technique \citep{pal09} was employed. In addition, we checked the result with the discrete Fourier transform algorithm \citep{len05}. In most cases for rotational periods, these two different techniques present similar results within the statistical errors. The F$\chi^2$ technique presented here represents the observed magnitudes as a Fourier time series truncated at the harmonic H:

\begin{equation}
\Phi_H(\{A_{0...2H},f\},t) = A0 + \sum_{h=1...H}A_{2h-1}\sin(h2{\pi}ft) + A_{2h}\cos(h2{\pi}ft)
\end{equation}

In practice, we fit the fourth-order Fourier function and also obtained the highest spectral power with the discrete Fourier transform algorithm. As a result, the final rotational period was determined assuming a double-peaked lightcurve. We present the resultant composite lightcurves of observed asteroids in Appendix Figs. A. 1 $\sbond$ 61, folded with their synodic periods.

In the approximation of a triaxial body with uniform albedo, the peak-to-peak variations in magnitude are caused by the change in apparent cross-section of the rotating body, with semi-axes a, b, and c, where $a > b > c$ (the body rotates along the c axis). According to \citet{bin89}, the lightcurve amplitude varies as a function of the polar aspect viewing angle $\theta$ (the angle between the rotation axis and the line of sight): 

\begin{equation}
{\Delta}m = 2.5log(\frac{a}{b}) - 1.25log(\frac{a^2\cos^2{\theta}+c^2\sin^2{\theta}}{b^2\cos^2{\theta}+c^2\sin^2{\theta}})
\end{equation}
\\
The lower limit of axis ratio a/b can be expressed as $a/b = 10^{0.4{\Delta}m}$, which corresponds to an equatorial view ($\theta = 90\arcdeg$). The peak-to-peak variation in lightcurve becomes larger when we increase the solar phase angle. The following empirical relationship between those two parameters was found by \citet{zap90}. 

\begin{equation}
A(0\arcdeg) = \frac{A(\alpha)}{(1+m\alpha)}
\end{equation}
\\
where $A(0\arcdeg)$ is the lghtcurve amplitude at zero phase angle and $A(\alpha)$ is the amplitude measured at solar phase angle $=\alpha$. These authors also found that average {\it m}-values are 0.030, 0.015, and 0.013 degree$^\mathrm{-1}$ for S, C, and M-type asteroids, respectively. We adopted the constant {\it m} of 0.03 for our analysis, as Maria family is known as a typical S-type asteroid family.

\subsection{Lightcurve inversion with dense and sparse data}

In order to derive the orientation of the spin axis using disk-integrated photometric data, dense lightcurves obtained over three or four apparitions are essential \citep{kaa02}. However, due to limited amount of dense datasets, the lightcurve inversion method using only sparse data \citep{kaa04} and a combination of sparse and dense data \citep{dur09,han11} has been improved during the past decades.

We obtained the available dense dataset by matching objects with the Maria family members from the following three data sources. In the MPC (Minor Planet Center) Light Curve Database\footnote{http://www.minorplanetcenter.net/light\_curve}, we downloaded lightcurves for the 17 Maria asteroids. From the online website of Courbes de rotation d'ast\'ero\"ides et de com\`etes (CdR\footnote{http://obswww.unige.ch/$\sim$behrend/page\_cou.html}, operated by R. Behrend), a total of 140 individual lightcurves for 13 object were acquired. Another 3 lightcurves were found in the Asteroid Photometric Catalog \citep[APC,][]{lag11} from the NASA's Planetary Data System (PDS). 

Most sparse photometric datasets are a by-product of world-wide astrometric surveys, such as the Catalina Sky Survey and the Siding Spring Survey. Detailed instructions for obtaining the sparse data from the AstDyS site are in \citet{han11}; we follow their methodology. According to very recent analysis by the same authors (priv. communication), only sparse datasets from the USNO in Flagstaff (MPC code 689), the Catalina Sky Survey (703), and the La Palma (950) were useful for determining the shape modeling of asteroids. Combining the dense and the sparse lightcurve datasets from the AstDyS, we computed the pole orientation for 16 objects, and determined 13 unique solutions.

\section{Results and discussions}

\subsection{Rotational properties of Maria asteroid family}

Our observations from 7 observatories yielded 218 individual lightcurves for 74 Maria asteroids. Among them we derived synodic rotational periods for the 51 members with a reliability code $\geq$ 2. For the 23 objects with a reliability code of 1, we set a lower bound to the periods except for 879 Ricarda for which the rotation period is known. The reliability parameters follow the definition by \citet{lag89}:\\
1. Very tentative result, may be completely wrong.\\
2. Reasonably secure result, based on over half coverage of the lightcurve.\\
3. Secure result, full lightcurve coverage, no ambiguity of period.\\
4. Multiple apparition coverage, pole position reported.

The information on rotational characteristics from our lightcurve datasets with other physical properties is summarized in Table 3; diameter and albedo information were mainly acquired from the WISE IR data \citep{mas11} with the exception of those marked with A \citep[AcuA, Asteroid catalog using AKARI;][]{usu11} and M (Mean albedo value of 0.254 for the Maria family asteroid), while the taxonomic information is compiled from \citet{nee10} and \citet{has12}. We used the datasets only with a reliability code $\geq$ 2 for our analysis. The lightcurves for 23 objects denoted with Q Notes of 1 did not cover even half-phased periods due to following three reasons: (1) very low amplitude of the lightcurve (smaller than 3-sigma); (2) variation of brightness much longer than observing time; (3) data interruption due to bad weather or instrument conditions. We examined a lower bound to the periods for those 23 objects to consider the influence of selection bias, i.e., favoring observations of faster rotators. Although all 23 objects rotate slower than 5 hours, they do not affect the overall tendency of our results, and besides, the peak-to-peak variation derived from completely wrong lightcurves could negatively affect the statistics. This is the reason why we only select periods only with a reliability code $\geq$ 2 for our analysis.

In order to improve the significance of the statistical analysis of the rotational properties in the Maria family, we searched the published lightcurve data from the Asteroid Lightcurve Database (LCDB) released in March 2013 and matched 58 objects with the Maria family asteroids. Out of 58 LCDB data sets, there are 17 objects overlapping with our observations. Therefore, we checked those original lightcurves from literatures, and then adopted lightcurves with minimum period gaps between the LCDB and our observations.

Figure 2 shows the distribution of rotational rates for 92 Maria family members used in this study and the corresponding best fit Maxwellian curve. The 51 objects obtained from our observations are marked with shaded bars. It can be obviously seen that asteroids rotating faster and slower considerably exceed the fitted distribution. From the Kolmogorov-Smirnov test, the compatibility between the observed distribution and the fitted Maxwellian is completely inconsistent at a 92\% confidence level. This non-Maxwellian distribution is also found in other old-type asteroid families, such as the Koronis family \citep{sli08} and the Flora family \citep{kry12}, with ages of 2.5 $\pm$ 1.0 Gyr and 1.0 $\pm$ 0.5 Gyr \citep{nes05}, respectively.

Laboratory experiments of catastrophic collisions \citep[and references therein]{hol02} and numerical simulations \citep{asp99,mic01} showed that the rotational frequency of fragments approximate the Maxwellian distribution. Accordingly, this inconsistency with our measurements suggests that the members belong to an old family and have had their rotational properties modified by non-gravitational forces operating over a long period of time; the spin rate change, in particular, can be attributed to the YORP effect. \citet{rub00} found that the spin states of pseudo-Gaspra and pseudo-Eros can significantly evolve on a 100 Myr timescale, which is far shorter when compared to their break-up and collisional processes. \citet{han11} found that the spin vectors of the Main-belt asteroids with D $< 30$ km were significantly affected by the YORP thermal effect. The size range of most of the 92 Maria members used in this study is distributed between 1.5 km and 30 km (see Fig. 3).

We examine the correlation between the rotational frequency (cycle/day) and diameter (km) in Fig. 4. Among the 92 Maria family asteroids used for this study, four objects might be regarded as interlopers. In terms of the visible geometric albedo $p_v$, two asteroids 3094 Chukokkala and 4860 Gubbio possess very low albedos; 0.068 and 0.037, respectively, provided that the mean value of albedo for the Maria family asteroids is 0.254 from the matched 1,152 WISE IR data. The other two members have a different spectral type of Xe for 1098 Hakone \citep{nee10}, C-type for 71145 (1999 XA183) \citep{has12}. Those possible interlopers are marked with filled boxes, however, do not affect the overall tendency in Fig. 4. 

In the results obtained from our observations, there is an apparent systematic trend toward larger dispersion of rotational frequency with decreasing size. The rotation of small asteroids can be easily accelerated or decelerated due to the YORP effect. Moreover, all asteroids larger than 22 km rotate more slowly than 4.8 hr (5 cycles per day), quite similar to what was found for the Flora family \citep{kry12}. 

Figure 5 shows the correlation between the rotational frequencies (cycle/day) and the amplitudes of lightcurve (peak-to-peak variation magnitude) that represents the overall shapes of asteroids. We derived amplitudes of lightcurves using the equation (3) for our observations. For lightcurve data from the LCDB, we adopted the maximum amplitudes of lightcurves. The various sizes of circles indicate the diameter of each asteroid relatively; asteroids larger than a diameter of 15 km are marked with blue circles.

We can see two features in Fig. 5: there is no object with both fast rotation (faster than 6 cycles per day) and large lightcurve amplitudes ($> 0.6$ mag). Despite insufficient number for large peak-to-peak variation, few highly elongated objects tend to rotate slowly. This could be explained by the break-up limit of elongated rubble pile; it makes sense that the more elongated an object is, the easier it can be shattered during its spin-up process. The colored curves in Fig. 5 approximate the critical rotational period ($P_c$) for bulk densities of 1.5, 2.0, and 2.5 g/cm$^3$, respectively, adopted from \citep{pra00}:

\begin{equation}
P_c \approx 3.3 \sqrt{\frac{1+\Delta m}{\rho}} 
\end{equation}
\\
where $P_c$ is the period in hours, $\rho$ρ is the bulk density in g/cm$^3$, $\Delta${\it m} is the amplitude of the lightcurve variation in magnitude. It is apparent from Fig. 5 that there is no object against ``spin barrier'' \citep{pra00} for bulk density of 2.5 g/cm$^3$. In addition, elongated objects are located far from the spin rate limit.

The other feature seen in Fig. 5 is that no objects larger than 15 km have amplitude of lightcurve larger than 0.5 magnitude except for 4860 Gubbio, regarded as an interloper. Small objects ($< 15$ km) with various shapes are spread out in this plot. If an elongated rubble pile asteroid is disrupted by a non-gravitational force such as the YORP effect, it might have less elongated, namely more spherical shape as a result. Figure 5 separates into two plots (Fig. A. 62 and Fig. A. 63) in Appendix A.

\subsection{Yarkovsky footprints on Maria asteroid family}

The Yarkovsky effect plays a significant role in the dynamical evolution of asteroid orbits. The study of this non-gravitational thermal force on the asteroid family has been improved dramatically during the past decades in several families, such as Koronis \citep{bot01}, Flora \citep{nes02}, and Eos \citep{vok06}. Semi-major axis drift, in accordance with either the sense of rotation or various sizes of asteroids, by the Yarkovsky effect results in a V-shape in the proper semi-major axis and absolute magnitude plane \citep{nes04,vok06,mil10}. The analysis for drift in semi-major axis allows us to estimate the age of an asteroid family \citep{bot01,nes05}. 

To find the Yarkovsky footprints in the Maria asteroid family, we investigate the pole orientation of the members. We examined sense of rotation for the 16 members which were observed during at least three apparitions, and successfully determined the pole orientations for the 13 objects by the combination with sparse data from the AstDyS site. The summary of pole solutions for the Maria family asteroids is listed in Table 4. Due to an intrinsic symmetry of the problem \citep{kaa02}, ground-based observations of objects with a small orbital inclination are affected by an ambiguity in the determination of the ecliptic longitude of the pole axis direction, resulting in two solutions that are placed 180 degrees apart and are statistically indistinguishable. We adopted the lower chi-squared solution marked with boldface for this study and estimated the uncertainty of the pole orientation to be 5 $\sbond$ 20 degrees, depending on the number of dense and sparse datasets. Out of 13 asteroids, the information on spin axis of 4 objects was also found in DAMIT\footnote{http://astro.troja.mff.cuni.cz/projects/asteroids3D} \citep[Database of Asteroid Models from Inversion Techniques;][]{dur10}. The result of our analysis for those 4 objects is quite consistent with that of DAMIT. 

Concerning the pole orientation, we concentrate on the determination of the sense of rotation. As a result, we found six prograde rotators (652 Jubilatrix, 787 Moskva, 875 Nymphe, 1158 Luda, 1996 Adams, and 3786 Yamada); while two objects (695 Bella and 897 Lysistrata) rotate in the opposite sense. We show the position of prograde and retrograde asteroids on a plane of proper semi-major axis (AU) versus absolute magnitude (H) in Fig. 6. In addition, we found five rotators with the pole along the ecliptic; 170 Maria, 575 Renate, 660 Crescentia, 727 Nipponia, and 6458 Nouda marked with open circles in Fig. 6. The spin axes of those five objects are very close to the ecliptic plane within their statistical errors, between 5 and 20 degrees. Although an observational bias cannot be excluded, the excess of prograde rotators with respect to retrograde rotators could be explained with a long-term evolution by the Yarkovsky effect. In case of retrograde rotators, the Yarkovsky drift causes the semi-major axis of family members to decrease; consequently, they could have been ejected by the 3:1 MMR to the inner Solar System. \citet{kry13} found a similar result for the Flora family, which is located near the outer border of the $\nu_6$ resonance area. 

Recent statistical studies by \citet{pao12} indicate that there is an excess of prograde versus retrograde rotators in the Main-belt for asteroids smaller than 100 km. However, these authors did not find any convincing explanations for this excess and hence this remains an open problem. On the contrary, a strong excess for retrograde rotation was found in NEAs that is completely consistent with a theoretical ratio of 2 $\pm$ 0.2 with respect to prograde rotators \citep{las04}. Regarding this prograde-retrograde asymmetry in MBAs and NEAs, they inferred that there is a connection between the excess of retrograde NEAs and the deficiency of retrograde MBA due to the Yarkovsky effect. However, it might be more complicated to generalize the trend to MBAs as they are too big to have been affected by the Yarkovsky drift. Another supporting example we can find is the largest NEA, 1036 Ganymed that is regarded to be originated from the Maria family. \citet{zap97} carried out an extensive analysis of the Maria family, and suggested that about 10 objects with the size between 15 and 30 km have probably been injected into the 3:1 mean-motion resonance with Jupiter. They proposed that a couple of giant S-type NEAs, 433 Eros and 1036 Ganymed have probably been originated from the Maria family. The spectral reflectance of this asteroid is quite similar to the average spectrum of the Maria family members \citep{fie11}; interestingly its spin orientation is retrograde \citep{kaa02}, moreover, it is on a high-inclination orbit of 26.7 degrees.

We can easily distinguish three interlopers (that is, two lower albedo objects of 3094 Chukokkala, 4860 Gubbio and Xe-type asteroid of 1098 Hakone) from Fig. 6, while 71145 (1999 XA183) is not close to the border of the family. One distinct property of the Maria family compared to other densely populated asteroid families such as Flora, Themis, and Eunomia, is that there is no prominent large body among the family members. The largest body in this family is not 170 Maria but 472 Roma with a diameter 50.3 km as calculated from the WISE IR data. In the recent paper by \citet{mas13}, the Roma family is substituted for the name of the Maria family. 

Figure 7 shows the spin vector obliquity of the Maria family members with respect to the rotational frequency (cycle/day), compared with the Koronis \citep{sli09} and Flora families \citep{kry13}. In those two families, there is a conspicuous group of prograde objects. Furthermore, retrograde objects share almost the same obliquities, which is referred to as the Slivan effect \citep{sli02}. In the Maria family, on the other hand, no prominent prograde groups being in the Slivan state were found, while the data for retrograde rotation is not sufficient for a conclusion. \citet{vok03} investigated several test prograde rotators chosen with various ranges of either semi-major axis or inclination in the Main-belt. Their preliminary results indicate that asteroids with low-inclination in the outer Main-belt, such as 24 Themis, might be trapped efficiently in the Slivan state. This is the reason why they suggest observations of either high-inclination or inner Main-belt asteroid and numerical simulations for their dynamical evolution. \citet{sko02} found that orbital inclination of Main-belt asteroids is the most important factor that affects the magnitude of the spin vector variation. Based on their dynamical spin evolution, if the orbital inclination is increased, also the maximum obliquity variation increases linearly. 

The other feature seen in Fig. 7 is that there are several objects with spin vector obliquities around 90 degrees, that is, spin axes approximately parallel to the orbital plane. The change of semi-major axis by the diurnal component of the Yarkovsky effect is proportional to the cosine of the obliquity of an asteroid, so it will vanish when the obliquity approaches 90 degrees. We can predict that the semi-major axis drift of these five objects plotted as open circles in Fig. 6 and Fig. 7 will not occur. However, we can expect the spin vector can be modified by the YORP effect. 

\subsection{NEA source region}

The 3:1 mean-motion resonance ($\sim$ 2.5 AU) is the most prominent Jovian resonance region in the Main-belt, and, as such, it is an important source region of NEAs and meteors. The Maria family is located quite close to the outer border of the 3:1 MMR with Jupiter. When members of the Maria family enter this unstable region due to the Yarkovsky drift, they can escape from its membership and become NEAs. For this reason, the distribution of proper semi-major axis vs. absolute magnitude is asymmetric as shown in Fig. 6. The \citet{bot02} model suggests that with 20 \% probability, the population of all near-Earth objects (with absolute magnitude H $< 18$) originated from the 3:1 MMR region. They estimated that the number of kilometer-sized NEOs (H $< 18$) escaped from the 3:1 MMR every million year is 100 $\pm$ 50 bodies. Similar results independently derived by \citet{mor03} show that 100 to 160 objects larger than 1 km enter the 3:1 MMR per million years. 

In order to make an order-of-magnitude estimation of the flux of the Maria members into the 3:1 MMR region, we try to determine the number of missing Maria family members. We do this by taking the difference of the original members (assuming the distribution was originally entirely symmetric) and the number of currently known objects (the absolute magnitude of all members is brighter than 18th mag). To obtain the number of original members, we assume that the center of the Maria family is at about 2.562 AU (denoted with the red dashed line) since the number density with respect to the semi-major axis is at a minimum. Then we double the number of family member, with proper semi-major axis $> 2.562$ AU as in Fig. 6. This results in a total of 5,368 original members. With this procedure, our preliminary result implies that 2,138 Maria asteroids had been injected into the 3:1 MMR region through its dynamical evolution age of 3 $\pm$ 1 Gyr. It is well known that the dynamical lifetime of objects placed in orbital resonances is only a few million years \citep{gla97}. Their numerical simulations for 3:1 MMR injected particles from the Maria family show that the fraction of particles experiencing the end-states of 1) being injected into the inner solar system and eventually colliding with the Sun or 2) being injected into a Jupiter-crossing orbit and being eventually removed from the solar system are 70 \% and 29 \%, respectively. Therefore, we may conclude that roughly 1,500 objects from the Maria family had been injected into the inner Solar System during 3 $\pm$ 1 Gyr, i.e., 37 to 75 Maria asteroids larger than 1 km every 100 Myr. 

We also look into the 3:1 resonance neighborhood region to find possible candidates among Maria members residing in this unstable area. According to \citet{mor03}, the boundary of the 3:1 MMR region can be defined approximately by the following formulas (only for the right side of the 3:1 resonance):

\begin{mathletters}
\begin{eqnarray}
a = 2.508 + \frac{e}{29.615} \ \ for\ \ e \leq 0.15936, \\
a = 2.485 + \frac{e}{5.615} \ \ for\ \ e > 0.15936
\end{eqnarray}
\end{mathletters}

They found that the number density of asteroids sharply increases up to $\sim$0.015 AU from the resonance boundary, yet it is not changed considerably over the next 0.025 AU. \citet{gui02} also pointed out that there is a chaotic diffusion region in the range of 0.01 $\sbond$ 0.02 AU in semi-major axis from the 3:1 resonance border. Asteroids residing in this region could enter the resonance on a time scale of 100 Myr. The oscillation of the border of the resonance could also have taken place \citep{mor95,morb95,rob01}. 

Figure 8 represents all 3,230 Maria asteroids projected onto a plane of proper semi-major axis and proper eccentricity. The 3:1 MMR boundary denoted with black dashed line is defined by Eq. (5a). The chaotic diffusion region that has been shifted to 0.015 AU from the resonance border is represented by the red dashed line. The number of objects residing in this region is 34; they could fall into the 3:1 resonance in the coming 100 Myr, especially 114123 (2002 VX49) and 137063 (1998 WK1) marked with green open circles are the most promising candidates for becoming new NEAs. The number of 34 is surprisingly consistent with the range 37 $\sbond$ 75 we obtained before, within an error range of age estimation. Large filled blue and red circles stand for prograde and retrograde rotators, respectively. Prograde objects tend to position outward the semi-major axis compared with retrograde, due to the Yarkovsky effect. This can be also seen in Fig. 6.

\section{Conclusions}

We performed observations of the Maria family asteroids from 2008 July to 2013 May. From our observations for 134 nights, synodic rotational periods for 51 objects were acquired including newly derived periods of 34 asteroids. The rotational rates distribution of the Maria family members obtained from our datasets in addition to published lightcurve data from the LCDB (Asteroid Lightcurve Database) indicates that there is a considerable excess of fast and slow rotators. The two-sample Kolmogorov-Smirnov test rejects the hypothesis that the spin distribution matches that of a Maxwellian at a confidence level of 92\%. 

From correlations among spin rates, sizes, and overall shapes of asteroids, we conclude that rotational properties of the Maria family have been altered significantly by the YORP effect. Such a substantial change of the rotational characteristic is also observed in old asteroid families such as Koronis and Flora \citep[e.g.][]{sli08,kry12}. On the other hand, however, we could not detect the Slivan state in the Maria family members that have relatively higher inclination orbits between 12 $\degr$ and 17 $\degr$ than those of Koronis and Flora families.

The YORP effect is most effective in the spin-up or spin-down of small asteroids. Consequently, there is an apparent trend toward larger dispersion of rotational frequency with decreasing size in Fig. 4. Regarding the correlation between rotational frequencies and the lightcurve amplitude (see Fig. 5), we confirm that there is no object being both fast rotating (faster than 6 cycles per day) and having a large lightcurve amplitude (more than 0.6 mag) thus verifying that no evidence of objects having high tensile strengths.

Based on the lightcurve inversion method, we obtained pole orientations for 13 members to trace non-gravitational forces such as the Yarkovsky effect. The excess of prograde rotators with respect to retrograde is found with a ratio ($N_p/N_r$) of 3. The retrograde objects might have been injected into the inner Solar System by entering the 3:1 mean-motion resonance with Jupiter. Therefore, the overall trend of proper semi-major axis and absolute magnitude plot is non-symmetric. From our simplest arithmetic of the flux for the 3:1 MMR resonant objects from the Maria family, we estimate that approximately 37 to 75 Maria asteroids larger than 1 km have entered the inner Solar System every 100 Myr. 

For understanding these non-gravitational forces that are affecting the Maria asteroid family in detail, we need more samples in order to improve our statistical result. This is the reason why we are planning to extend our observational campaign with a network of telescopes in both hemispheres. The Yarkovsky/YORP model based on high accurate parameters such as spin status, size, albedo, density, and thermal conductivity is expected to shed new light on the dynamical evolution in the history of the inner Solar System.



\acknowledgments

We thank the refree of this paper, Alberto Cellino, for his valuable comments and suggestions that significantly improved the quality of this work.
M.J.K. would like to thank Josef Hanu\u{s} for his valuable comments in the asteroid shape modeling using sparse datasets. 
M.J.K. would like to thank Fumihiko Usui for matching family members with the WISE data. 
M.J.K. was partly supported by Korea Research Council of Fundamental Science \& Technology. 
Work at Yonsei was supported by the NRF grant 2011-0030875. 
Y.J.C. was partly supported by the collaborative research project between NRF-JSPS (2012010270).
T.C.H. gratefully acknowledges financial support from the Korea Research Council for Fundamental Science and Technology (KRCF) through the Young Research Scientist Fellowship Program, and also the support of the Korea Astronomy and Space Science Institute (KASI) grant 2013-9-400-00. We also thank the following (mostly amateur) astronomers for their observational material collected in the CdR-database and used in this paper : P. Antonini, M. Audejean, L. Bernasconi, J.-G. Bosch, C. Cavadore, M. Conjat, M. Diebold, J.-P. Godard, F. Hormuth, M. Husarik, L. Kurtze, A. Leroy, J. Linder, F. Mallman, F. Manzini, A. Martin, E. Morelle, R. Montaigut, R. Naves, R. Poncy, C. Rinner, G. Rousseau, R. Roy, O. Thizy, M. Tlouzeau, and B. Warner.






\clearpage

\appendix

\renewcommand{\thefigure}{A.\arabic{figure}}
\setcounter{figure}{0}

\section{Lightcurves of the individual object in the Maria asteroid family}

We present 61 individual lightcurves (with a reliability code $\geq$ 2) for a total of 51 Maria family members (Fig. A.1 $\sbond$ 61). All figures were drawn as composite lightcurve folded with their synodic rotational period. Horizontal axes of each lightcurve represent rotational phase and vertical axes are differential magnitude between instrumental magnitude of the asteroid and an average magnitude of comparison stars.

\begin{figure}
\epsscale{.63}
\plotone{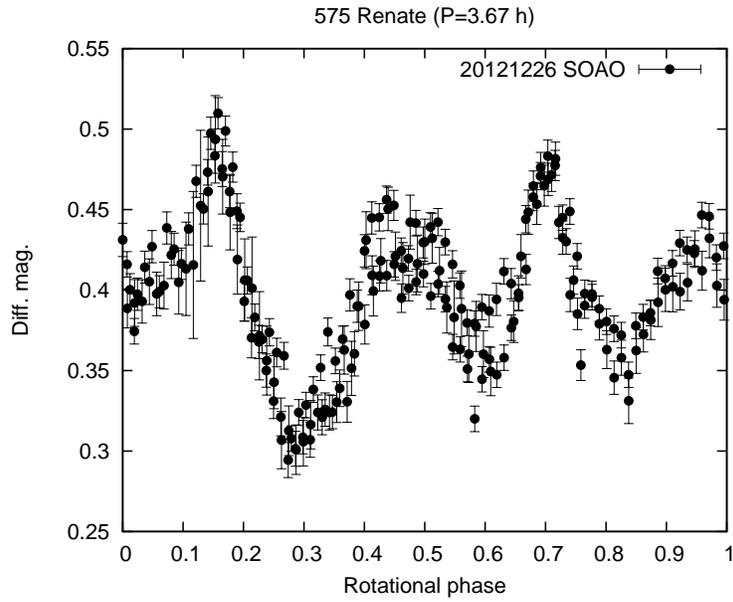}
\caption{Composite lightcurve of 575 Renate folded with the period of 3.67 h at the zero epoch of JD 2456287.9665778. \label{figA1}}
\end{figure}

\begin{figure}
\epsscale{.63}
\plotone{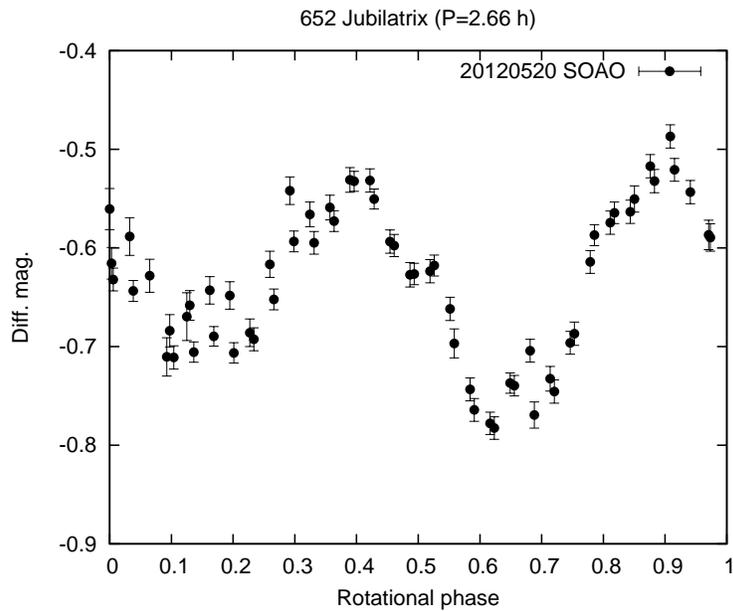}
\caption{Composite lightcurve of 652 Jubilatrix folded with the period of 2.66 h at the zero epoch of JD 2456068.0645664. \label{figA2}}
\end{figure}

\clearpage

\begin{figure}
\epsscale{.63}
\plotone{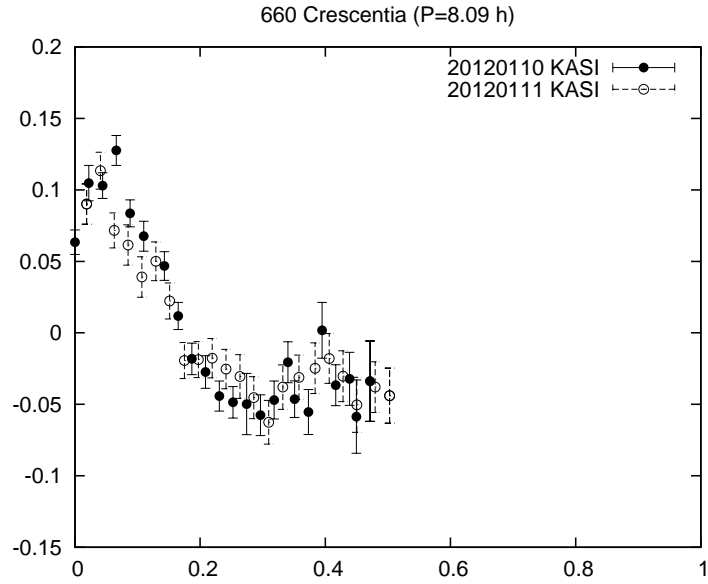}
\caption{Composite lightcurve of 660 Crescentia folded with the period of 8.09 h at the zero epoch of JD 2455936.90016201. \label{figA3}}
\end{figure}

\begin{figure}
\epsscale{.63}
\plotone{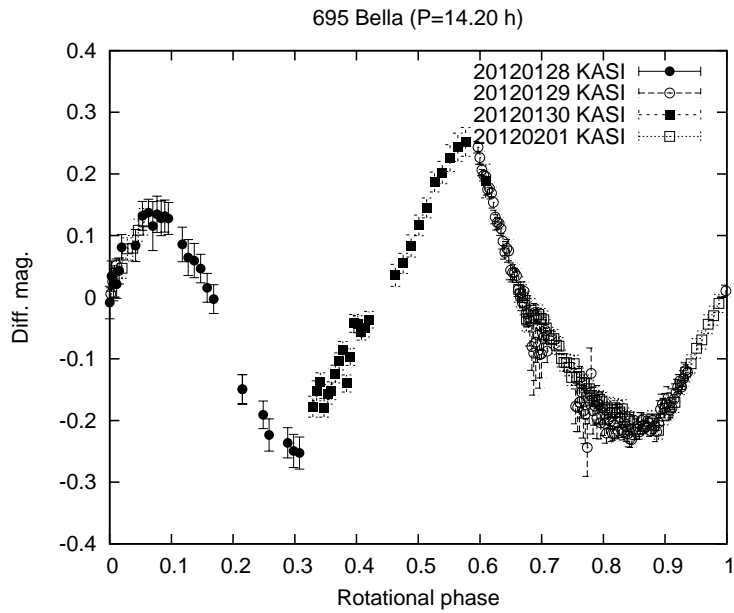}
\caption{Composite lightcurve of 695 Bella folded with the period of 14.20 h at the zero epoch of JD 2455955.15314811. \label{figA4}}
\end{figure}

\clearpage

\begin{figure}
\epsscale{.63}
\plotone{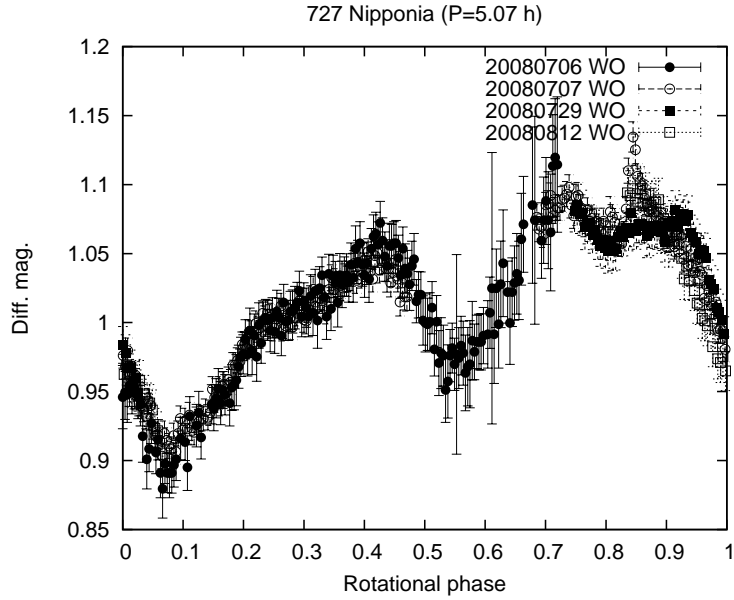}
\caption{Composite lightcurve of 727 Nipponia folded with the period of 5.07 h at the zero epoch of JD 2454654.271546. \label{figA5}}
\end{figure}

\begin{figure}
\epsscale{.63}
\plotone{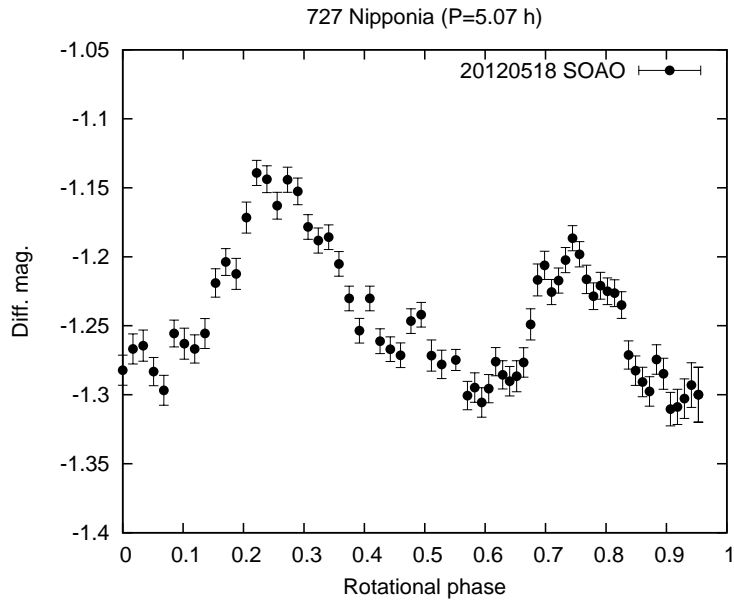}
\caption{Composite lightcurve of 727 Nipponia folded with the period of 5.07 h at the zero epoch of JD 2456066.1004983. \label{figA6}}
\end{figure}

\clearpage

\begin{figure}
\epsscale{.63}
\plotone{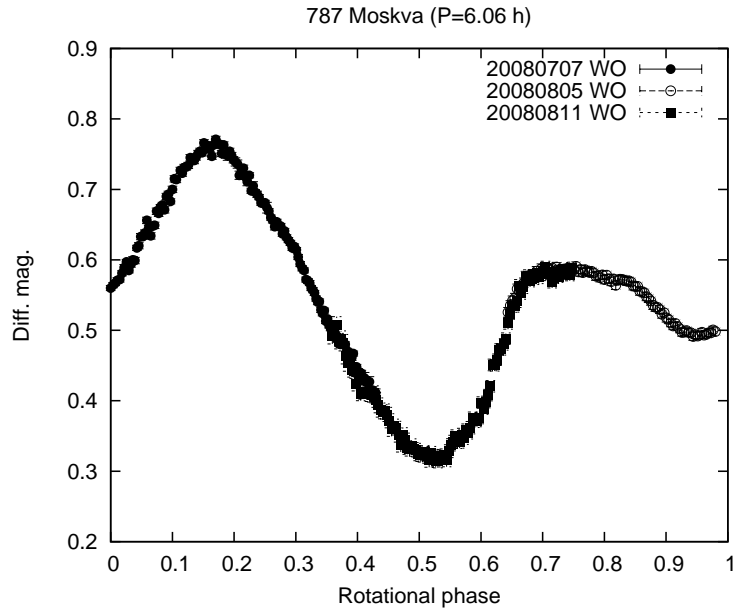}
\caption{Composite lightcurve of 787 Moskva folded with the period of 6.06 h at the zero epoch of JD 2454655.46699649. \label{figA7}}
\end{figure}

\begin{figure}
\epsscale{.63}
\plotone{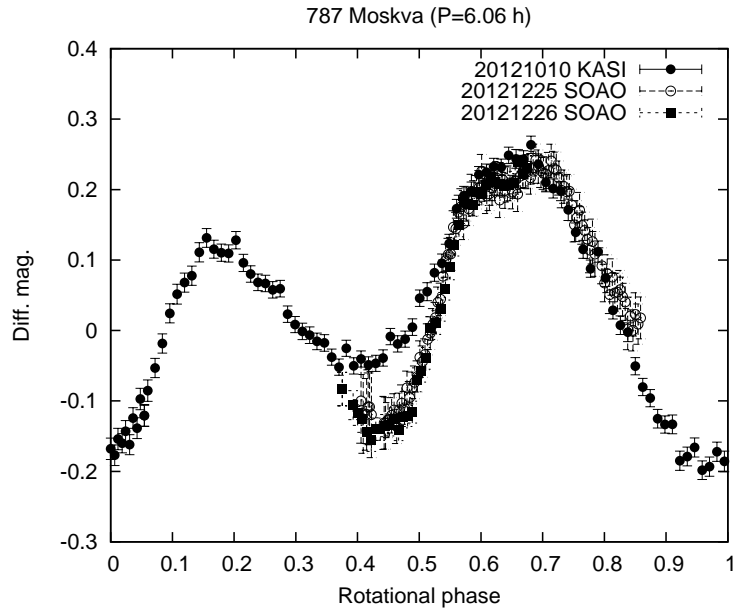}
\caption{Composite lightcurve of 787 Moskva folded with the period of 6.05 h at the zero epoch of JD 2456210.93379629. \label{figA8}}
\end{figure}

\clearpage

\begin{figure}
\epsscale{.63}
\plotone{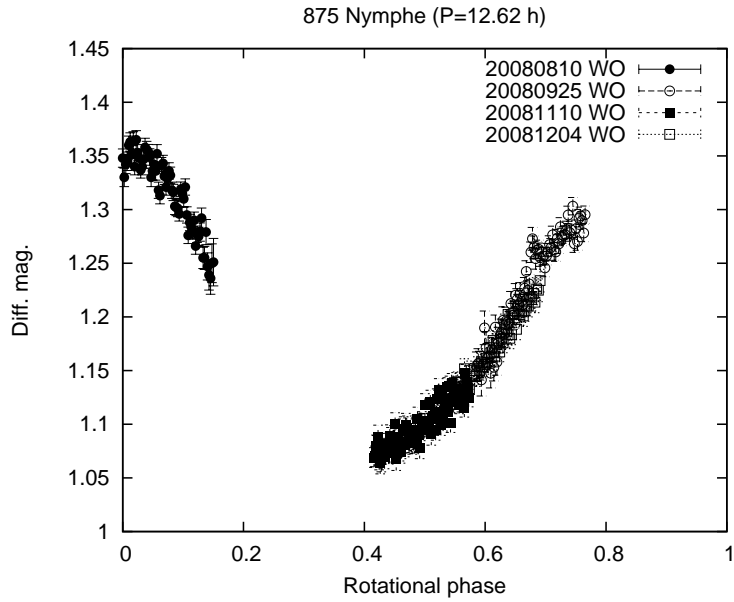}
\caption{Composite lightcurve of 875 Nymphe folded with the period of 12.62 h at the zero epoch of JD 2454689.52207176. \label{figA9}}
\end{figure}

\begin{figure}
\epsscale{.63}
\plotone{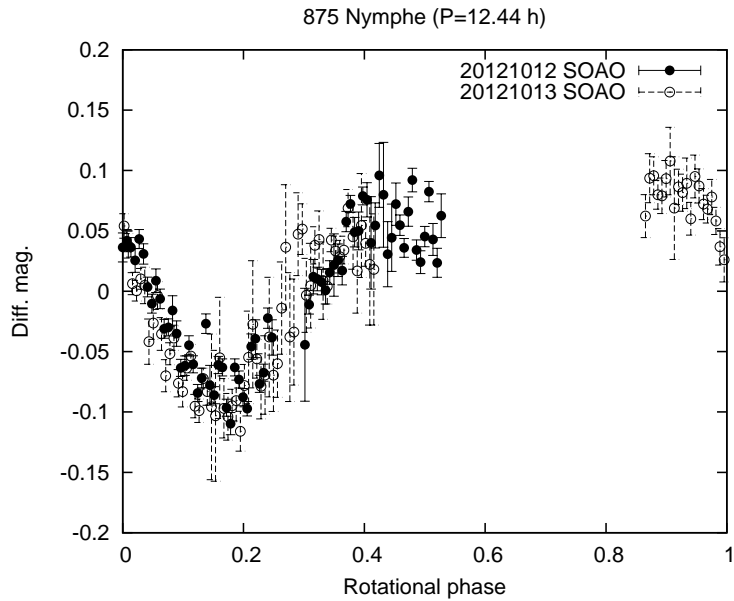}
\caption{Composite lightcurve of 875 Nymphe folded with the period of 12.62 h at the zero epoch of JD 2456213.0821109. \label{figA10}}
\end{figure}

\clearpage

\begin{figure}
\epsscale{.63}
\plotone{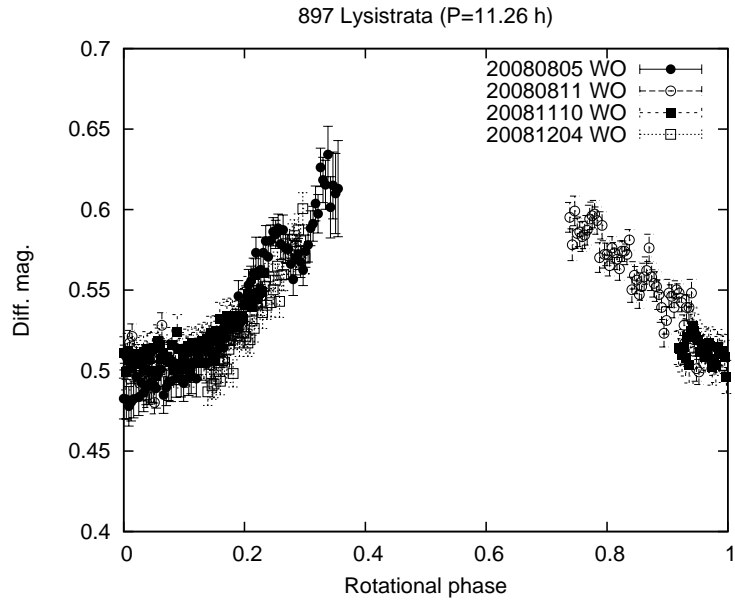}
\caption{Composite lightcurve of 897 Lysistrata folded with the period of 11.26 h at the zero epoch of JD 2454684.48635417. \label{figA11}}
\end{figure}

\begin{figure}
\epsscale{.63}
\plotone{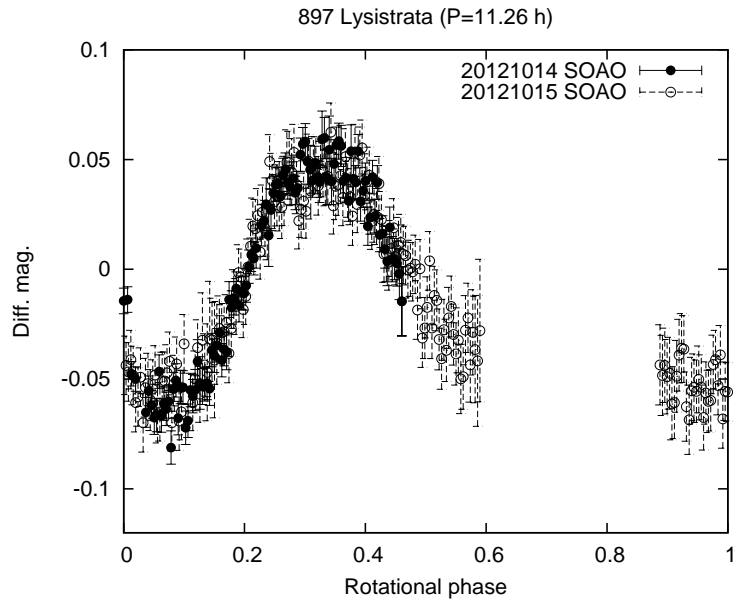}
\caption{Composite lightcurve of 897 Lysistrata folded with the period of 11.26 h at the zero epoch of JD 2456215.1489845. \label{figA12}}
\end{figure}

\clearpage

\begin{figure}
\epsscale{.63}
\plotone{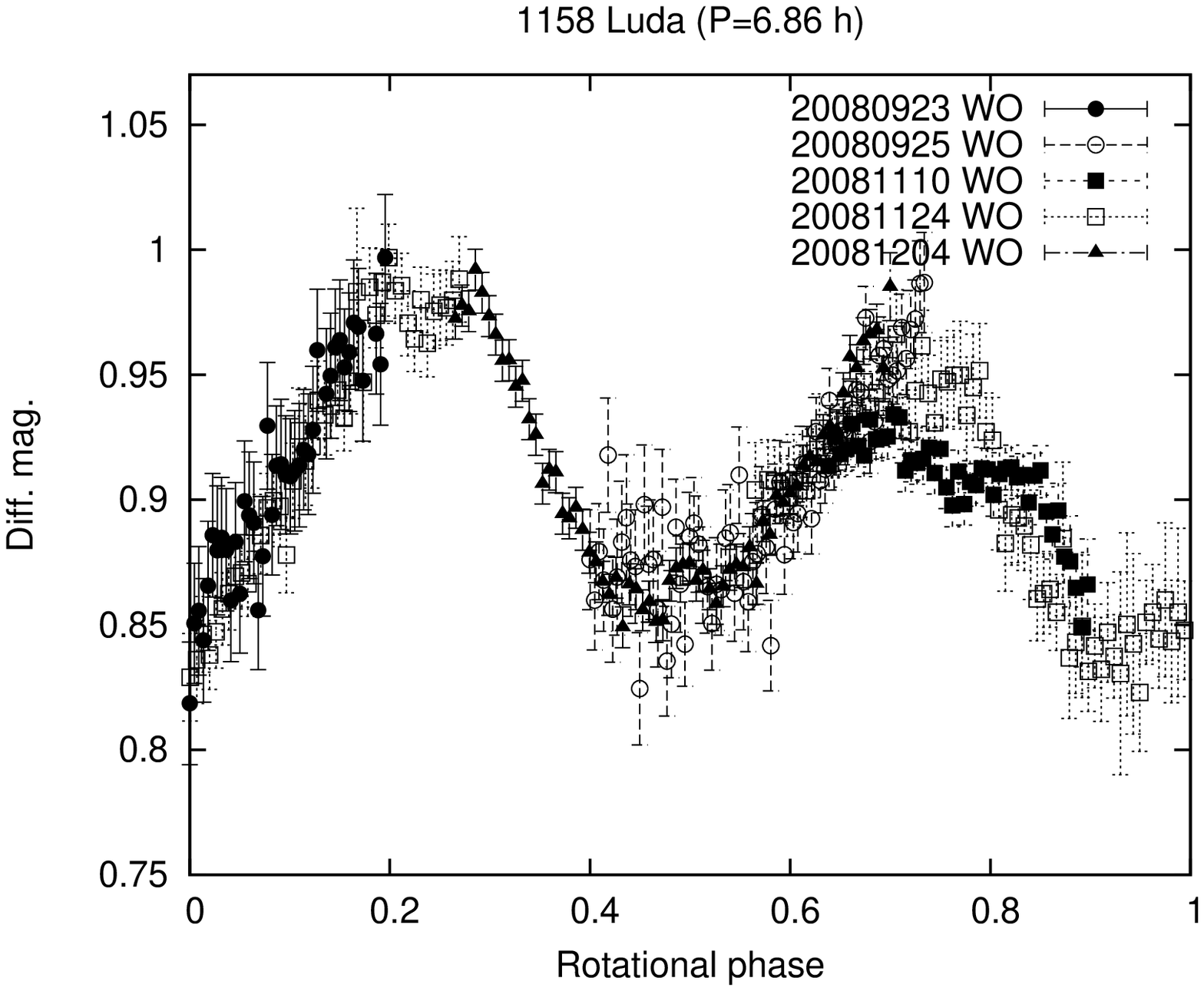}
\caption{Composite lightcurve of 1158 Luda folded with the period of 6.86 h at the zero epoch of JD 2454733.5603125. \label{figA13}}
\end{figure}

\begin{figure}
\epsscale{.63}
\plotone{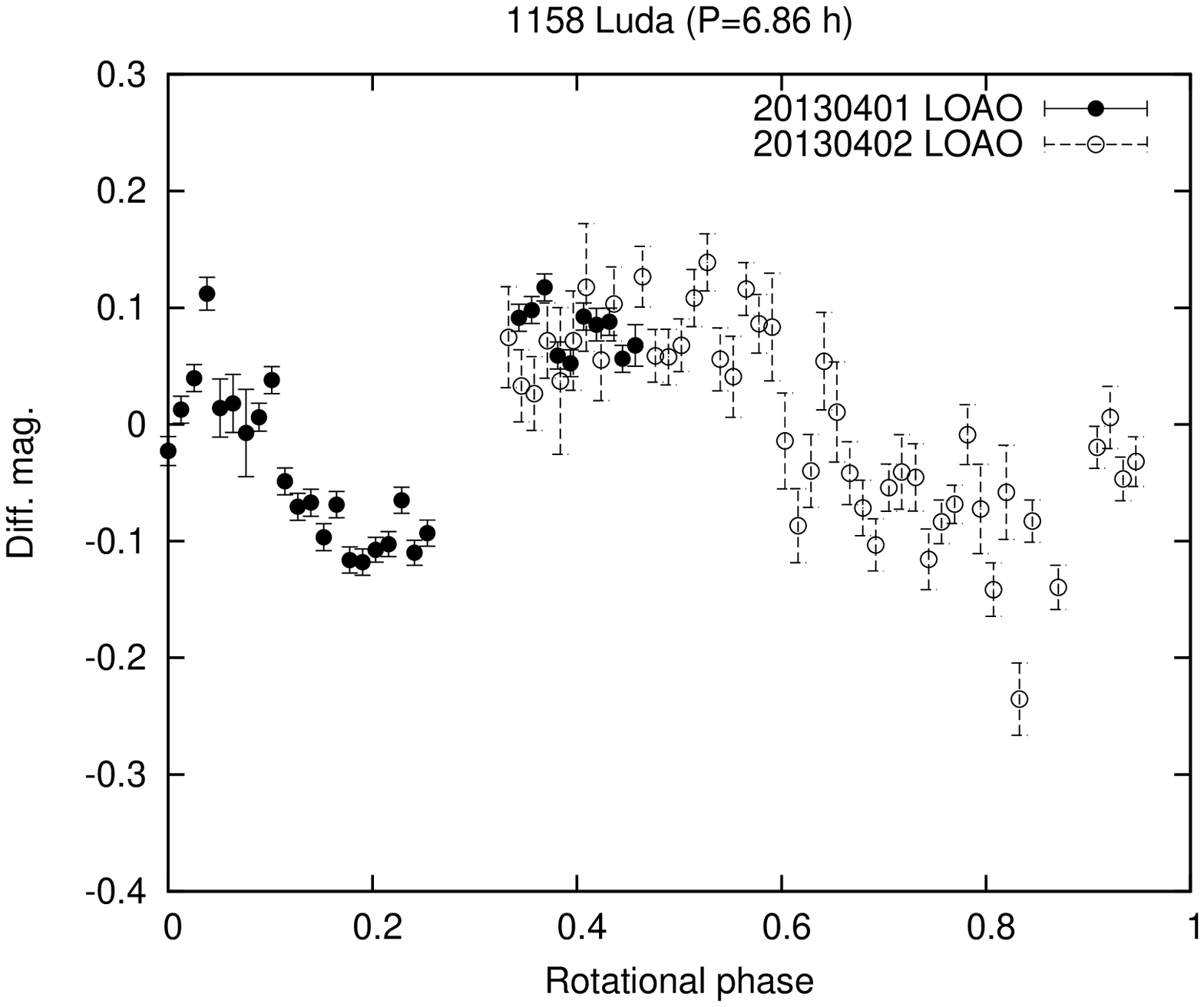}
\caption{Composite lightcurve of 1158 Luda folded with the period of 6.86 h at the zero epoch of JD 2456383.65252. \label{figA14}}
\end{figure}

\clearpage

\begin{figure}
\epsscale{.63}
\plotone{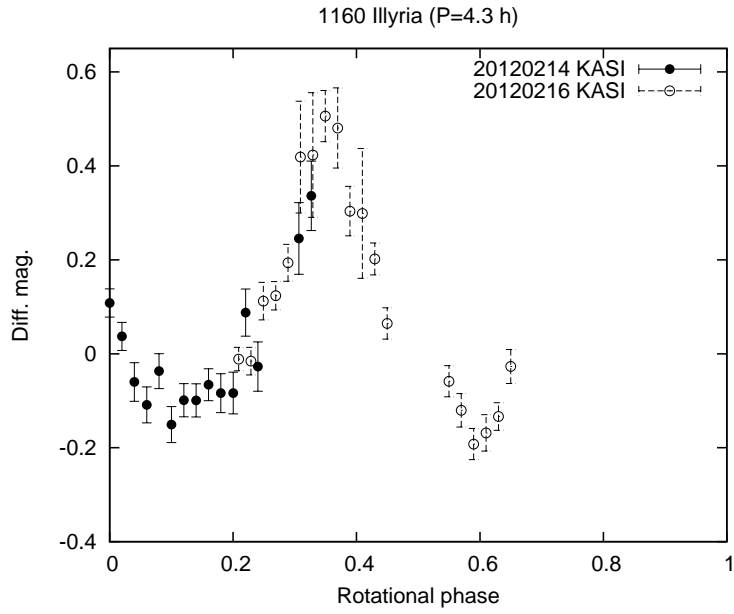}
\caption{Composite lightcurve of 1160 Illyria folded with the period of 4.3 h at the zero epoch of JD 2455971.92340279. \label{figA15}}
\end{figure}

\begin{figure}
\epsscale{.63}
\plotone{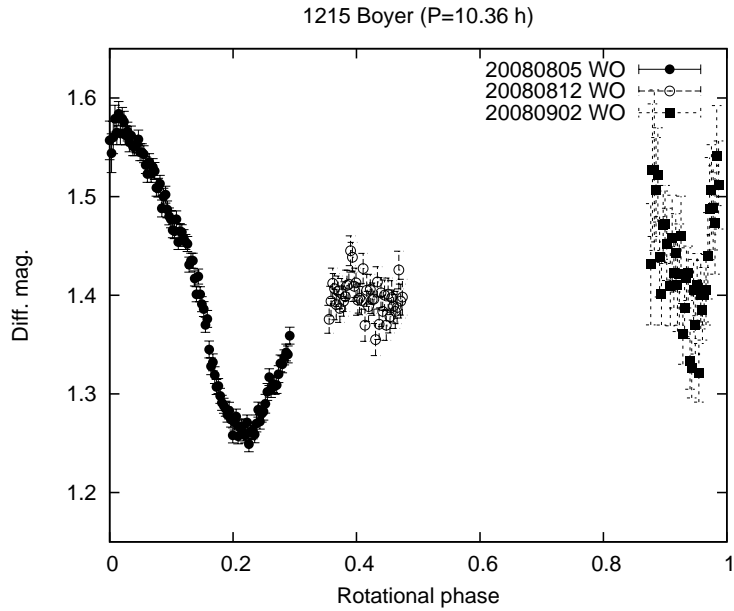}
\caption{Composite lightcurve of 1215 Boyer folded with the period of 10.36 h at the zero epoch of JD 2454684.22641206. \label{figA16}}
\end{figure}

\clearpage

\begin{figure}
\epsscale{.63}
\plotone{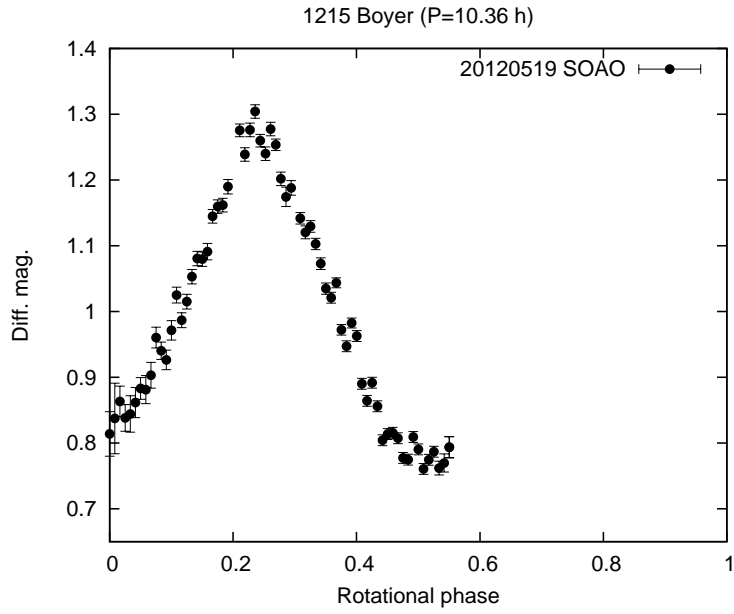}
\caption{Composite lightcurve of 1215 Boyer folded with the period of 10.36 h at the zero epoch of JD 2456067.0625117. \label{figA17}}
\end{figure}

\begin{figure}
\epsscale{.63}
\plotone{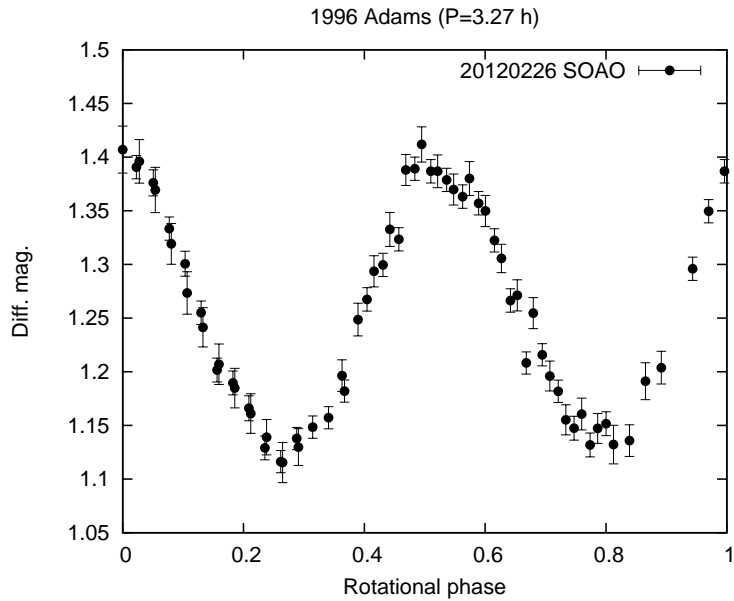}
\caption{Composite lightcurve of 1996 Adams folded with the period of 3.27 h at the zero epoch of JD 2455983.9871538. \label{figA18}}
\end{figure}

\clearpage

\begin{figure}
\epsscale{.63}
\plotone{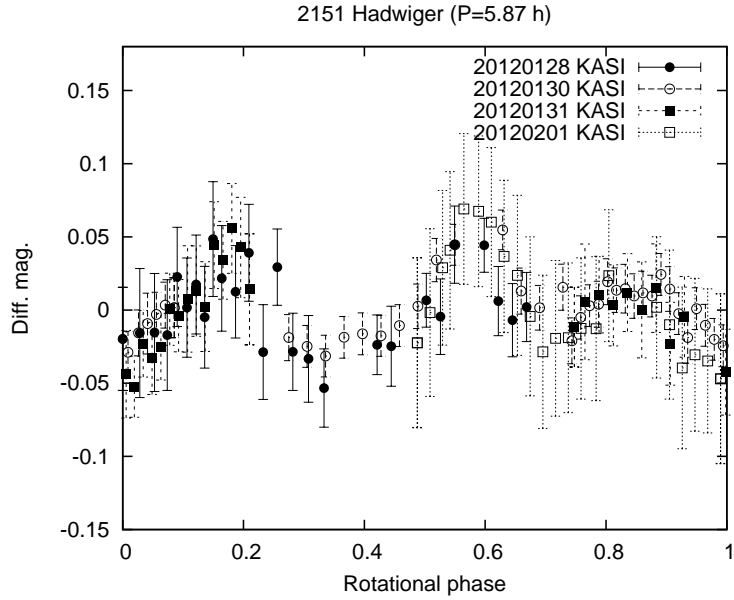}
\caption{Composite lightcurve of 2151 Hadwiger folded with the period of 5.87 h at the zero epoch of JD 2455955.174537. \label{figA19}}
\end{figure}

\begin{figure}
\epsscale{.63}
\plotone{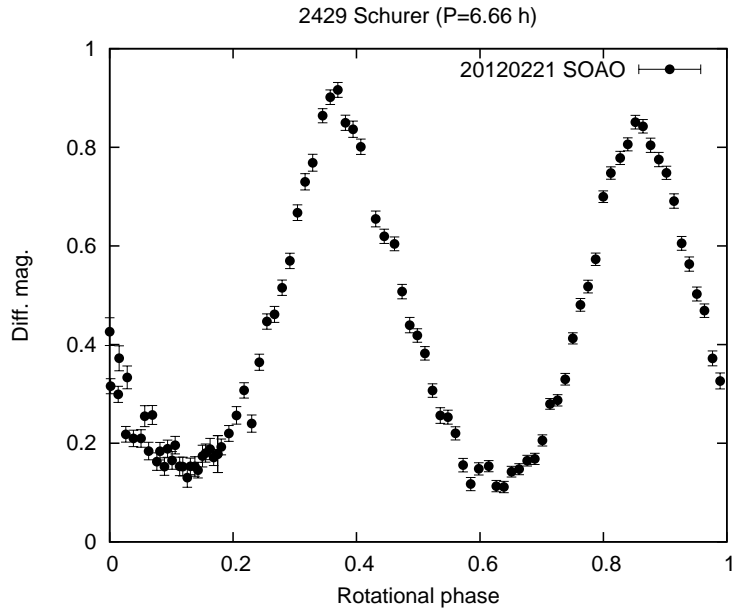}
\caption{Composite lightcurve of 2429 Schurer folded with the period of 6.66 h at the zero epoch of JD 2455979.0384352. \label{figA20}}
\end{figure}

\clearpage

\begin{figure}
\epsscale{.63}
\plotone{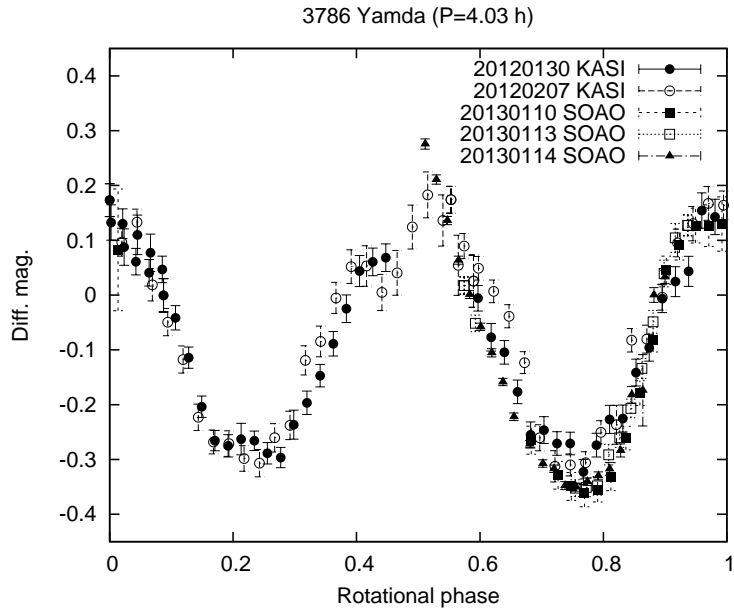}
\caption{Composite lightcurve of 3786 Yamada folded with the period of 4.03 h at the zero epoch of JD 2455956.92598379. \label{figA21}}
\end{figure}

\begin{figure}
\epsscale{.63}
\plotone{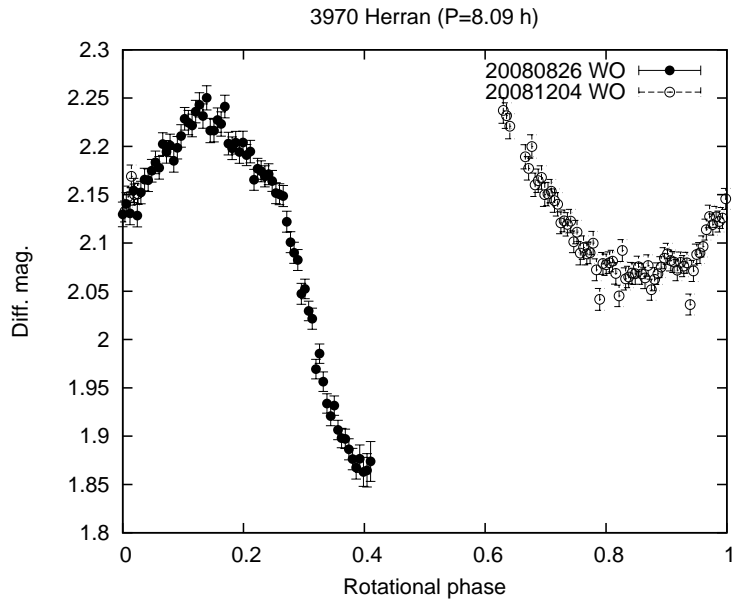}
\caption{Composite lightcurve of 3970 Herran folded with the period of 8.09 h at the zero epoch of JD 2454705.49471068. \label{figA22}}
\end{figure}

\clearpage

\begin{figure}
\epsscale{.63}
\plotone{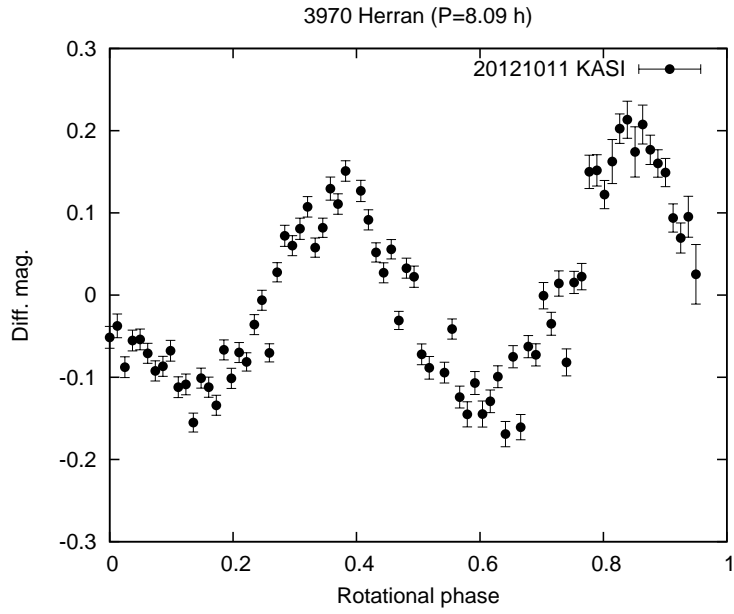}
\caption{Composite lightcurve of 3970 Herran folded with the period of 8.09 h at the zero epoch of JD 2456212.020706. \label{figA23}}
\end{figure}

\begin{figure}
\epsscale{.63}
\plotone{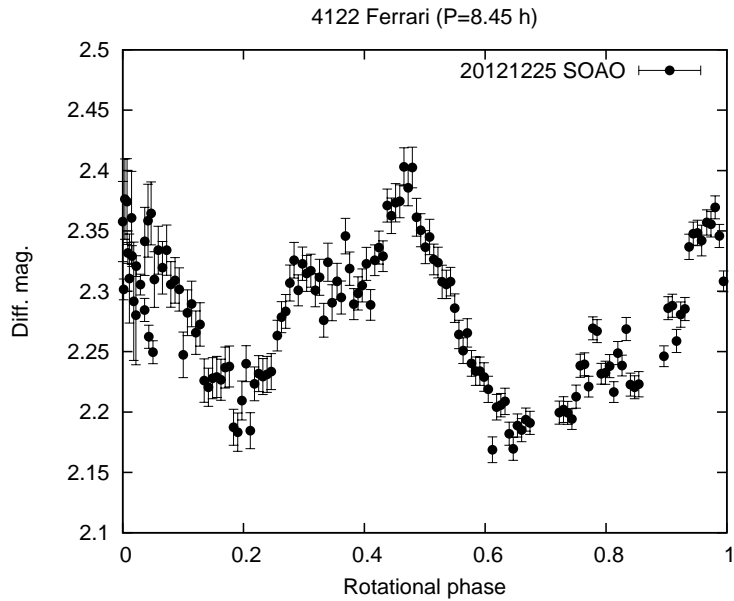}
\caption{Composite lightcurve of 4122 Ferrari folded with the period of 8.45 h at the zero epoch of JD 2456287.0039468. \label{figA24}}
\end{figure}

\clearpage

\begin{figure}
\epsscale{.63}
\plotone{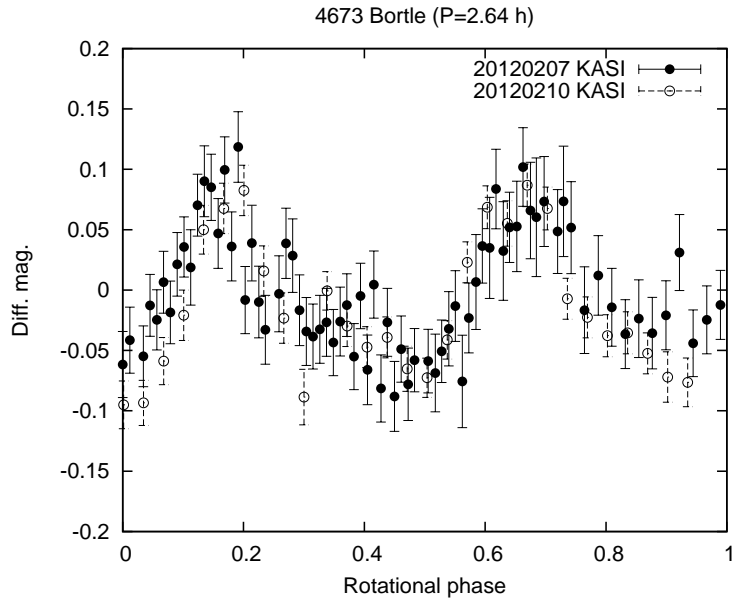}
\caption{Composite lightcurve of 4673 Bortle folded with the period of 2.64 h at the zero epoch of JD 2455965.21822912. \label{figA25}}
\end{figure}

\begin{figure}
\epsscale{.63}
\plotone{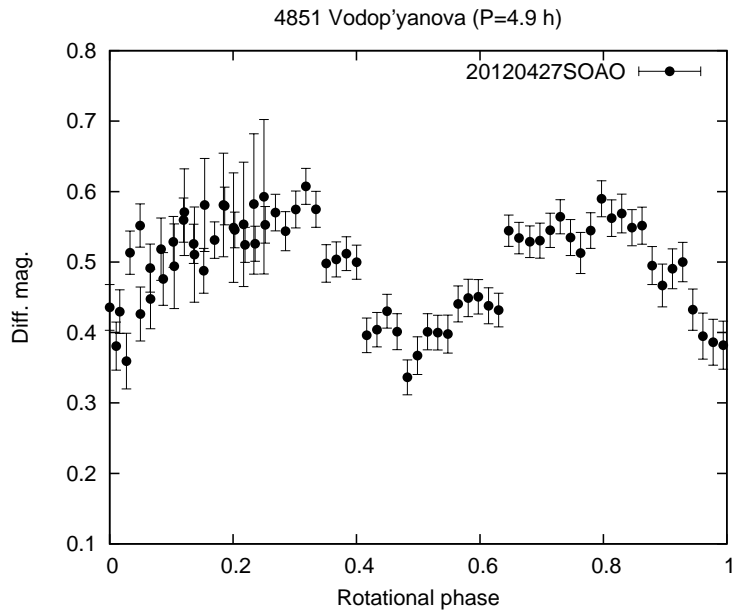}
\caption{Composite lightcurve of 4851 Vodop'yanova folded with the period of 4.9 h at the zero epoch of JD 2456044.9942261. \label{figA26}}
\end{figure}

\clearpage

\begin{figure}
\epsscale{.63}
\plotone{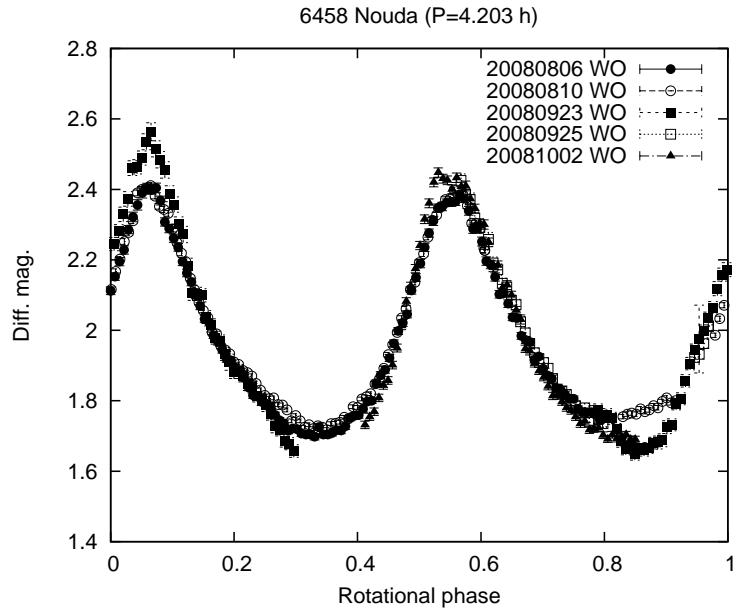}
\caption{Composite lightcurve of 6458 Nouda folded with the period of 4.203 h at the zero epoch of JD 2454685.3756149918. \label{figA27}}
\end{figure}

\begin{figure}
\epsscale{.63}
\plotone{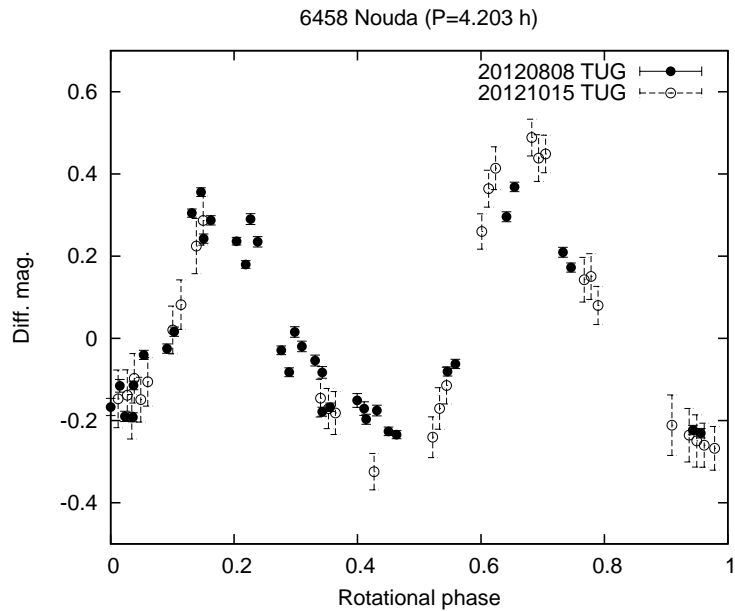}
\caption{Composite lightcurve of 6458 Nouda folded with the period of 4.203 h at the zero epoch of JD 2456148.3075578702. \label{figA28}}
\end{figure}

\clearpage

\begin{figure}
\epsscale{.63}
\plotone{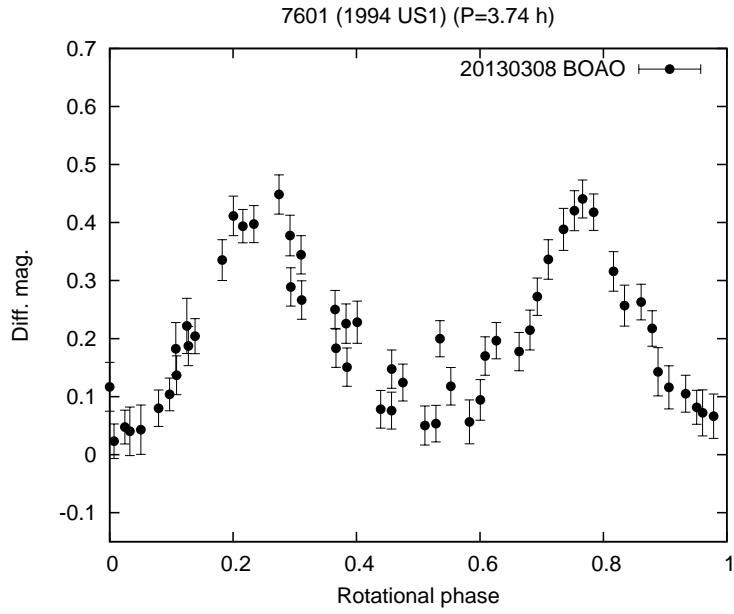}
\caption{Composite lightcurve of 7601 (1994 US1) folded with the period of 3.74 h at the zero epoch of JD 2456360.01997. \label{figA29}}
\end{figure}

\begin{figure}
\epsscale{.63}
\plotone{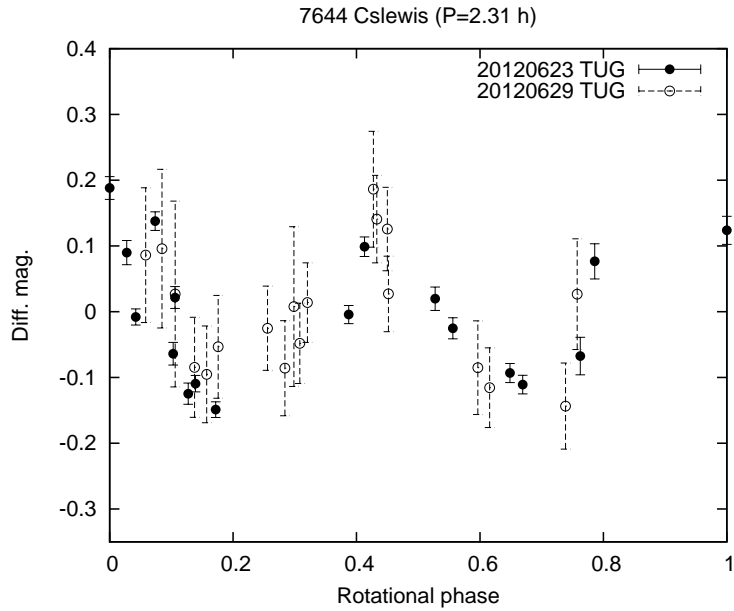}
\caption{Composite lightcurve of 7644 Cslewis folded with the period of 2.31 h at the zero epoch of JD 2456102.3593865740. \label{figA30}}
\end{figure}

\clearpage

\begin{figure}
\epsscale{.63}
\plotone{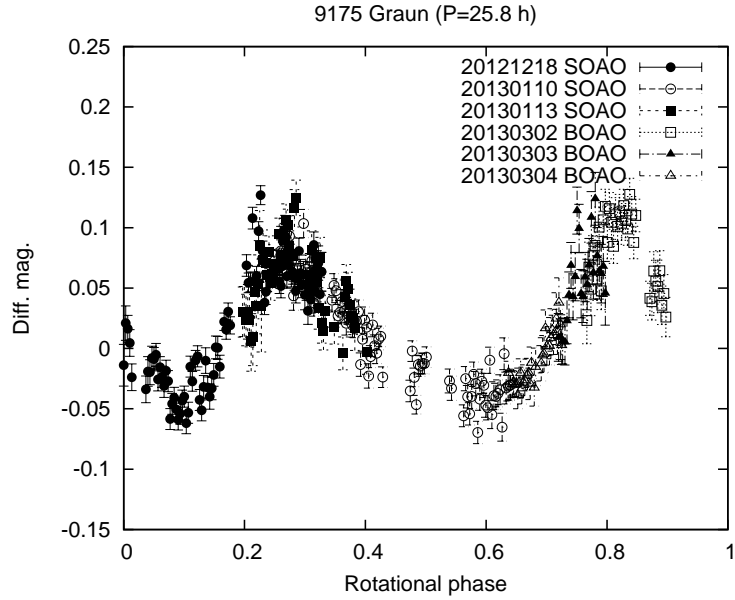}
\caption{Composite lightcurve of 9175 Graun folded with the period of 25.8 h at the zero epoch of JD 2456280.0405648. \label{figA31}}
\end{figure}

\begin{figure}
\epsscale{.63}
\plotone{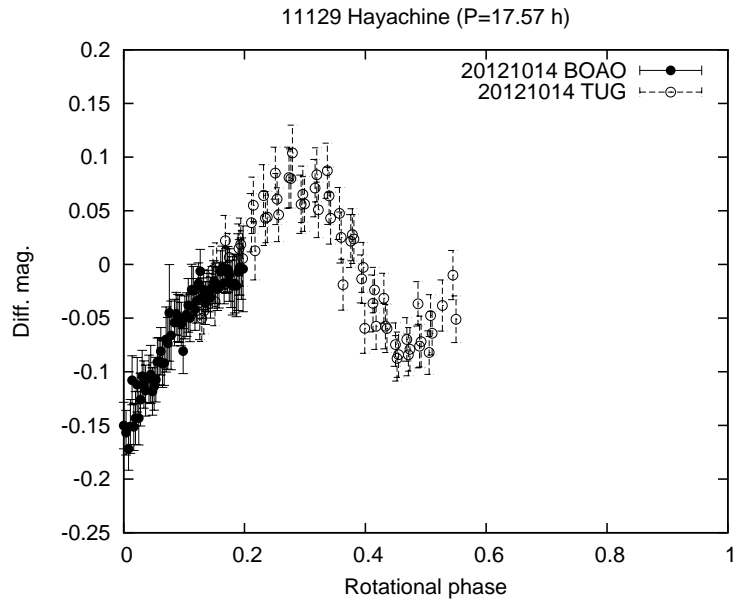}
\caption{Composite lightcurve of 11129 Hayachine folded with the period of 17.57 h at the zero epoch of JD 2456215.21134. \label{figA32}}
\end{figure}

\clearpage

\begin{figure}
\epsscale{.63}
\plotone{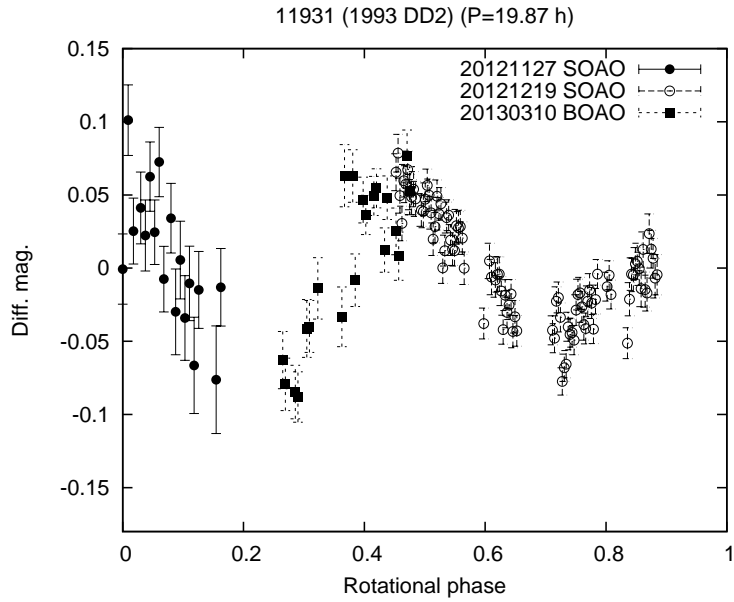}
\caption{Composite lightcurve of 11931 (1993 DD2) folded with the period of 19.87 h at the zero epoch of JD 2456259.0628221. \label{figA33}}
\end{figure}

\begin{figure}
\epsscale{.63}
\plotone{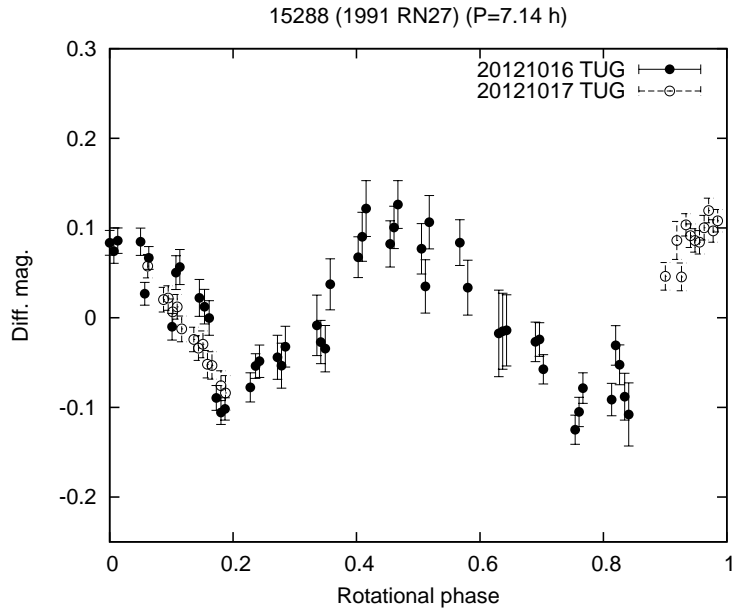}
\caption{Composite lightcurve of 15288 (1991 RN27) folded with the period of 7.14 h at the zero epoch of JD 2456217.3856597221. \label{figA34}}
\end{figure}

\clearpage

\begin{figure}
\epsscale{.63}
\plotone{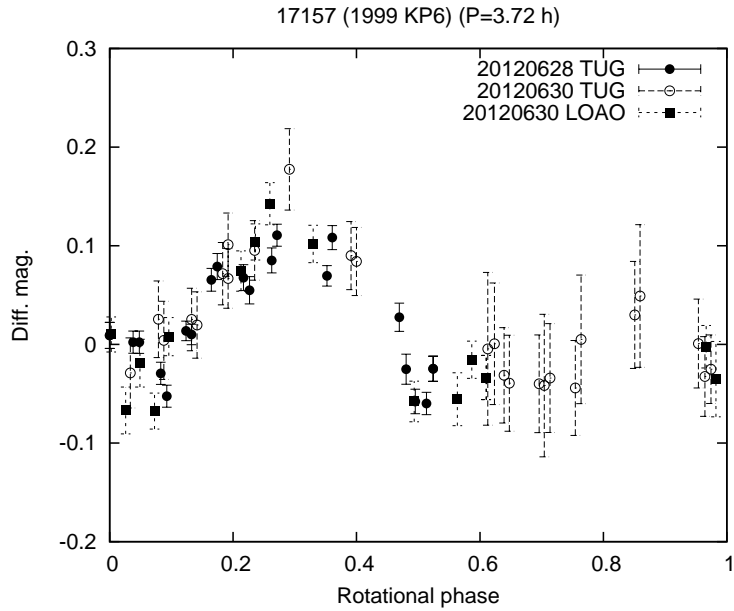}
\caption{Composite lightcurve of 17157 (1999 KP6) folded with the period of 3.72 h at the zero epoch of JD 2456107.4715162036. \label{figA35}}
\end{figure}

\begin{figure}
\epsscale{.63}
\plotone{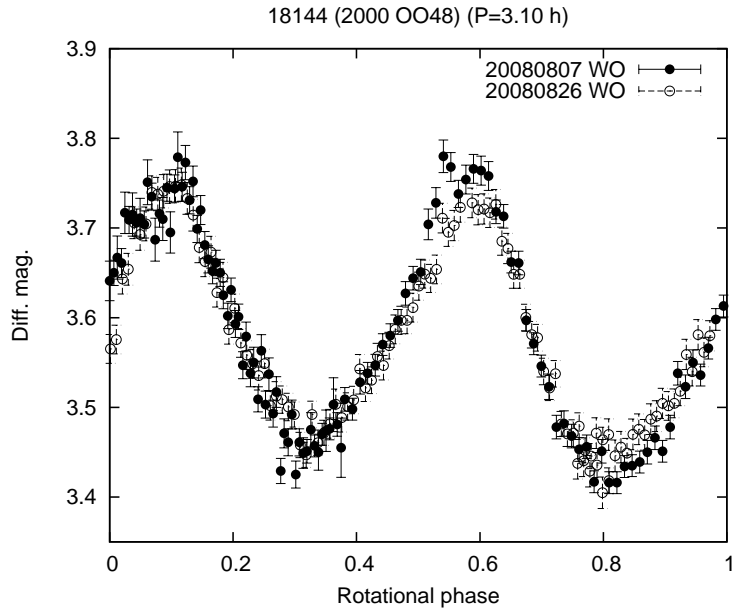}
\caption{Composite lightcurve of 18144 (2000 OO48) folded with the period of 3.10 h at the zero epoch of JD 2454686.41185188. \label{figA36}}
\end{figure}

\clearpage

\begin{figure}
\epsscale{.63}
\plotone{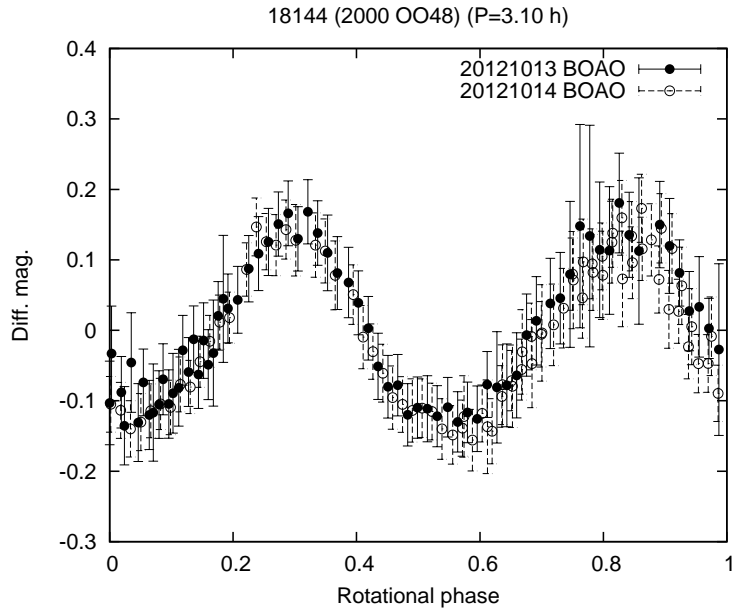}
\caption{Composite lightcurve of 18144 (2000 OO48) folded with the period of 3.10 h at the zero epoch of JD 2456213.94409. \label{figA37}}
\end{figure}

\begin{figure}
\epsscale{.63}
\plotone{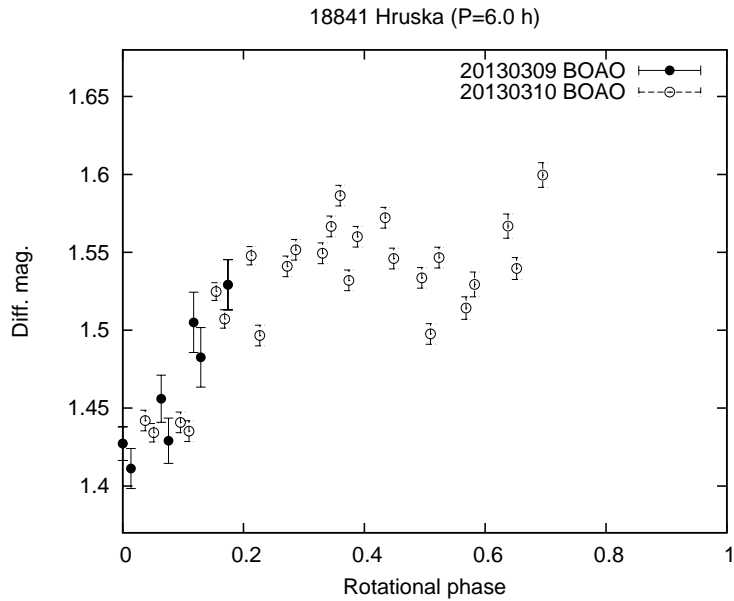}
\caption{Composite lightcurve of 18841 Hruska folded with the period of 6.0 h at the zero epoch of JD 2456361.13958. \label{figA38}}
\end{figure}

\clearpage

\begin{figure}
\epsscale{.63}
\plotone{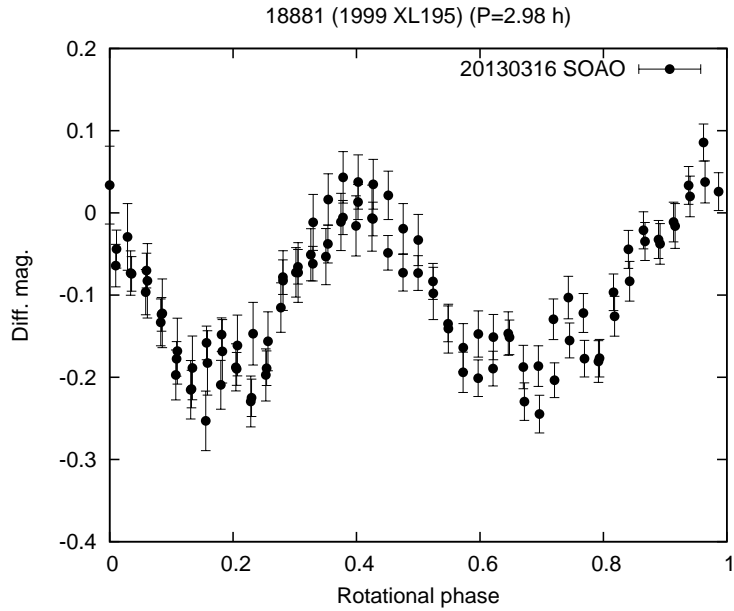}
\caption{Composite lightcurve of 18881 (1999 XL195) folded with the period of 2.98 h at the zero epoch of JD 2456367.9451630. \label{figA39}}
\end{figure}

\begin{figure}
\epsscale{.63}
\plotone{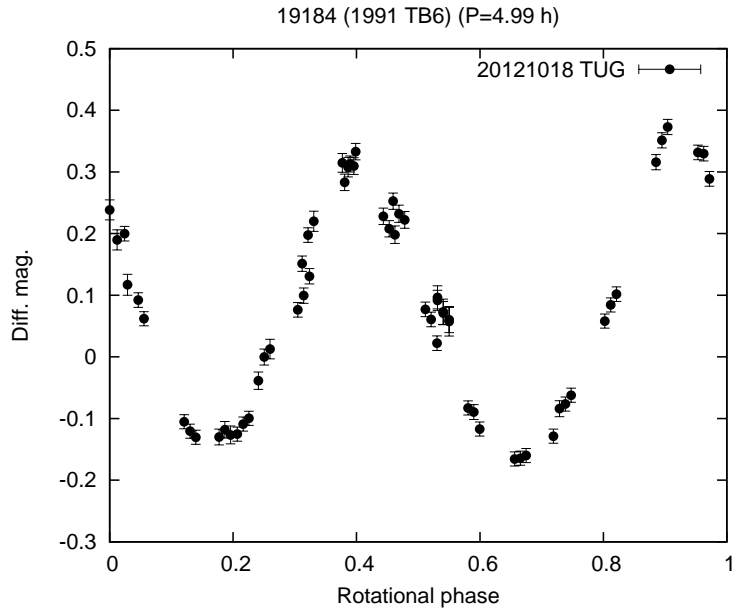}
\caption{Composite lightcurve of 19184 (1991 TB6) folded with the period of 4.99 h at the zero epoch of JD 2456219.3113541668. \label{figA40}}
\end{figure}

\clearpage

\begin{figure}
\epsscale{.63}
\plotone{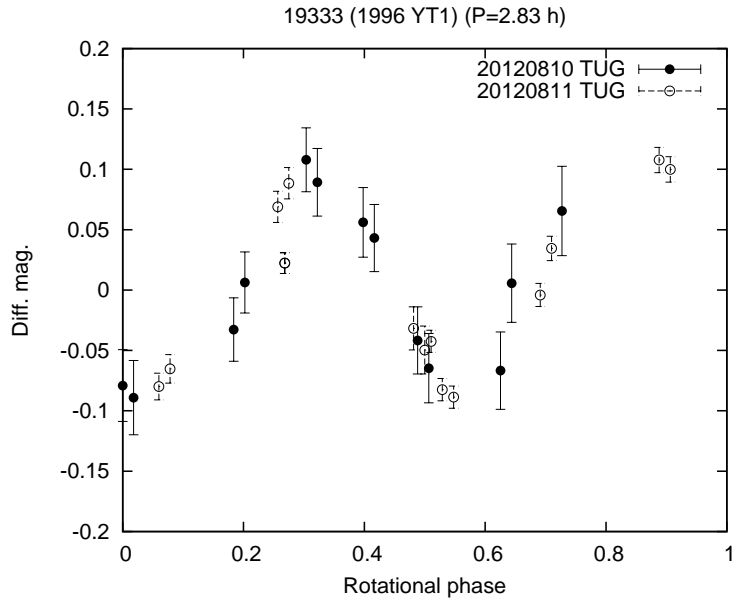}
\caption{Composite lightcurve of 19333 (1996 YT1) folded with the period of 2.83 h at the zero epoch of JD 2456150.4649768518. \label{figA41}}
\end{figure}

\begin{figure}
\epsscale{.63}
\plotone{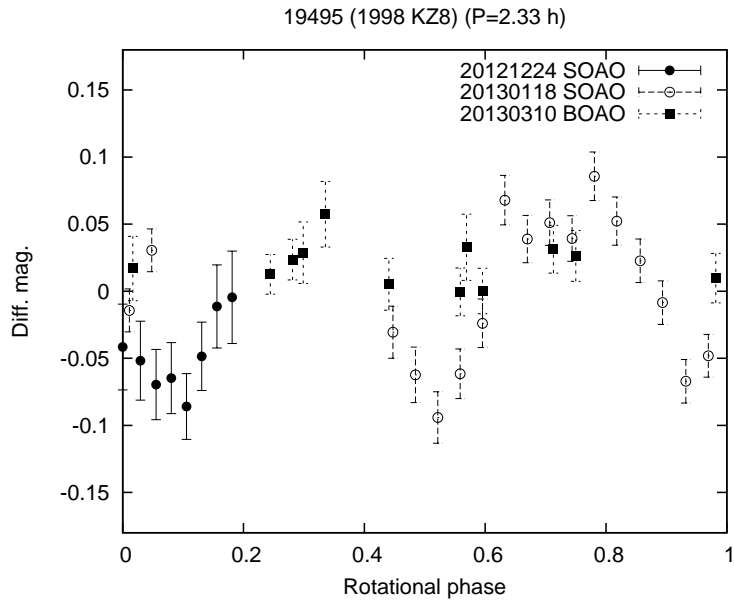}
\caption{Composite lightcurve of 19495 (1998 KZ8) folded with the period of 2.33 h at the zero epoch of JD 2456285.9586143. \label{figA42}}
\end{figure}

\clearpage

\begin{figure}
\epsscale{.63}
\plotone{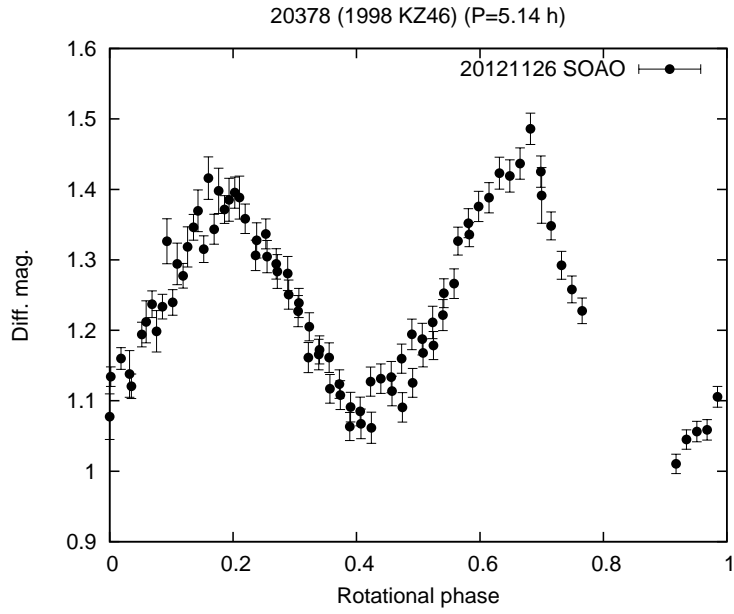}
\caption{Composite lightcurve of 20378 (1998 KZ46) folded with the period of 5.14 h at the zero epoch of JD 2456258.0254846. \label{figA43}}
\end{figure}

\begin{figure}
\epsscale{.63}
\plotone{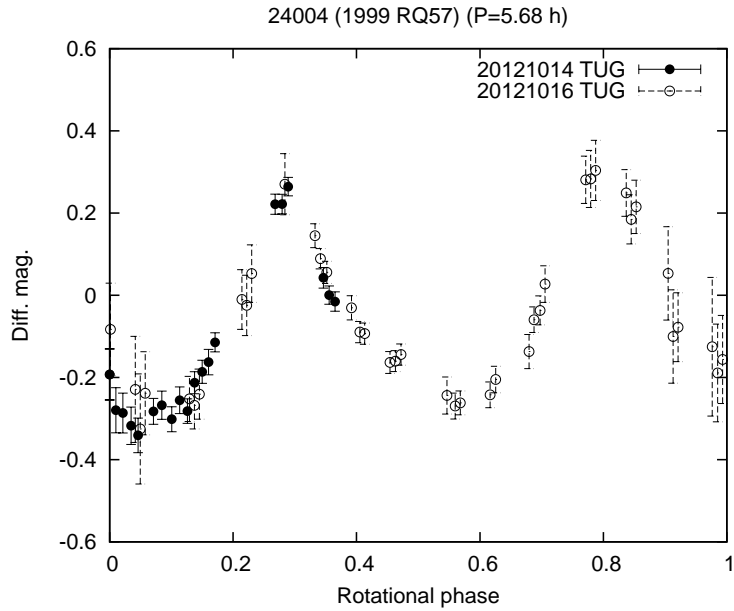}
\caption{Composite lightcurve of 24004 (1999 RQ57) folded with the period of 5.68 h at the zero epoch of JD 2456215.1844212962. \label{figA44}}
\end{figure}

\clearpage

\begin{figure}
\epsscale{.63}
\plotone{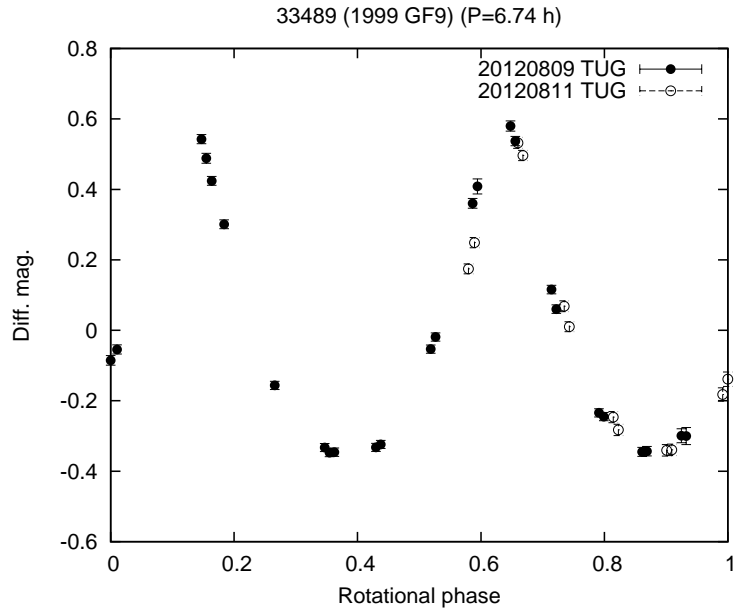}
\caption{Composite lightcurve of 33489 (1999 GF9) folded with the period of 6.74 h at the zero epoch of JD 2456149.3320601853. \label{figA45}}
\end{figure}

\begin{figure}
\epsscale{.63}
\plotone{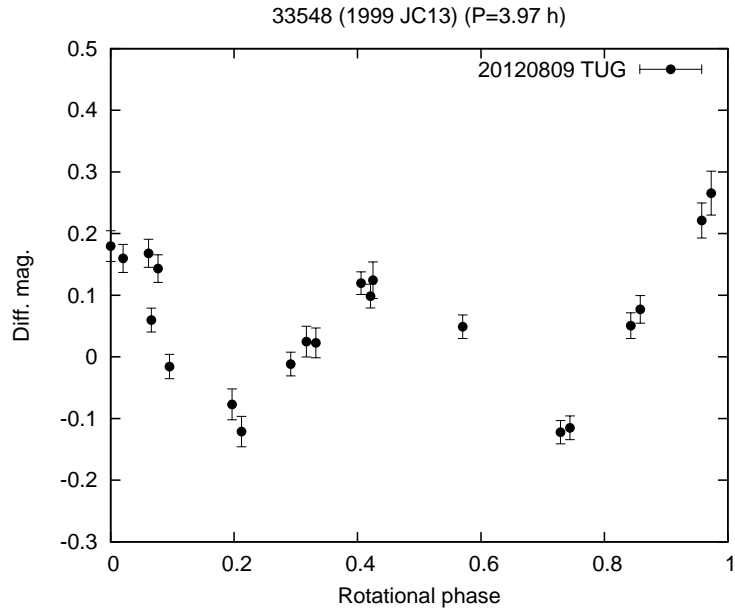}
\caption{Composite lightcurve of 33548 (1999 JC13) folded with the period of 3.97 h at the zero epoch of JD 2456149.3438657406. \label{figA46}}
\end{figure}

\clearpage

\begin{figure}
\epsscale{.63}
\plotone{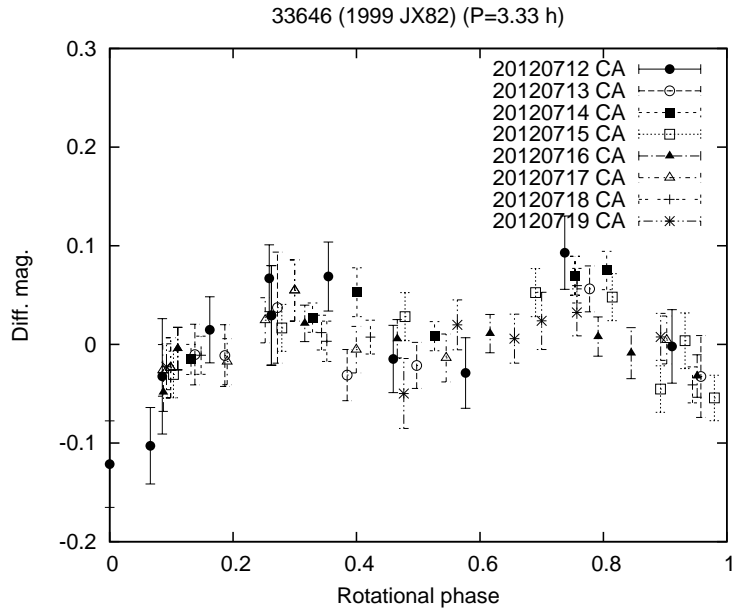}
\caption{Composite lightcurve of 33646 (1999 JX82) folded with the period of 3.33 h at the zero epoch of JD 56120.8596605. \label{figA47}}
\end{figure}

\begin{figure}
\epsscale{.63}
\plotone{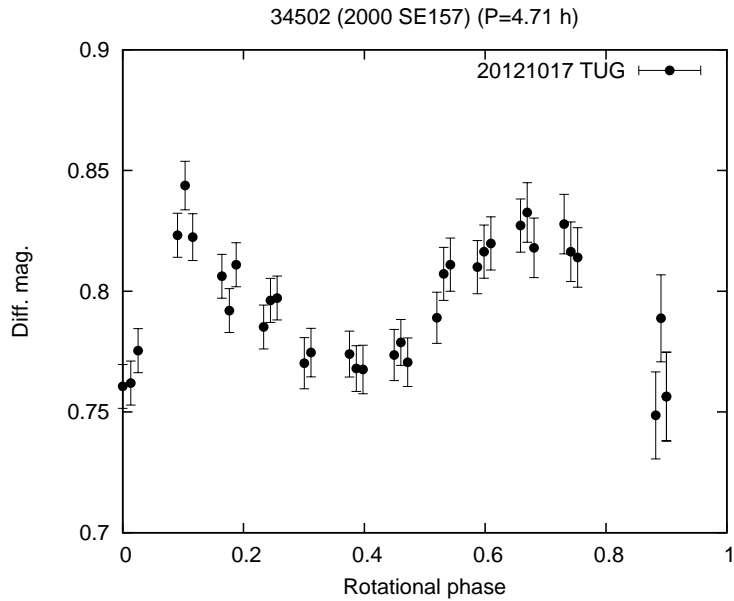}
\caption{Composite lightcurve of 34502 (2000 SE157) folded with the period of 4.71 h at the zero epoch of JD 2456218.3401041669. \label{figA48}}
\end{figure}

\clearpage

\begin{figure}
\epsscale{.63}
\plotone{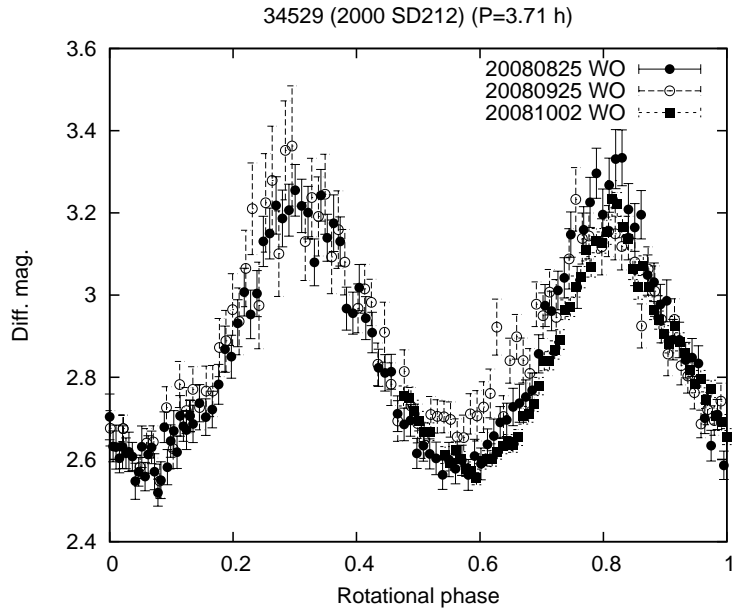}
\caption{Composite lightcurve of 34529 (2000 SD212) folded with the period of 3.71 h at the zero epoch of JD 2454704.38197917. \label{figA49}}
\end{figure}

\begin{figure}
\epsscale{.63}
\plotone{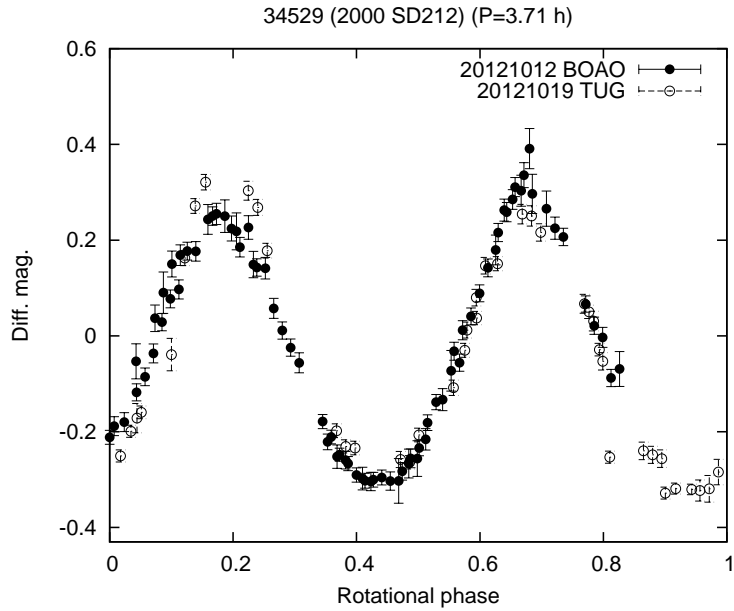}
\caption{Composite lightcurve of 34529 (2000 SD212) folded with the period of 3.71 h at the zero epoch of JD 2456212.97750. \label{figA50}}
\end{figure}

\clearpage

\begin{figure}
\epsscale{.63}
\plotone{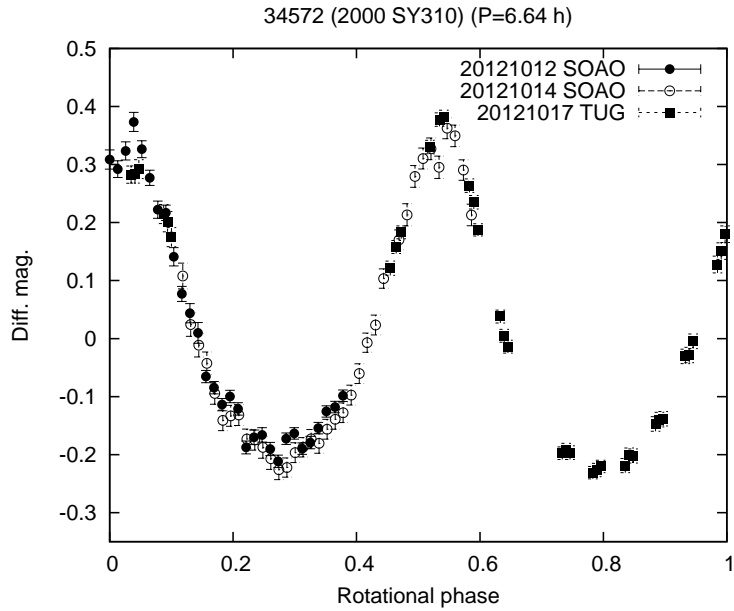}
\caption{Composite lightcurve of 34572 (2000 SY310) folded with the period of 6.64 h at the zero epoch of JD 2456212.9511140. \label{figA51}}
\end{figure}

\begin{figure}
\epsscale{.63}
\plotone{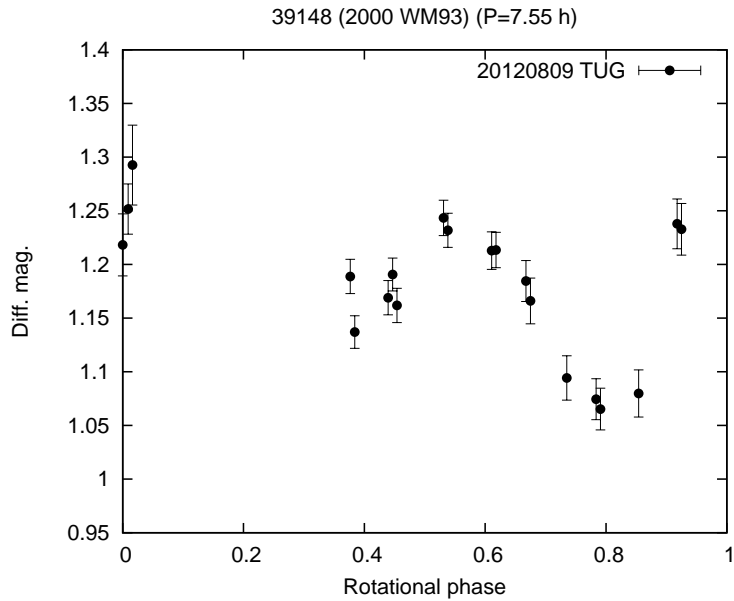}
\caption{Composite lightcurve of 39148 (2000 WM93) folded with the period of 7.55 h at the zero epoch of JD 2456149.2793981479. \label{figA52}}
\end{figure}

\clearpage

\begin{figure}
\epsscale{.63}
\plotone{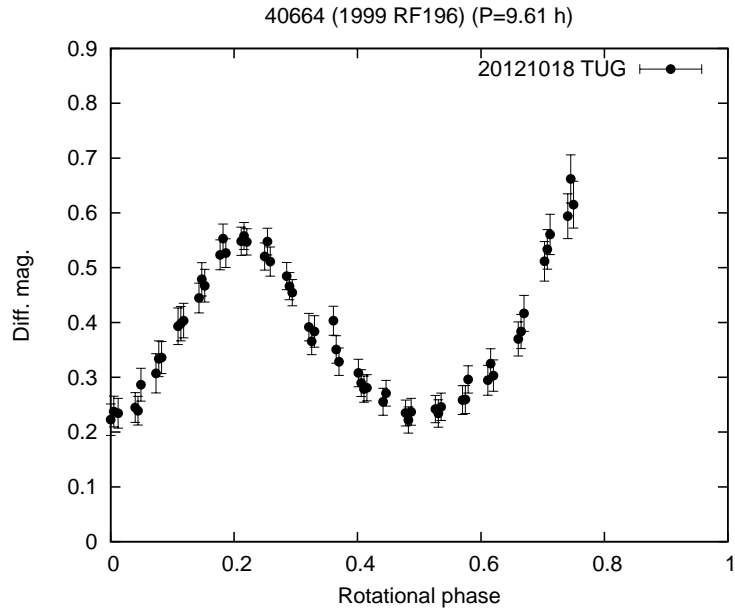}
\caption{Composite lightcurve of 40664 (1999 RF196) folded with the period of 9.61 h at the zero epoch of JD 2456219.3257870371. \label{figA53}}
\end{figure}

\begin{figure}
\epsscale{.63}
\plotone{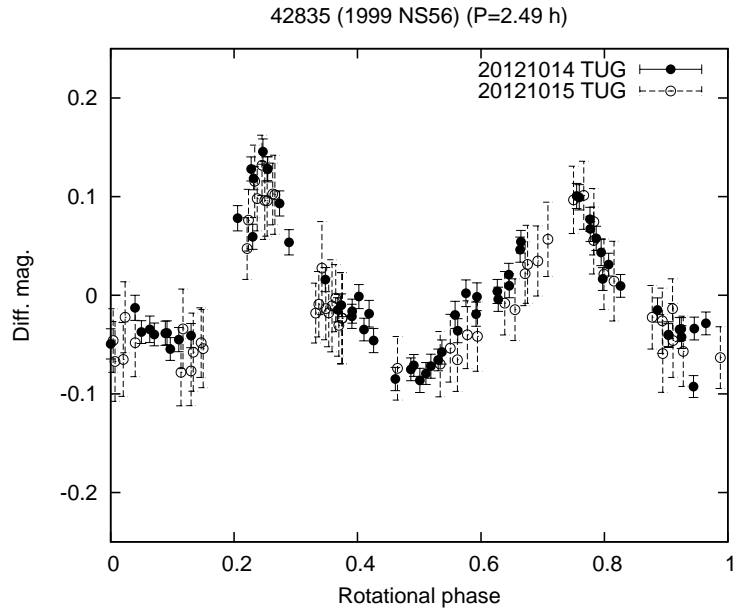}
\caption{Composite lightcurve of 42835 (1999 NS56) folded with the period of 2.49 h at the zero epoch of JD 2456215.2331712963. \label{figA54}}
\end{figure}

\clearpage

\begin{figure}
\epsscale{.63}
\plotone{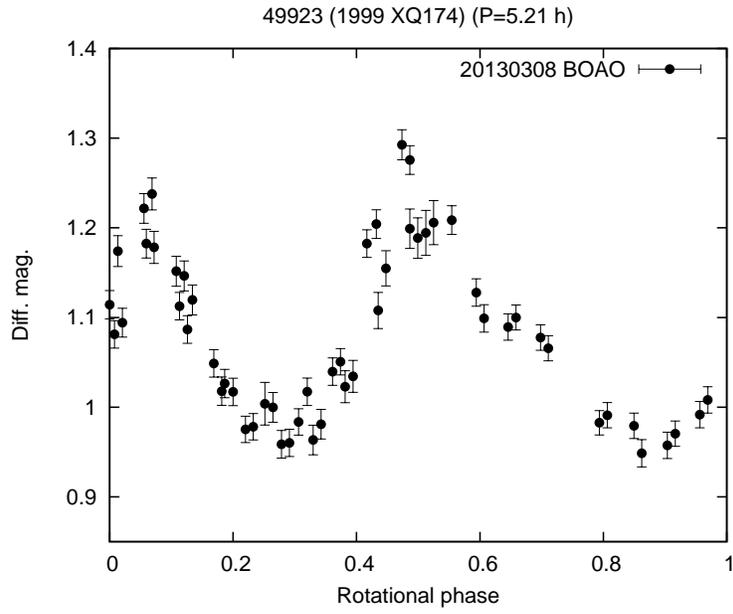}
\caption{Composite lightcurve of 49923 (1999 XQ174) folded with the period of 5.21 h at the zero epoch of JD 2456360.03059. \label{figA55}}
\end{figure}

\begin{figure}
\epsscale{.63}
\plotone{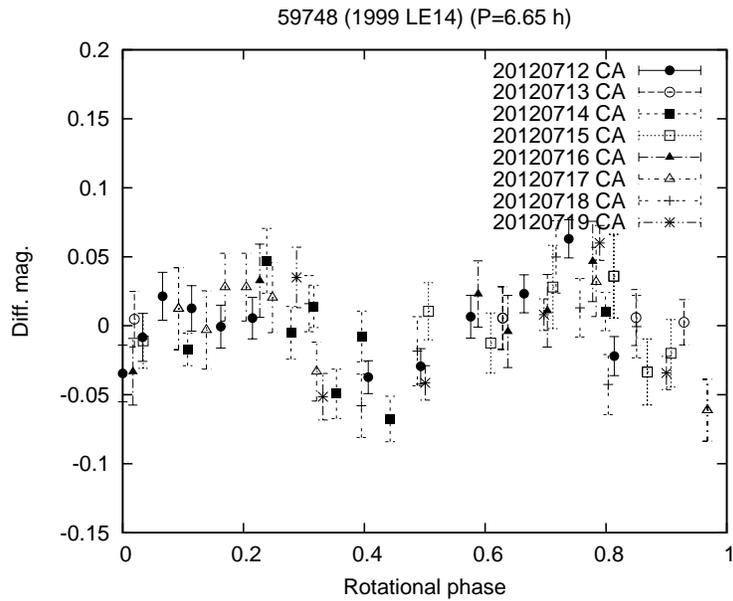}
\caption{Composite lightcurve of 59748 (1999 LE14) folded with the period of 6.65 h at the zero epoch of JD 56120.8554302. \label{figA56}}
\end{figure}

\clearpage

\begin{figure}
\epsscale{.63}
\plotone{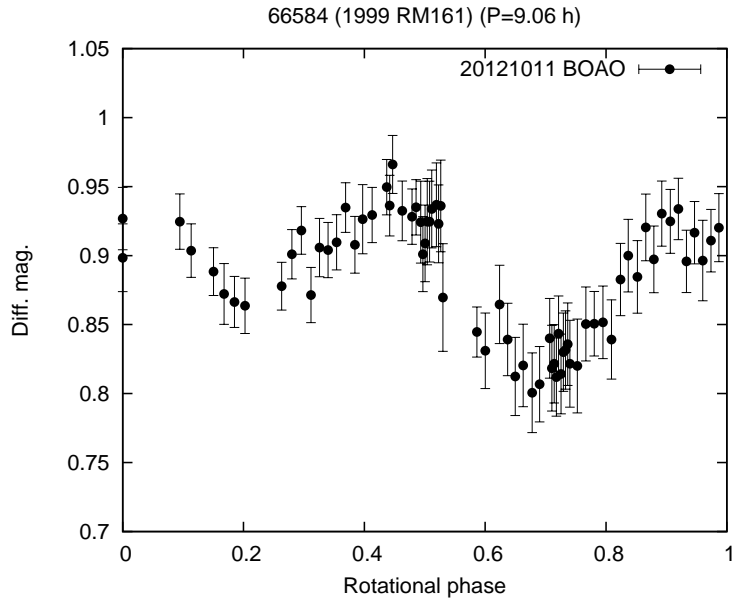}
\caption{Composite lightcurve of 66584 (1999 RM161) folded with the period of 9.06 h at the zero epoch of JD 2456211.96289. \label{figA57}}
\end{figure}

\begin{figure}
\epsscale{.63}
\plotone{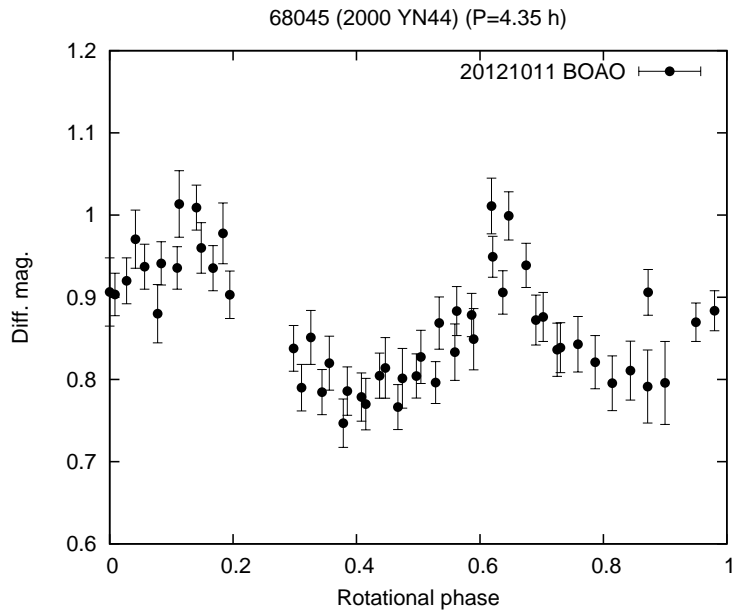}
\caption{Composite lightcurve of 68045 (2000 YN44) folded with the period of 4.35 h at the zero epoch of JD 2456212.00895. \label{figA58}}
\end{figure}

\clearpage

\begin{figure}
\epsscale{.63}
\plotone{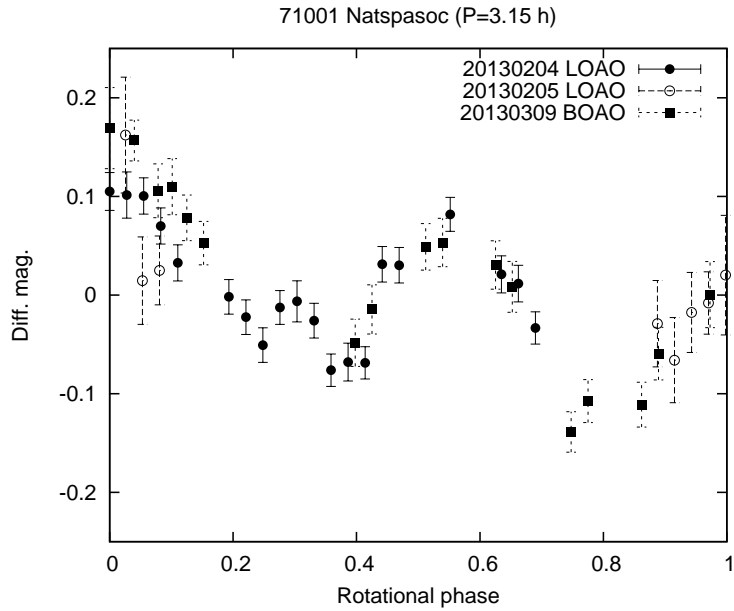}
\caption{Composite lightcurve of 71001 Natspasoc folded with the period of 3.15 h at the zero epoch of JD 2456328.67165. \label{figA59}}
\end{figure}

\begin{figure}
\epsscale{.63}
\plotone{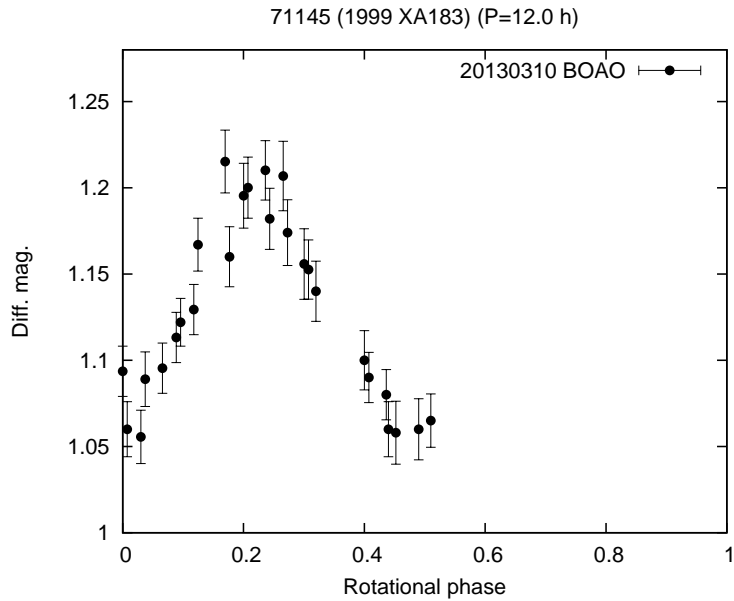}
\caption{Composite lightcurve of 71145 (1999 XA183) folded with the period of 12.0 h at the zero epoch of JD 2456362.15605. \label{figA60}}
\end{figure}

\clearpage

\begin{figure}
\epsscale{.63}
\plotone{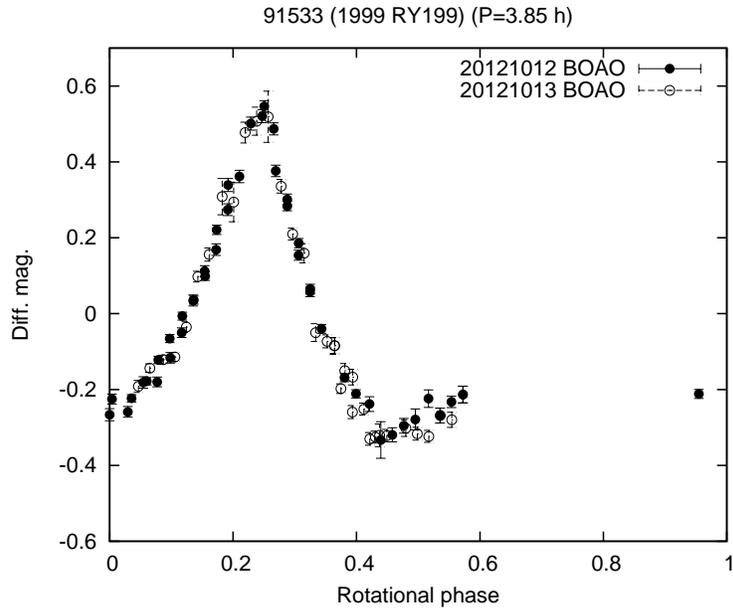}
\caption{Composite lightcurve of 91533 (1999 RY199) folded with the period of 3.85 h at the zero epoch of JD 2456212.92043. \label{figA61}}
\end{figure}

\clearpage

\begin{figure}
\epsscale{.99}
\plotone{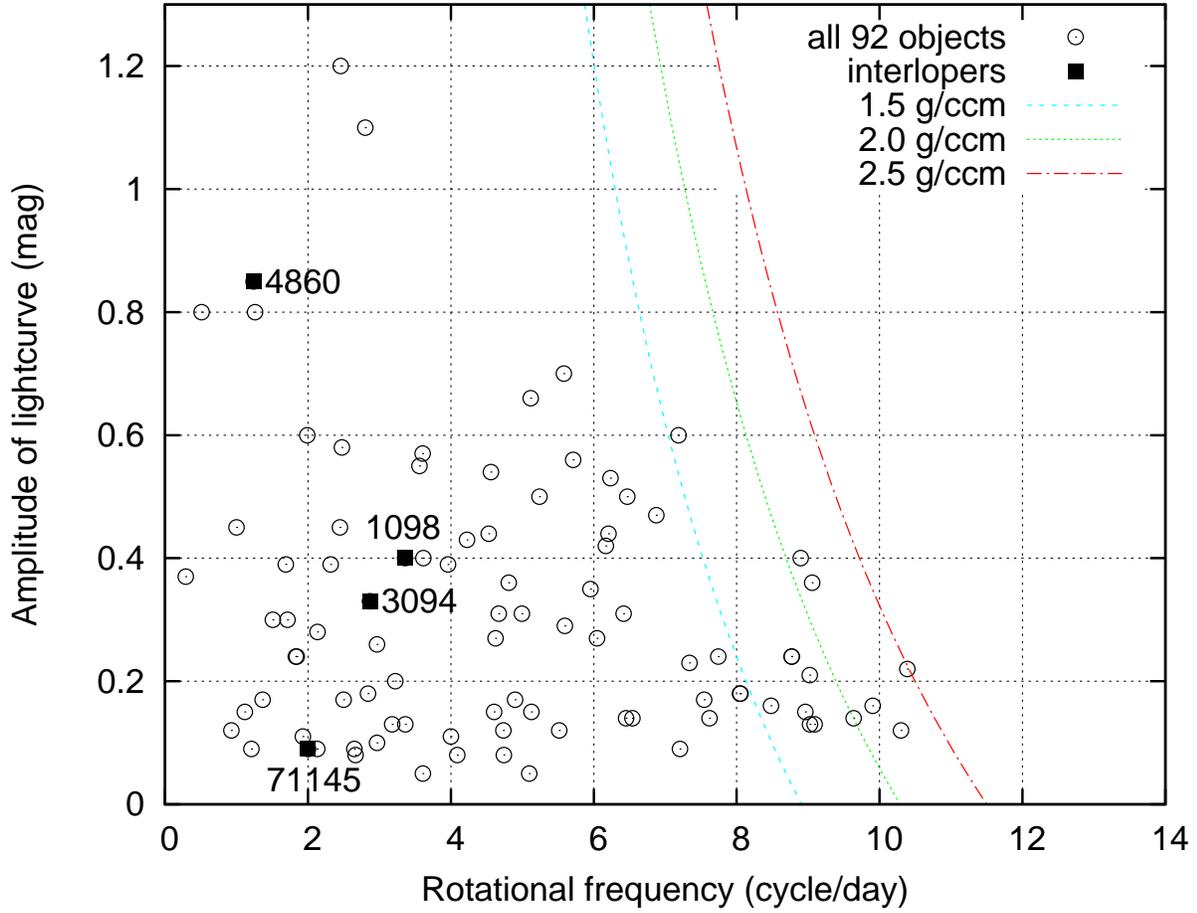}
\caption{Correlation between rotational frequencies (cycle/day) and the amplitude of lightcurve (peak-to-peak variation magnitude). The colored curves are approximate critical rotational period for bulk densities of 1.5, 2.0, and 2.5 g/cm$^3$. \label{figA62}}
\end{figure}

\clearpage

\begin{figure}
\epsscale{.99}
\plotone{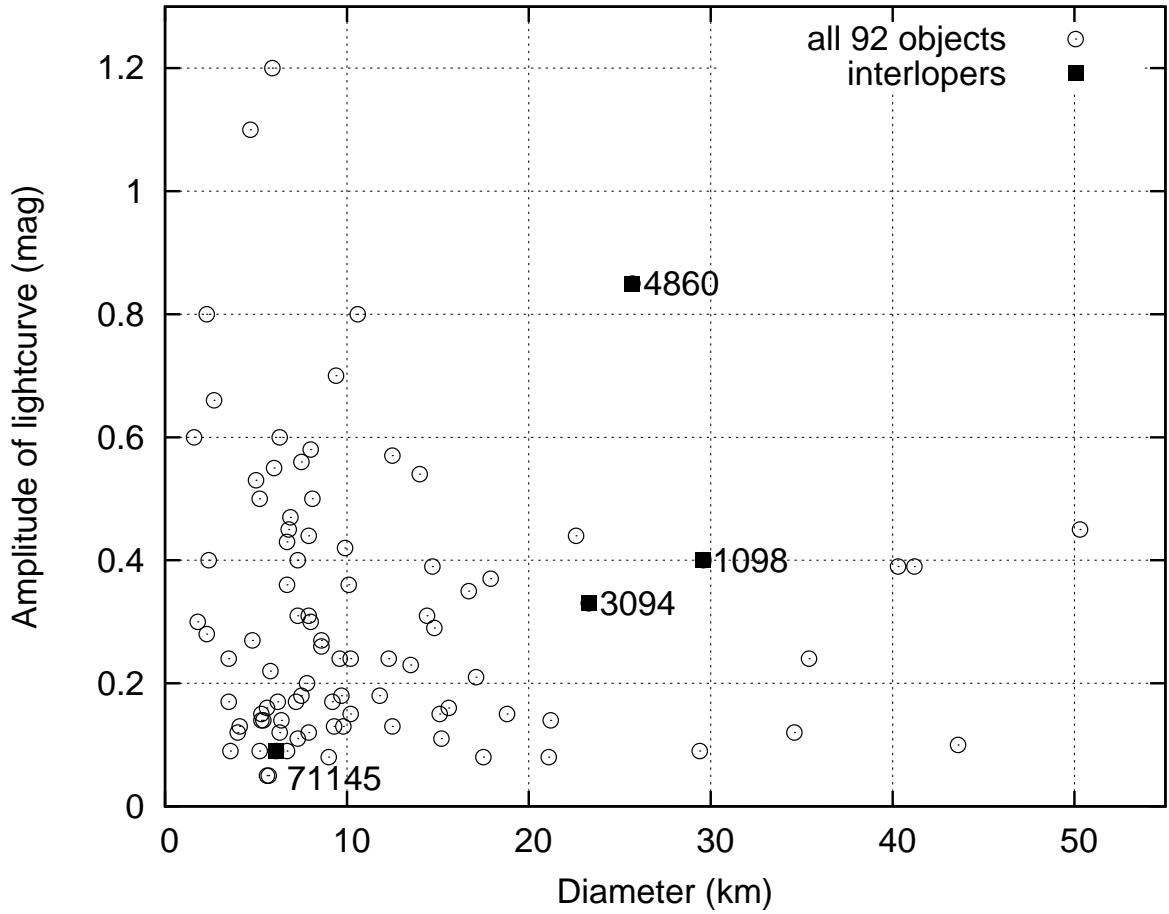}
\caption{Correlation between rotational frequencies (cycle/day) and the size of asteroid (diameter in km). No objects larger than 15 km have the amplitude of lightcurve more than 0.5 magnitude except for 4860 Gubbio regarded as the interloper. Small objects ($< 15$ km) spread out with various shape. \label{figA63}}
\end{figure}




\clearpage



\renewcommand{\thefigure}{\arabic{figure}}
\setcounter{figure}{0}

\begin{figure}
\plotone{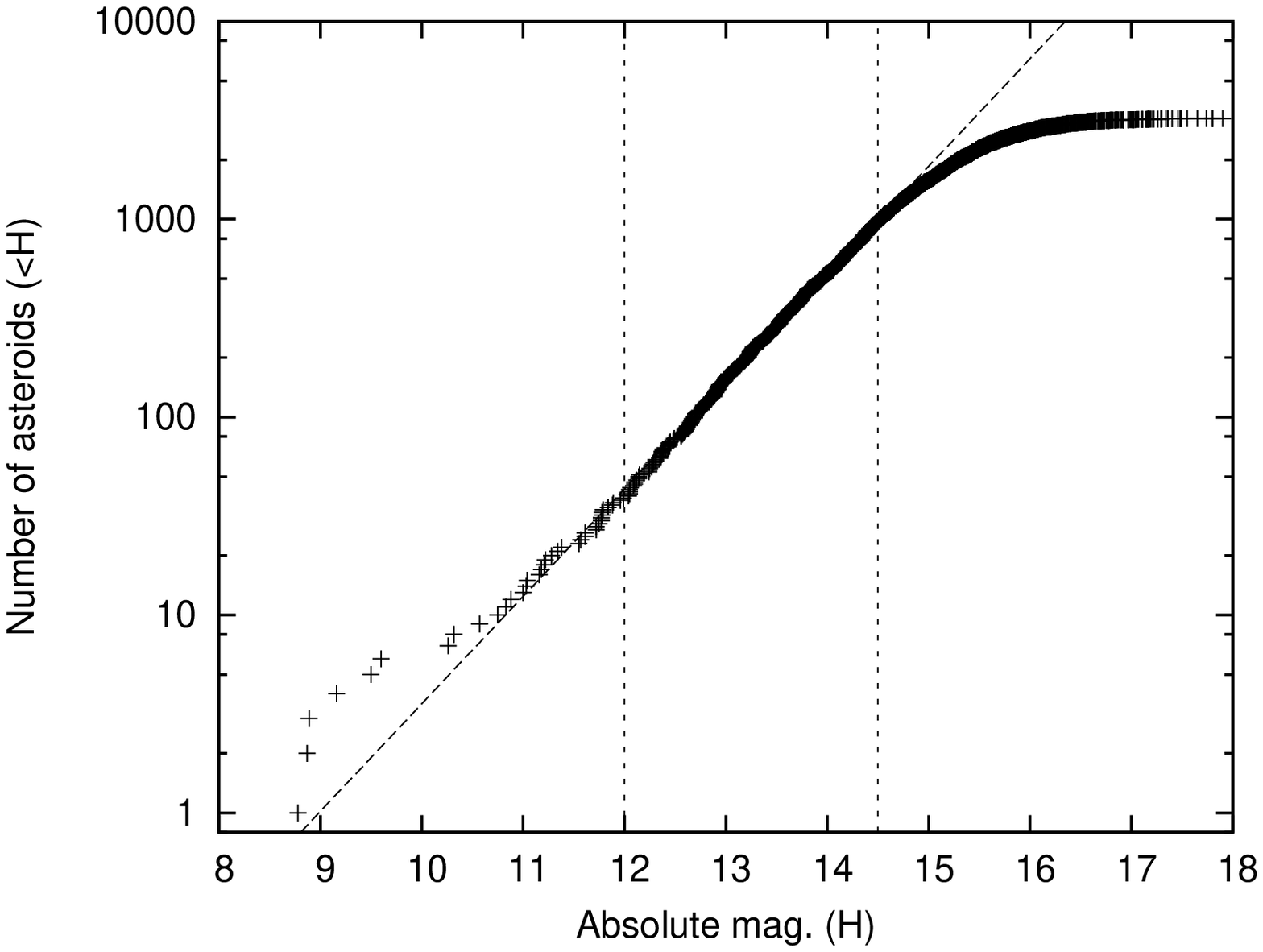}
\caption{Cumulative distributions N ($< H$) for the Maria family members. A power-law best fit to the data in the magnitude range (12 $< H <$ 14.5) results in a coefficient $\gamma \sim 0.54$. \label{fig1}}
\end{figure}

\clearpage


\begin{figure}
\plotone{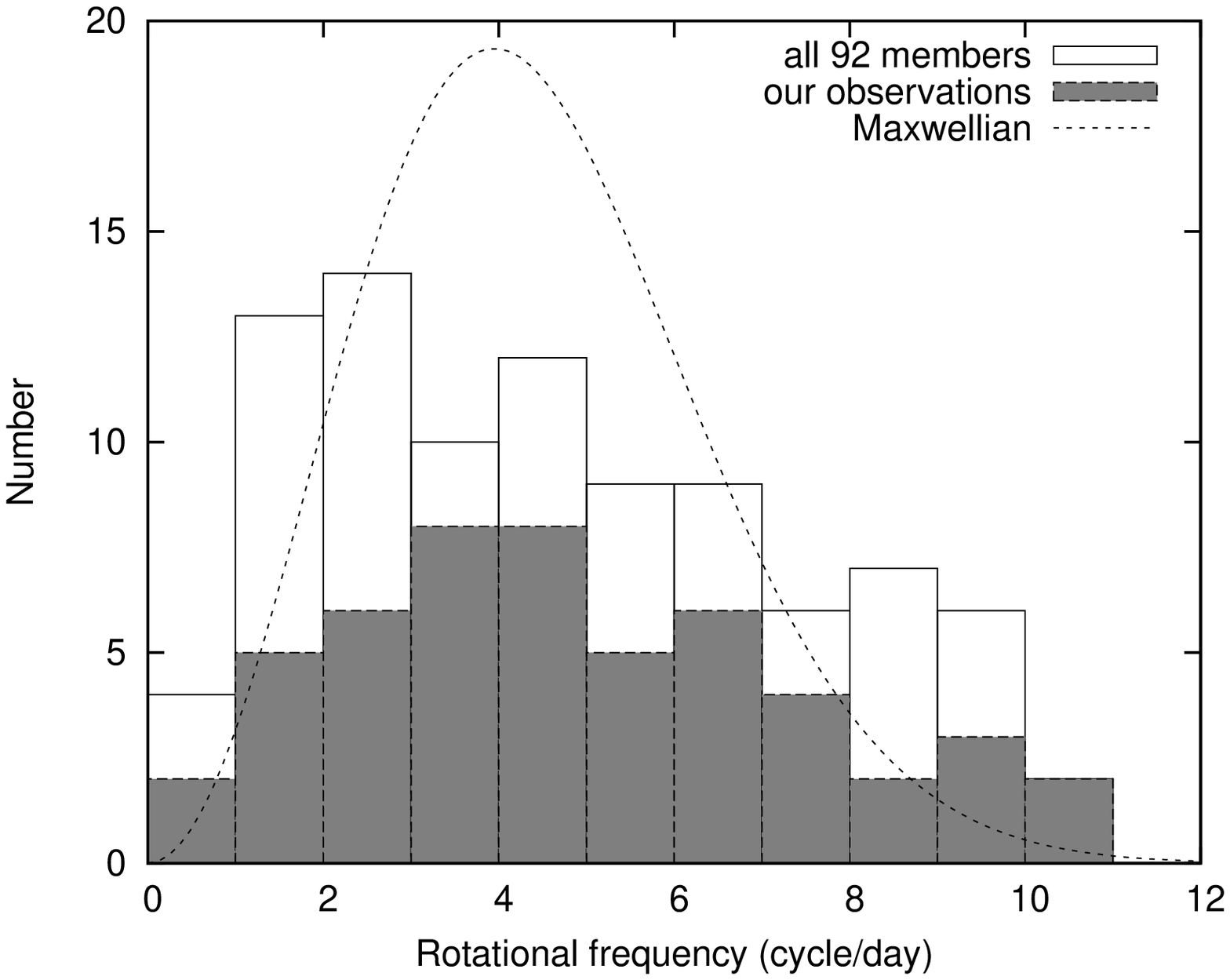}
\caption{Rotational rates distribution for the 92 Maria family members and the best fit Maxwellian curve (dashed line). The 51 objects from our observations are marked with shaded bars. If we added to this graph the upper bounds for the rotational frequencies for the 23 objects with a reliability code of 1 (see Table 3), the new objects would populate the bins 1 to 5, thereby increasing the excess of slow rotators. \label{fig2}}
\end{figure}

\clearpage


\begin{figure}
\plotone{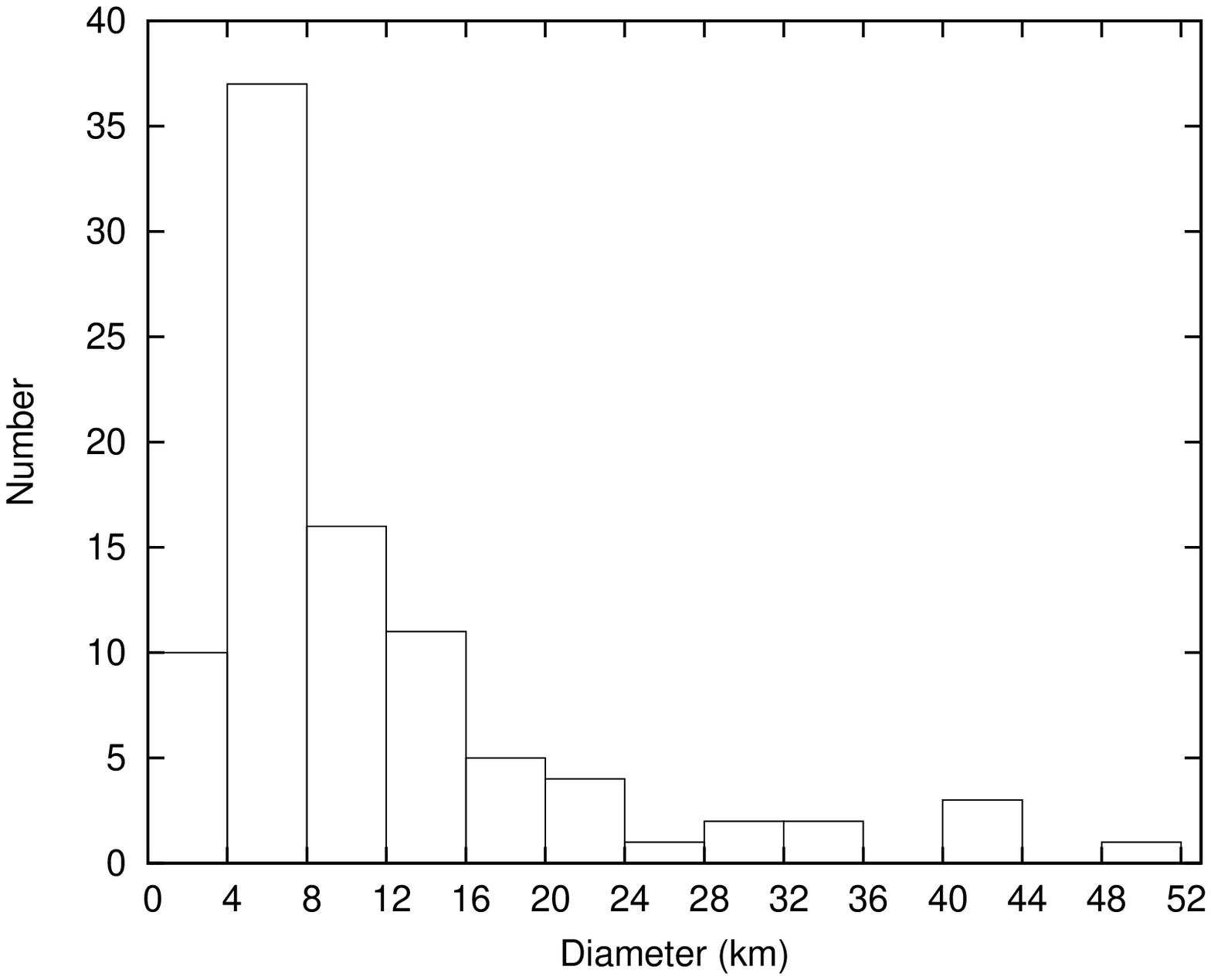}
\caption{Histogram of diameters for the 92 Maria family members. Most objects are smaller than 30 km ($\sim$ 93.5 \%). \label{fig3}}
\end{figure}

\clearpage


\begin{figure}
\plotone{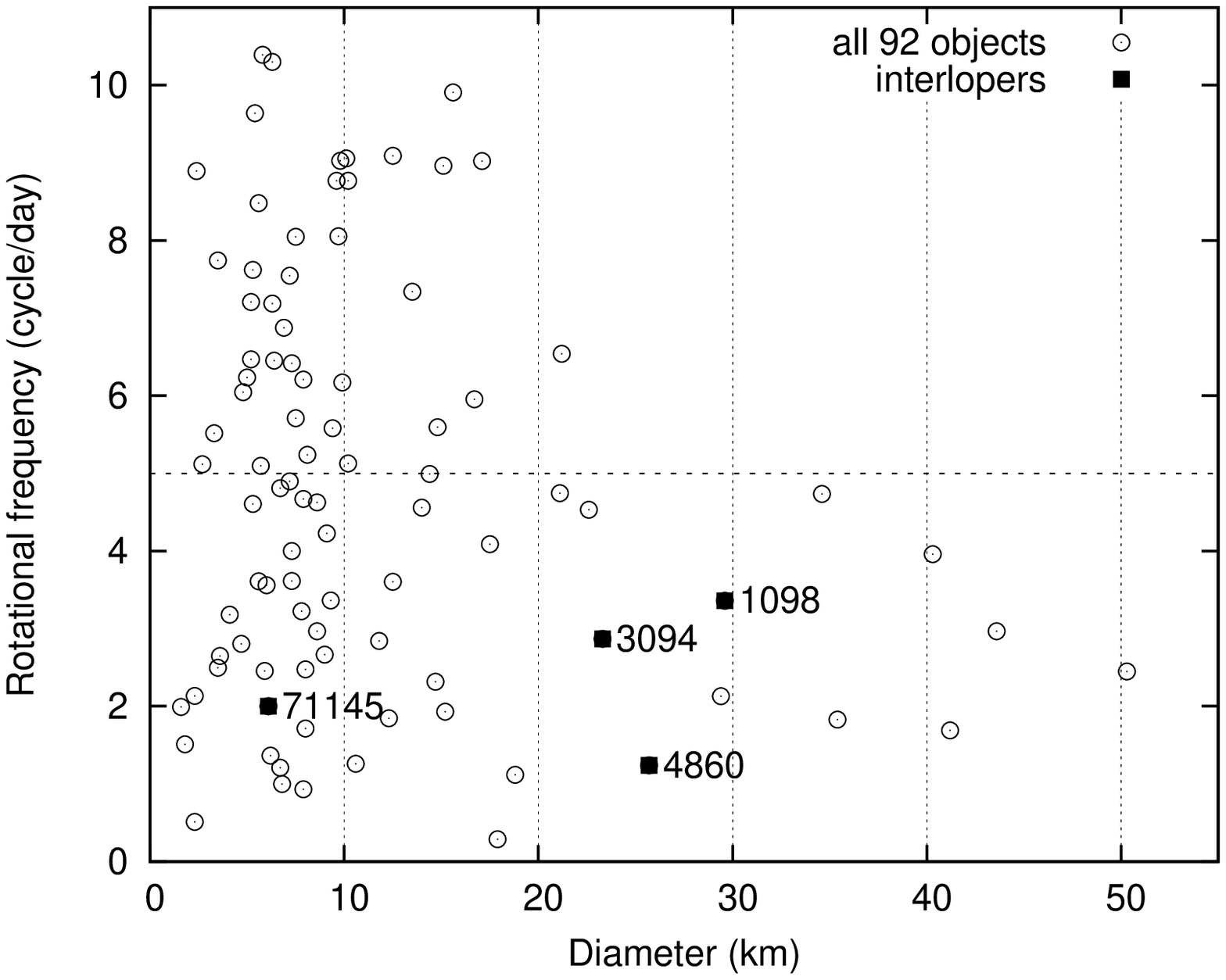}
\caption{Rotational frequency (cycle/day) versus diameter for the 92 Maria members. The dashed horizontal line represents the rotational frequency of 5 (cycle/day), that corresponds to spin rate of 4.8 hr. All asteroids larger than 22 km rotate slower than 4.8 hr. \label{fig4}}
\end{figure}

\clearpage


\begin{figure}
\plotone{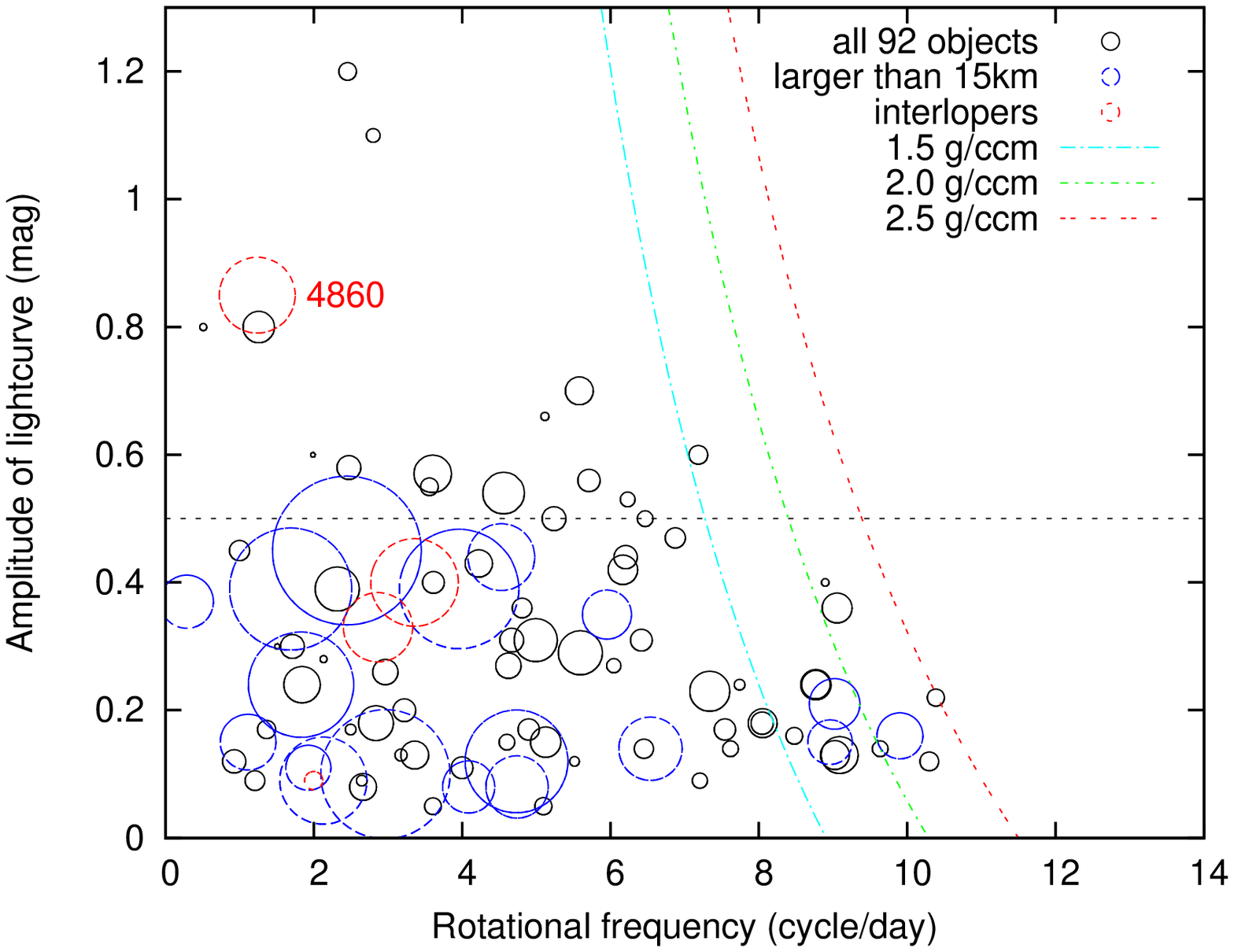}
\caption{Correlation between rotational frequencies (cycle/day) and lightcurve amplitude (magnitude), as well as compared with the size of asteroid. The sizes of various circles indicate diameter of each asteroid. Asteroids larger than 15 km in size are marked with the blue circles. The dashed horizontal line represents amplitude of lightcurve of 0.5 (magnitude). The colored curves represent approximate critical rotational frequencies for bulk densities of 1.5, 2.0, and 2.5 g/cm$^3$, respectively. This figure separates into two plots in Appendix A. \label{fig5}}
\end{figure}

\clearpage


\begin{figure}
\plotone{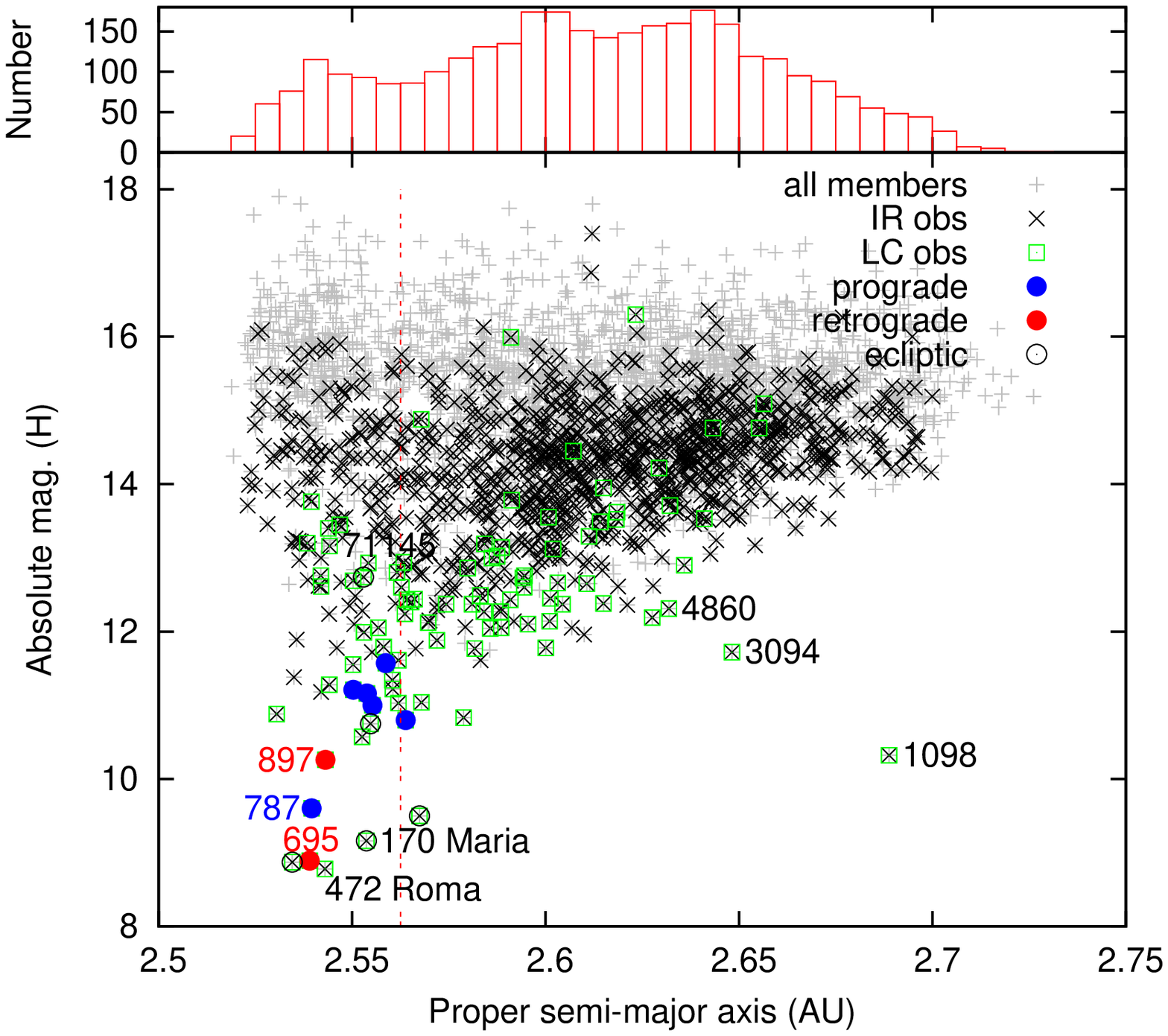}
\caption{Total of 3,230 known members of the Maria asteroid family (grey crosses) projected onto the proper semi-major axis (AU) versus absolute magnitude (H) plane. 1,152 members matched with the WISE IR observation are marked with the black ``$\times$ symbols'' and the empty green square represents all 92 objects we used in this study from the lightcurve observation. The filled blue and red circles stand for prograde and retrograde rotators, respectively, while the five rotators with the pole along the ecliptic are marked with open circles. The upper part of this figure is the density histogram with respect to the semi-major axis. \label{fig6}}
\end{figure}

\clearpage


\begin{figure}
\plotone{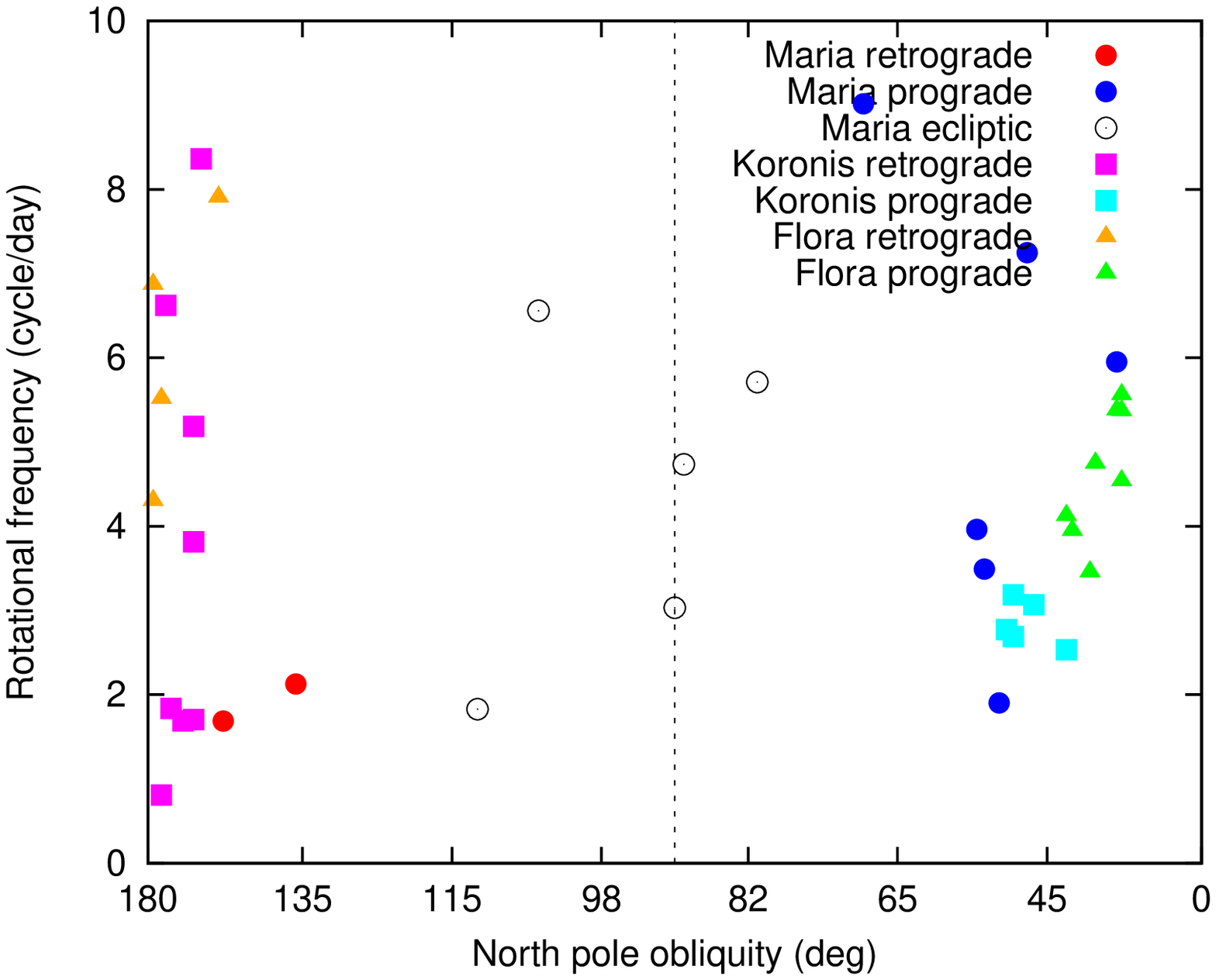}
\caption{Spin vector obliquity of the Maria family members (red and blue circles) with respect to rotational frequency (cycle/day) compared with those of Koronis (violet and cyan squares) and Flora family (orange and green triangle). The five open circles (same as Fig. 6) denote Maria objects for which the pole axes lie close to the ecliptic plane. The abscissa in this figure follows the definition given in \citet{sli02}. \label{fig7}}
\end{figure}

\clearpage


\begin{figure}
\plotone{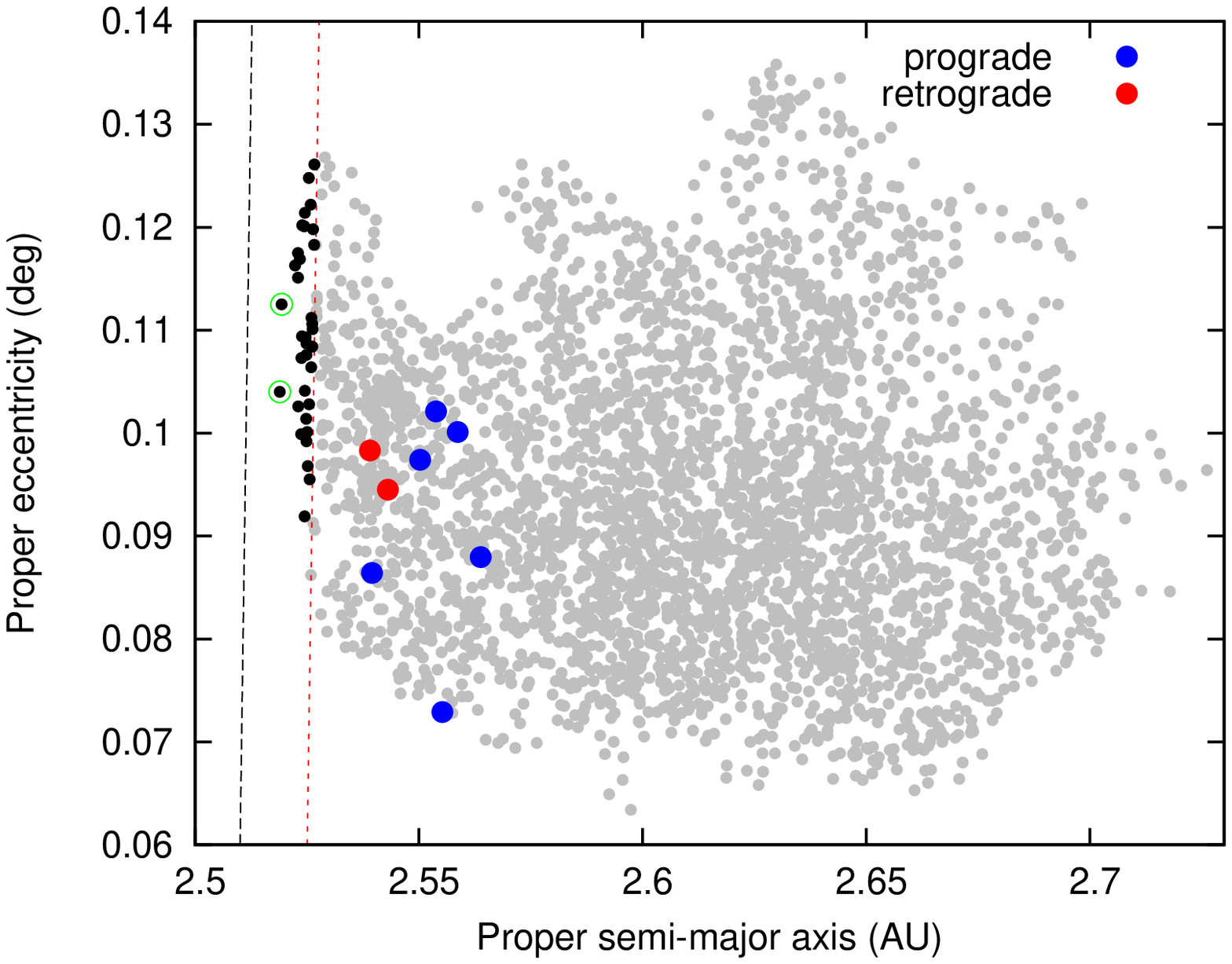}
\caption{All 3,230 members of the Maria asteroid family (grey) projected onto the proper semi-major axis (AU) versus proper eccentricity (deg) plane. The black and red dashed lines denote the 3:1 resonance boundary and chaotic diffusion region, respectively. The 37 asteroids placed within 0.015 AU from the resonance border are marked with the filled back circles. Two objects marked with green open circles, 114123 (2002 VX49) (lower one) and 137063 (1998 WK1) (upper one), are the most promising candidates for becoming new NEAs. Prograde (blue) and retrograde (red) rotators are the same as Fig. 6. \label{fig8}}
\end{figure}

\clearpage

\begin{deluxetable}{ccccccc}
\tabletypesize{\scriptsize}
\tablecaption{Observatory and Instrument details}
\tablewidth{0pt}
\tablehead{
\colhead{Telescope\tablenotemark{a}} & \colhead{Observing date range} & \colhead{$\lambda$\tablenotemark{b}} & \colhead{$\varphi$\tablenotemark{b}} & \colhead{Altitude} &
\colhead{Instrument\tablenotemark{c}} & \colhead{Pixel scale} \\
\colhead{} & \colhead{} & \colhead{} & \colhead{} & \colhead{(m)} &
\colhead{} & \colhead{(" pix$^{-1}$)}
}
\startdata
WO 0.46 m  & 2008 Jul. 6 $\sbond$ Dec. 4   & 34:45:48  & +30:35:45 & 875    & SBIG ST-10 & 1.10 \\ 
TUG 1.0 m  & 2012 Jun. 20 $\sbond$ 2013 Mar. 6  & 30:19:59  & +36:49:31 & 2538.6 & SI 4K CCD  & 0.62 \\
BOAO 1.8 m & 2012 Apr. 15 $\sbond$ 2013 Mar. 10 & 128:58:36 & +36:09:53 & 1143   & e2v 4K CCD & 0.43 \\
SOAO 0.6 m & 2012 Feb. 4  $\sbond$ 2013 May 8   & 128:27:27 & +36:56:04 & 1354   & e2v 2K CCD & 1.02 \\
KASI 0.6 m & 2012 Jan. 10 $\sbond$ Oct. 12 & 127:28:31 & +36:46:53 & 87     & SBIG ST-8  & 1.06 \\
LOAO 1.0 m & 2012 Jun. 28 $\sbond$ 2013 Apr. 3  & 249:12:41 & +32:26:32 & 2776   & e2v 4K CCD & 0.80 \\
CA 1.2 m   & 2012 Jul. 12 $\sbond$ 19      & 2:32:45   & +37:13:25 & 2173.1 & e2v 4K CCD & 0.63 \\
\enddata
\tablenotetext{a}{Abbreviations: WO = Wise Observatory, TUG = Tubitak Ulusal Gozlemevi (Turkish National Observatory), BOAO = Bohyunsan Optical Astronomy Observatory, SOAO = Sobaeksan Optical Astronomy Observatory, KASI = Korea Astronomy and Space Science Institute, LOAO = Lemmonsan Optical Astronomy Observatory, CA = Calar Alto}
\tablenotetext{b}{Eastern longitude and geocentric latitude of each observatory}
\tablenotetext{c}{SI 4K CCD, e2v 4K CCD, and e2v 2K CCD were configured with 2$\times$2 binning.}
\end{deluxetable}


\clearpage

\begin{deluxetable}{ccccccccc}
\tabletypesize{\scriptsize}
\tablecaption{Observational circumstances of the Maria asteroid family members}
\tablewidth{0pt}
\tablehead{
\colhead{Asteroid} & \colhead{UT date} & \colhead{RA} & \colhead{DEC} & \colhead{$r$} &
\colhead{$\Delta$} & \colhead{$\alpha$} & \colhead{$V$} & \colhead{Telescope} \\
\colhead{} & \colhead{(YY/MM/DD)} & \colhead{(hr)} & \colhead{(deg)} & \colhead{(AU)} &
\colhead{(AU)} & \colhead{(deg)} & \colhead{(mag)} & \colhead{} 
}
\startdata
575 Renate     & 2012/12/26.7 & 6.99  & +44.61 & 2.760 & 1.826 & 7.86  & 14.97 & SOAO \\
652 Jubilatrix & 2012/05/20.7 & 16.53 & -12.46 & 2.675 & 1.678 & 4.68  & 15.07 & SOAO \\
660 Crescentia & 2012/01/10.4 & 2.07  & -4.80  & 2.750 & 2.444 & 20.76 & 14.30 & KASI \\
               & 2012/01/11.4 & 2.08  & -4.66  & 2.751 & 2.458 & 20.79 & 14.32 & KASI \\
695 Bella      & 2012/01/28.7 & 10.17 & -5.59  & 2.885 & 2.004 & 10.51 & 13.78 & KASI \\
                    & 2012/01/29.7 &  10.16 &  -5.61  &  2.886  & 1.999  & 10.22  & 13.76 & KASI   \\
                    & 2012/01/30.7 &  10.14 &  -5.63  &  2.887  & 1.994  & 9.93   & 13.75 & KASI   \\
                    & 2012/02/01.7 &  10.11 &  -5.64  &  2.889  & 1.984  & 9.35   & 13.71 & KASI   \\
727 Nipponia        & 2008/07/06.8 &  17.07 &  -6.07  &  2.770  & 1.864  & 11.67  & 13.90 & WO     \\
                    & 2008/07/07.8 &  17.06 &  -6.17  &  2.769  & 1.870  & 11.96  & 13.92 & WO     \\
                    & 2008/07/29.7 &  16.91 &  -8.59  &  2.753  & 2.040  & 17.53  & 14.29 & WO     \\
                    & 2008/08/12.7 &  16.91 &  -10.32 &  2.742  & 2.185  & 19.90  & 14.51 & WO     \\
                    & 2012/05/18.7 &  16.97 &  -1.59  &  2.817  & 1.871  & 8.98   & 13.84 & SOAO   \\
787 Moskva          & 2008/07/07.0 &  0.10  &  +9.41  &  2.240  & 1.829  & 26.55  & 14.16 & WO     \\
                    & 2008/08/05.0 &  0.38  &  +9.56  &  2.263  & 1.538  & 21.92  & 13.67 & WO     \\
                    & 2008/08/11.0 &  0.40  &  +9.17  &  2.268  & 1.486  & 20.23  & 13.55 & WO     \\
                    & 2012/10/10.5 &  23.34 &  -2.97  &  2.313  & 1.382  & 11.59  & 13.13 & KASI   \\
                    & 2012/12/25.4 &  24.00 &  -5.31  &  2.407  & 2.301  & 23.97  & 14.74 & SOAO   \\
                    & 2012/12/26.4 &  0.02  &  -5.22  &  2.408  & 2.316  & 23.92  & 14.75 & SOAO   \\
875 Nymphe          & 2008/08/10.0 &  3.51  &  +15.37 &  2.414  & 2.308  & 24.65  & 16.37 & WO     \\
                    & 2008/09/25.0 &  4.07  &  +12.12 &  2.489  & 1.826  & 20.34  & 15.80 & WO     \\
                    & 2008/11/10.9 &  3.71  &  +5.25  &  2.565  & 1.596  & 5.81   & 15.02 & WO     \\
                    & 2008/12/04.8 &  3.37  &  +2.81  &  2.603  & 1.692  & 10.31  & 15.38 & WO     \\
                    & 2012/10/12.7 &  3.32  &  +9.27  &  2.467  & 1.560  & 12.15  & 15.16 & SOAO   \\
                    & 2012/10/13.7 &  3.31  &  +9.10  &  2.469  & 1.556  & 11.75  & 15.14 & SOAO   \\
879 Ricarda         & 2012/02/22.3 &  9.48  &  -3.95  &  2.818  & 1.865  & 6.59   & 16.00 & LOAO   \\
                    & 2012/02/25.6 &  9.42  &  -3.74  &  2.822  & 1.875  & 6.07   & 16.05 & KASI   \\
897 Lysistrata      & 2008/08/05.0 &  3.32  &  +29.02 &  2.604  & 2.618  & 22.39  & 15.61 & WO     \\
                    & 2008/08/11.0 &  3.44  &  +29.49 &  2.610  & 2.548  & 22.62  & 15.57 & WO     \\
                    & 2008/11/10.9 &  3.49  &  +28.71 &  2.688  & 1.715  & 4.85   & 14.10 & WO     \\
                    & 2008/12/04.8 &  3.10  &  +25.08 &  2.706  & 1.772  & 8.16   & 14.35 & WO     \\
                    & 2012/10/14.7 &  3.39  &  +31.17 &  2.650  & 1.782  & 12.97  & 14.50 & SOAO   \\
                    & 2012/10/15.7 &  3.37  &  +31.10 &  2.651  & 1.775  & 12.63  & 14.48 & SOAO   \\
1158 Luda           & 2008/09/23.0 &  7.90  &  +34.29 &  2.354  & 2.556  & 23.11  & 15.79 & WO     \\
                    & 2008/09/25.0 &  7.96  &  +34.22 &  2.356  & 2.536  & 23.28  & 15.78 & WO     \\
                    & 2008/11/10.1 &  9.04  &  +33.02 &  2.402  & 2.036  & 24.01  & 15.37 & WO     \\
                    & 2008/11/24.1 &  9.21  &  +33.10 &  2.417  & 1.885  & 22.46  & 15.17 & WO     \\
                    & 2008/12/04.1 &  9.27  &  +33.37 &  2.428  & 1.785  & 20.66  & 15.01 & WO     \\
                    & 2013/04/01.1 &  7.34  &  +31.85 &  2.523  & 2.218  & 23.19  & 15.64 & LOAO   \\
                    & 2013/04/02.1 &  7.36  &  +31.69 &  2.524  & 2.232  & 23.22  & 15.66 & LOAO   \\
1160 Illyria        & 2012/02/14.4 &  1.71  &  +22.27 &  2.336  & 2.532  & 22.95  & 16.05 & KASI   \\
                    & 2012/02/16.4 &  1.76  &  +22.56 &  2.338  & 2.556  & 22.72  & 16.07 & KASI   \\
1215 Boyer          & 2008/08/05.7 &  16.88 &  -11.37 &  2.701  & 2.058  & 19.15  & 15.84 & WO     \\
                    & 2008/08/12.7 &  16.89 &  -12.27 &  2.692  & 2.131  & 20.28  & 15.94 & WO     \\
                    & 2008/09/02.7 &  17.05 &  -14.93 &  2.664  & 2.369  & 22.13  & 16.21 & WO     \\
                    & 2012/05/19.6 &  16.77 &  -1.47  &  2.838  & 1.881  & 8.16   & 15.35 & SOAO   \\
1996 Adams          & 2012/02/26.6 &  9.38  &  +30.63 &  2.832  & 1.918  & 9.32   & 16.39 & SOAO   \\
2151 Hadwiger       & 2012/01/28.7 &  12.79 &  +10.30 &  2.466  & 1.818  & 20.17  & 15.36 & KASI   \\
                    & 2012/01/30.7 &  12.79 &  +10.31 &  2.467  & 1.797  & 19.75  & 15.33 & KASI   \\
                    & 2012/01/31.7 &  12.80 &  +10.32 &  2.467  & 1.787  & 19.53  & 15.31 & KASI   \\
                    & 2012/02/01.7 &  12.80 &  +10.33 &  2.468  & 1.776  & 19.31  & 15.29 & KASI   \\
2429 Schurer        & 2012/02/20.7 &  11.41 &  +20.27 &  2.641  & 1.688  & 7.17   & 15.97 & SOAO   \\
3055 Annapavlova    & 2013/03/20.6 &  12.05 &  +1.11  &  2.647  & 1.651  & 0.53   & 15.81 & SOAO   \\
3158 Anga           & 2012/06/30.9 &  19.38 &  +0.68  &  2.698  & 1.747  & 9.50   & 16.49 & TUG    \\
                    & 2012/07/01.9 &  19.36 &  +0.69  &  2.697  & 1.743  & 9.31   & 16.48 & TUG    \\
3786 Yamada         & 2012/01/30.4 &  3.56  &  +20.93 &  2.754  & 2.310  & 20.06  & 16.22 & KASI   \\
                    & 2012/02/07.4 &  3.63  &  +20.64 &  2.755  & 2.418  & 20.69  & 16.34 & KASI   \\
                    & 2013/01/10.8 &  10.87 &  -12.00 &  2.599  & 1.968  & 19.19  & 15.72 & KASI   \\
                    & 2013/01/13.8 &  10.87 &  -12.41 &  2.597  & 1.933  & 18.66  & 15.66 & SOAO   \\
                    & 2013/01/14.8 &  10.86 &  -12.53 &  2.596  & 1.922  & 18.46  & 15.64 & SOAO   \\
3970 Herran         & 2008/08/26.0 &  3.09  &  +18.29 &  2.351  & 1.892  & 24.65  & 16.78 & WO     \\
                    & 2008/12/04.8 &  2.04  &  +24.51 &  2.488  & 1.632  & 16.07  & 16.24 & WO     \\
                    & 2012/10/11.6 &  2.11  &  +15.88 &  2.367  & 1.393  & 6.93   & 15.51 & KASI   \\
4104 Alu            & 2013/01/22.9 &  9.70  &  +38.46 &  2.800  & 1.882  & 8.84   & 16.71 & TUG    \\
                    & 2013/02/12.9 &  9.32  &  +40.38 &  2.793  & 1.880  & 9.40   & 16.72 & TUG    \\
4122 Ferrari        & 2012/12/25.6 &  7.24  &  +14.73 &  2.597  & 1.637  & 5.96   & 15.91 & SOAO   \\
4673 Bortle         & 2012/02/07.7 &  10.72 &  +27.46 &  2.470  & 1.529  & 8.71   & 15.28 & KASI   \\
                    & 2012/02/10.7 &  10.68 &  +28.03 &  2.472  & 1.524  & 8.08   & 15.25 & KASI   \\
4851 Vodop'yanova   & 2012/04/27.6 &  13.51 &  -10.76 &  2.593  & 1.602  & 4.89   & 16.31 & SOAO   \\
5326 (1988 RT6)     & 2012/06/24.9 &  19.77 &  +2.51  &  2.229  & 1.312  & 14.79  & 16.06 & TUG    \\
                    & 2012/06/28.9 &  19.73 &  +2.46  &  2.228  & 1.294  & 13.61  & 15.99 & TUG    \\
5977 (1992 TH1)     & 2012/07/01.1 &  22.30 &  -6.23  &  2.373  & 1.636  & 20.47  & 16.26 & TUG    \\
                    & 2012/07/04.0 &  22.31 &  -6.43  &  2.368  & 1.602  & 19.76  & 16.19 & TUG    \\
6458 Nouda          & 2008/08/06.9 &  22.27 &  +10.86 &  2.193  & 1.265  & 14.17  & 16.12 & WO     \\
                    & 2008/08/10.9 &  22.22 &  +10.53 &  2.190  & 1.246  & 12.79  & 16.04 & WO     \\
                    & 2008/09/23.8 &  21.81 &  +2.32  &  2.176  & 1.269  & 14.86  & 16.14 & WO     \\
                    & 2008/09/25.8 &  21.80 &  +1.90  &  2.175  & 1.280  & 15.57  & 16.18 & WO     \\
                    & 2008/10/02.7 &  21.80 &  +0.46  &  2.175  & 1.326  & 17.96  & 16.34 & WO     \\
                    & 2012/08/08.9 &  21.13 &  +7.75  &  2.214  & 1.247  & 10.60  & 15.98 & TUG    \\
                    & 2012/10/15.7 &  21.01 &  -4.02  &  2.178  & 1.584  & 24.86  & 16.94 & TUG    \\
7601 (1994 US1)     & 2013/03/08.6 &  11.11 &  +27.54 &  2.925  & 1.985  & 7.59   & 16.97 & BOAO   \\
7644 Cslewis        & 2012/06/23.9 &  17.55 &  -6.51  &  2.548  & 1.566  & 7.558  & 17.25 & TUG    \\
                    & 2012/06/29.8 &  17.46 &  -6.20  &  2.539  & 1.575  & 9.243  & 17.33 & TUG    \\
8653 (1990 KE)      & 2012/08/10.9 &  23.00 &  +8.34  &  2.371  & 1.459  & 13.69  & 16.48 & TUG    \\
                    & 2012/08/11.9 &  22.99 &  +8.25  &  2.370  & 1.453  & 13.32  & 16.46 & TUG    \\
9175 Graun          & 2012/12/18.6 &  6.52  &  +42.69 &  2.792  & 1.853  & 7.38   & 16.51 & SOAO   \\
                    & 2013/01/10.6 &  6.04  &  +41.46 &  2.819  & 1.904  & 8.96   & 16.65 & SOAO   \\
                    & 2013/01/13.5 &  5.99  &  +41.17 &  2.822  & 1.922  & 9.75   & 16.71 & SOAO   \\
                    & 2013/03/02.4 &  5.81  &  +35.59 &  2.870  & 2.442  & 19.43  & 17.61 & BOAO   \\
                    & 2013/03/03.4 &  5.82  &  +35.47 &  2.871  & 2.456  & 19.51  & 17.63 & BOAO   \\
                    & 2013/03/04.4 &  5.83  &  +35.37 &  2.872  & 2.470  & 19.59  & 17.64 & BOAO   \\
11129 Hayachine     & 2012/10/14.5 &  1.92  &  +22.20 &  2.687  & 1.715  & 5.95   & 16.79 & BOAO   \\
                    & 2012/10/14.9 &  1.91  &  +22.17 &  2.688  & 1.715  & 5.85   & 16.78 & TUG    \\
11931 (1993 DD2)    & 2012/11/27.6 &  5.34  &  +44.68 &  2.406  & 1.485  & 10.74  & 16.24 & SOAO   \\
                    & 2012/12/19.6 &  4.86  &  +45.39 &  2.407  & 1.470  & 9.96   & 16.20 & SOAO   \\
                    & 2013/03/31.4 &  5.66  &  +37.81 &  2.429  & 2.476  & 23.47  & 17.80 & BOAO   \\
12740 (1992 EX8)    & 2013/03/16.7 &  14.41 &  +6.09  &  2.579  & 1.736  & 14.33  & 16.57 & SOAO   \\
13679 Shinanogawa   & 2013/05/31.9 &  16.58 &  +3.81  &  2.350  & 1.396  & 10.82  & 16.36 & TUG    \\
15288 (1991 RN27)   & 2012/10/16.9 &  3.42  &  +39.12 &  2.606  & 1.759  & 14.14  & 16.81 & TUG    \\
                    & 2012/10/17.9 &  3.41  &  +39.19 &  2.605  & 1.752  & 13.86  & 16.79 & TUG    \\
17157 (1999 KP6)    & 2012/06/28.9 &  19.41 &  -1.00  &  2.282  & 1.323  & 11.09  & 15.99 & TUG    \\
                    & 2012/06/30.3 &  19.39 &  -1.08  &  2.282  & 1.318  & 10.71  & 15.97 & LOAO   \\
                    & 2012/06/30.9 &  19.39 &  -1.12  &  2.282  & 1.317  & 10.55  & 15.96 & TUG    \\
18144 (2000 OO48)   & 2008/08/07.0 &  23.44 &  -6.17  &  2.185  & 1.277  & 15.48  & 16.98 & WO     \\
                    & 2008/08/26.9 &  23.32 &  -10.68 &  2.178  & 1.183  & 6.31   & 16.44 & WO     \\
                    & 2012/10/13.5 &  22.29 &  -20.03 &  2.175  & 1.413  & 21.17  & 17.38 & BOAO   \\
                    & 2012/10/14.5 &  22.29 &  -20.08 &  2.175  & 1.423  & 21.45  & 17.40 & BOAO   \\
18841 Hruska        & 2013/03/09.6 &  11.14 &  +28.02 &  2.463  & 1.521  & 9.36   & 16.49 & BOAO   \\
                    & 2013/03/10.6 &  11.13 &  +28.00 &  2.464  & 1.524  & 9.57   & 16.50 & BOAO   \\
18881 (1999 XL195)  & 2013/03/16.6 &  10.37 &  -4.14  &  2.485  & 1.533  & 8.44   & 16.09 & SOAO   \\
19184 (1991 TB6)    & 2012/10/18.9 &  3.52  &  +21.92 &  2.357  & 1.441  & 12.22  & 17.09 & TUG    \\
19333 (1996 YT1)    & 2012/08/10.9 &  21.58 &  -2.86  &  2.416  & 1.417  & 5.29   & 16.41 & TUG    \\
                    & 2012/08/11.9 &  21.56 &  -2.83  &  2.415  & 1.415  & 5.11   & 16.40 & TUG    \\
19495 (1998 KZ8)    & 2012/12/24.6 &  6.11  &  +21.39 &  2.396  & 1.413  & 1.04   & 15.11 & SOAO   \\
                    & 2013/01/18.5 &  5.71  &  +24.42 &  2.407  & 1.518  & 12.66  & 15.87 & SOAO   \\
                    & 2013/03/10.4 &  5.84  &  +28.49 &  2.436  & 2.090  & 23.80  & 16.95 & BOAO   \\
19557 (1999 JC79)   & 2012/06/23.9 &  19.82 &  -7.05  &  2.297  & 1.352  & 12.22  & 16.50 & TUG    \\
                    & 2012/06/29.9 &  19.76 &  -7.68  &  2.289  & 1.316  & 9.84   & 16.34 & TUG    \\
20378 (1998 KZ46)   & 2012/11/26.6 &  4.70  &  +26.29 &  2.337  & 1.357  & 3.73   & 15.76 & SOAO   \\
21816 (1999 TE31)   & 2012/10/19.9 &  1.37  &  +16.35 &  2.356  & 1.365  & 3.07   & 16.85 & TUG    \\
24004 (1999 RQ57)   & 2012/10/14.8 &  1.03  &  +22.64 &  2.552  & 1.575  & 5.74   & 16.78 & TUG    \\
                    & 2012/10/16.8 &  1.00  &  +22.59 &  2.554  & 1.578  & 5.78   & 16.79 & TUG    \\
29393 (1996 NA3)    & 2012/04/28.6 &  15.24 &  +7.03  &  2.535  & 1.585  & 9.56   & 16.65 & SOAO   \\
31554 (1999 EJ2)    & 2012/06/24.9 &  18.13 &  +1.58  &  2.393  & 1.433  & 10.35  & 16.34 & TUG    \\
                    & 2012/06/30.9 &  18.05 &  +1.55  &  2.396  & 1.442  & 10.83  & 16.37 & TUG    \\
32116 (2000 LD4)    & 2012/06/21.9 &  17.05 &  -4.46  &  2.709  & 1.746  & 8.53   & 17.16 & TUG    \\
                    & 2012/06/29.8 &  16.91 &  -5.28  &  2.710  & 1.788  & 11.16  & 17.32 & TUG    \\
33229 (1998 FC124)  & 2012/10/16.8 &  23.85 &  -6.34  &  2.280  & 1.355  & 12.08  & 16.18 & TUG    \\
33489 (1999 GF9)    & 2012/08/09.9 &  21.98 &  +10.12 &  2.200  & 1.250  & 12.24  & 16.13 & TUG    \\
                    & 2012/08/11.9 &  21.96 &  +9.93  &  2.200  & 1.244  & 11.63  & 16.10 & TUG    \\
33548 (1999 JC13)   & 2012/08/09.9 &  21.25 &  +4.05  &  2.337  & 1.357  & 8.39   & 16.99 & TUG    \\
33646 (1999 JX82)   & 2012/07/12.9 &  16.92 &  -5.27  &  2.394  & 1.530  & 16.11  & 17.29 & CA     \\
                    & 2012/07/13.9 &  16.92 &  -5.41  &  2.393  & 1.537  & 16.40  & 17.31 & CA     \\
                    & 2012/07/14.9 &  16.91 &  -5.56  &  2.393  & 1.544  & 16.69  & 17.33 & CA     \\
                    & 2012/07/15.9 &  16.90 &  -5.69  &  2.393  & 1.550  & 16.98  & 17.35 & CA     \\
                    & 2012/07/16.9 &  16.90 &  -5.85  &  2.392  & 1.557  & 17.26  & 17.37 & CA     \\
                    & 2012/07/17.9 &  16.89 &  -6.00  &  2.392  & 1.565  & 17.54  & 17.39 & CA     \\
                    & 2012/07/18.9 &  16.89 &  -6.14  &  2.392  & 1.572  & 17.82  & 17.41 & CA     \\
                    & 2012/07/19.9 &  16.89 &  -6.29  &  2.391  & 1.580  & 18.09  & 17.43 & CA     \\
34035 (2000 OV27)   & 2012/07/01.0 &  21.92 &  -3.71  &  2.221  & 1.434  & 20.76  & 16.74 & TUG    \\
                    & 2012/07/04.0 &  21.93 &  -3.92  &  2.219  & 1.406  & 19.89  & 16.67 & TUG    \\
34502 (2000 SE157)  & 2012/10/17.8 &  23.15 &  +17.93 &  2.640  & 1.765  & 12.65  & 17.09 & TUG    \\
34529 (2000 SD212)  & 2008/08/25.0 &  1.27  &  +0.90  &  2.647  & 1.844  & 15.99  & 17.91 & WO     \\
                    & 2008/09/25.9 &  1.02  &  -4.42  &  2.667  & 1.683  & 5.09   & 17.29 & WO     \\
                    & 2008/10/02.9 &  0.93  &  -5.64  &  2.672  & 1.682  & 3.99   & 17.23 & WO     \\
                    & 2012/10/12.6 &  0.30  &  -8.39  &  2.664  & 1.708  & 7.66   & 17.44 & BOAO   \\
                    & 2012/10/19.8 &  0.23  &  -9.29  &  2.669  & 1.753  & 10.38  & 17.61 & TUG    \\
34572 (2000 SY310)  & 2012/10/12.5 &  23.42 &  +21.14 &  2.403  & 1.481  & 11.61  & 16.17 & SOAO   \\
                    & 2012/10/14.5 &  23.40 &  +20.92 &  2.401  & 1.488  & 12.15  & 16.20 & SOAO   \\
                    & 2012/10/17.8 &  23.37 &  +20.53 &  2.399  & 1.501  & 13.07  & 16.25 & TUG    \\
39148 (2000 WM93)   & 2012/08/09.9 &  21.13 &  +4.03  &  2.492  & 1.515  & 7.93   & 16.65 & TUG    \\
40664 (1999 RF196)  & 2012/10/18.9 &  2.74  &  +7.24  &  2.291  & 1.320  & 7.36   & 17.14 & TUG    \\
41510 (2000 QU171)  & 2012/08/05.9 &  20.55 &  +4.59  &  2.400  & 1.433  & 9.37   & 17.25 & TUG    \\
                    & 2012/08/08.9 &  20.51 &  +4.47  &  2.397  & 1.432  & 9.60   & 17.25 & TUG    \\
42704 (1998 MB32)   & 2013/01/14.2 &  7.45  &  +29.02 &  2.532  & 1.556  & 3.39   & 16.49 & SOAO   \\
42835 (1999 NS56)   & 2012/10/14.9 &  1.91  &  +25.92 &  2.301  & 1.336  & 8.20   & 16.71 & TUG    \\
                    & 2012/10/15.9 &  1.90  &  +25.76 &  2.302  & 1.334  & 7.81   & 16.69 & TUG    \\
43174 (1999 XF180)  & 2013/01/22.9 &  9.82  &  +35.25 &  2.513  & 1.586  & 9.54   & 16.81 & TUG    \\
                    & 2013/02/12.9 &  9.48  &  +38.28 &  2.506  & 1.576  & 9.61   & 16.79 & TUG    \\
49653 (1999 JO85)   & 2012/06/24.8 &  17.34 &  -12.18 &  2.439  & 1.449  & 7.07   & 16.76 & TUG    \\
49923 (1999 XQ174)  & 2013/03/08.6 &  11.44 &  +27.97 &  2.700  & 1.760  & 8.37   & 17.09 & BOAO   \\
50511 (2000 DZ101)  & 2013/05/31.9 &  16.46 &  +4.41  &  2.398  & 1.447  & 10.86  & 17.51 & TUG    \\
59748 (1999 LE14)   & 2012/07/12.9 &  19.06 &  -3.24  &  2.374  & 1.392  & 8.28   & 16.67 & CA     \\
                    & 2012/07/13.9 &  19.04 &  -3.20  &  2.373  & 1.393  & 8.41   & 16.68 & CA     \\
                    & 2012/07/14.9 &  19.03 &  -3.17  &  2.373  & 1.394  & 8.56   & 16.69 & CA     \\
                    & 2012/07/15.9 &  19.02 &  -3.15  &  2.373  & 1.395  & 8.74   & 16.70 & CA     \\
                    & 2012/07/16.9 &  19.00 &  -3.12  &  2.372  & 1.396  & 8.93   & 16.71 & CA     \\
                    & 2012/07/17.9 &  18.99 &  -3.10  &  2.372  & 1.398  & 9.15   & 16.72 & CA     \\
                    & 2012/07/18.9 &  18.97 &  -3.08  &  2.372  & 1.400  & 9.38   & 16.73 & CA     \\
                    & 2012/07/19.9 &  18.96 &  -3.07  &  2.371  & 1.402  & 9.628  & 16.74 & CA     \\
66584 (1999 RM161)  & 2012/10/11.6 &  1.95  &  +34.61 &  2.298  & 1.377  & 12.35  & 17.64 & BOAO   \\
68045 (2000 YN44)   & 2012/10/11.6 &  1.51  &  +30.88 &  2.732  & 1.790  & 8.60   & 18.54 & BOAO   \\
71001 Natspasoc     & 2013/02/04.2 &  6.97  &  +18.32 &  2.460  & 1.567  & 12.19  & 17.26 & LOAO   \\
                    & 2013/02/05.2 &  6.96  &  +18.47 &  2.461  & 1.575  & 12.58  & 17.29 & LOAO   \\
                    & 2013/03/09.5 &  6.93  &  +22.12 &  2.486  & 1.911  & 21.39  & 18.03 & BOAO   \\
71145 (1999 XA183)  & 2013/03/10.6 &  11.21 &  +27.61 &  2.861  & 1.923  & 8.04   & 17.53 & BOAO   \\
91533 (1999 RY199)  & 2012/10/12.6 &  1.10  &  +30.27 &  2.821  & 1.873  & 7.83   & 17.87 & BOAO   \\
                    & 2013/10/13.6 &  1.08  &  +30.17 &  2.820  & 1.872  & 7.71   & 17.86 & BOAO   \\
109792 (2001 RN91)  & 2012/02/20.6 &  9.84  &  +29.42 &  2.865  & 1.912  & 6.48   & 19.11 & BOAO   \\
114819 (2003 OU11)  & 2012/10/14.6 &  1.29  &  +27.89 &  2.725  & 1.764  & 7.02   & 18.64 & BOAO

\enddata
\tablecomments{UT date corresponding to the mid time of the observation, J2000 coordinates of asteroid (RA and Dec), the heliocentric ($r$) and the topocentric distances ($\Delta$), the solar phase angle ($\alpha$), the apparent predicted magnitude ($V$), and observed telescopes}
\end{deluxetable}


\clearpage

\begin{deluxetable}{ccccccc}
\tabletypesize{\scriptsize}
\tablecaption{Physical properties of the Maria asteroid family members}
\tablewidth{0pt}
\tablehead{
\colhead{Asteroid} & \colhead{$P_{rot}$} & \colhead{$A(0)$} & \colhead{Q Notes} & \colhead{Diameter} &
\colhead{Albedo} & \colhead{Type} \\
\colhead{} & \colhead{(hr)} & \colhead{(mag)} & \colhead{} & \colhead{(km)} &
\colhead{} & \colhead{}
}
\startdata
575 Renate         & 3.67  & 0.14 & 3 & 21.2  & 0.175  & LS$^{2}$ \\
652 Jubilatrix     & 2.66  & 0.21 & 3 & 17.1  & 0.168  & -   \\
660 Crescentia     & 8.09  & 0.10 & 3 & 43.6  & 0.205  & S$^{2}$  \\
695 Bella          & 14.20 & 0.39 & 4 & 41.2  & 0.245  & S$^{2}$  \\
727 Nipponia       & 5.07  & 0.12 & 3 & 34.6$^{A}$ & 0.212$^{A}$ & DT$^{1}$ \\
787 Moskva         & 6.06  & 0.39 & 4 & 40.3  & 0.120  & S$^{1}$  \\
875 Nymphe         & 12.44 & 0.11 & 4 & 15.2  & 0.192  & -   \\
879 Ricarda        & 82.9  & 0.37 & 1 & 17.9  & 0.243  & S$^{1}$  \\
897 Lysistrata     & 11.26 & 0.09 & 2 & 29.4  & 0.146  & S$^{1}$  \\
1158 Luda          & 6.86  & 0.08 & 3 & 21.1  & 0.192  & S$^{2}$  \\
1160 Illyria       & 4.29  & 0.29 & 3 & 14.8  & 0.224  & -   \\
{\bf 1215 Boyer}         & {\bf 10.36} & 0.39 & 2 & 14.7  & 0.301  & S$^{1}$  \\
1996 Adams         & 3.27  & 0.23 & 4 & 13.5  & 0.141  & -   \\
2151 Hadwiger      & 5.87  & 0.08 & 3 & 17.5$^{A}$ & 0.209$^{A}$ & S$^{2}$  \\
2429 Schurer       & 6.66  & 0.57 & 3 & 12.5  & 0.198  & -   \\
3055 Annapavlova   & $>$ 7 & -    & 1 & 9.4   & 0.199  & -   \\
3158 Anga          & $>$ 8 & -    & 1 & 8.1   & 0.269  & S$^{1}$  \\
3786 Yamada        & 4.03  & 0.35 & 3 & 16.7  & 0.234  & S$^{1}$  \\
{\bf 3970 Herran}        & {\bf 8.09}  & 0.26 & 3 & 8.6   & 0.264  & -   \\
4104 Alu           & $>$ 6 & -    & 1 & 8.8   & 0.230  & S$^{1}$  \\
{\bf 4122 Ferrari}       & {\bf 8.45}  & 0.18 & 3 & 11.8  & 0.221  & -   \\
4673 Bortle        & 2.64  & 0.13 & 3 & 12.5  & 0.290  & S$^{1}$  \\
{\bf 4851 Vodop'yanova}  & {\bf 4.90}   & 0.17 & 3 & 7.2$^{M}$  & 0.254$^{M}$  & -   \\
5326 (1988 RT6)    & $>$ 6 & -    & 1 & 7.4   & 0.296  & LS$^{2}$ \\
5977 (1992 TH1)    & $>$ 12 & -   & 1 & 11.8  & 0.154  & -   \\
6458 Nouda         & 4.203 & 0.56 & 3 & 7.5   & 0.181  & -   \\
{\bf 7601 (1994 US1)}    & {\bf 3.74}  & 0.31 & 3 & 7.3   & 0.190  & S$^{2}$  \\
{\bf 7644 Cslewis}       & {\bf 2.31}  & 0.22 & 3 & 5.8   & 0.174  & -   \\
8653 (1990 KE)     & $>$ 14 & -   & 1 & 6.6$^{M}$  & 0.254$^{M}$  & S$^{2}$  \\
9175 Graun         & 25.8  & 0.12 & 3 & 7.9   & 0.308  & -   \\
{\bf 11129 Hayachine}    & {\bf 17.57} & 0.17 & 2 & 6.2   & 0.350  & -   \\
{\bf 11931 (1993 DD2)}   & {\bf 19.87} & 0.09 & 3 & 6.7   & 0.363  & -   \\
12740 (1992 EX8)   & $>$ 10 & -   & 1 & 9.1   & 0.285  & -   \\
13679 Shinanogawa  & $>$ 8 & -    & 1 & 5.7   & 0.308  & S$^{2}$  \\
{\bf 15288 (1991 RN27)}  & {\bf 7.14}  & 0.13 & 3 & 9.3   & 0.188  & L$^{2}$  \\
{\bf 17157 (1999 KP6)}   & {\bf 3.72}  & 0.14 & 3 & 6.4   & 0.296  & L$^{2}$  \\
{\bf 18144 (2000 OO48)}  & {\bf 3.10}  & 0.24 & 3 & 3.5   & 0.437  & SQ$^{2}$ \\
{\bf 18841 Hruska}       & {\bf 6}   & 0.11 & 2 & 7.3   & 0.212  & -   \\
{\bf 18881 (1999 XL195)} & {\bf 2.98}  & 0.18 & 3 & 9.7   & 0.173  & -   \\
{\bf 19184 (1991 TB6)}   & {\bf 4.99}  & 0.36 & 3 & 6.7   & 0.129  & -   \\
{\bf 19333 (1996 YT1)}   & {\bf 2.83}  & 0.16 & 3 & 5.6   & 0.268  & -   \\
{\bf 19495 (1998 KZ8)}   & {\bf 2.33}  & 0.12 & 3 & 6.3   & 0.342  & -   \\
19557 (1999 JC79)  & $>$ 5 & -    & 1 & 8.4   & 0.230  & S$^{2}$  \\
{\bf 20378 (1998 KZ46)}  & {\bf 5.14}  & 0.31 & 3 & 7.9   & 0.236  & LS$^{2}$ \\
21816 (1999 TE31)  & $>$ 20 & -   & 1 & 4.6   & 0.206  & -   \\
{\bf 24004 (1999 RQ57)}  & {\bf 5.68}  & 0.43 & 3 & 9.1$^{M}$  & 0.254$^{M}$  & -   \\
29393 (1996 NA3)   & $>$ 20 & -   & 1 & 6.6$^{M}$  & 0.254$^{M}$  & LS$^{2}$ \\
31554 (1999 EJ2)   & $>$ 12 & -   & 1 & 7.5   & 0.240  & LS$^{2}$ \\
32116 (2000 LD4)   & $>$ 8 & -    & 1 & 6.0$^{M}$  & 0.254$^{M}$ & -   \\
33229 (1998 FC124) & $>$ 5 & -    & 1 & 5.4   & 0.377  & -   \\
{\bf 33489 (1999 GF9)}   & {\bf 6.74}  & 0.55 & 3 & 6.0   & 0.198  & L$^{2}$  \\
{\bf 33548 (1999 JC13)}  & {\bf 3.97}  & 0.27 & 3 & 4.8   & 0.227  & L$^{2}$  \\
{\bf 33646 (1999 JX82)}  & {\bf 3.33}  & 0.09 & 3 & 5.2   & 0.218  & L$^{2}$  \\
34035 (2000 OV27)  & $>$ 10 & -   & 1 & 5.8   & 0.276  & -   \\
{\bf 34502 (2000 SE157)} & {\bf 4.71}  & 0.05 & 3 & 5.7   & 0.344  & Q$^{2}$  \\
{\bf 34529 (2000 SD212)} & {\bf 3.71}  & 0.50 & 3 & 5.2   & 0.261  & -   \\
{\bf 34572 (2000 SY310)} & {\bf 6.64}  & 0.40 & 3 & 7.3   & 0.252  & -   \\
{\bf 39148 (2000 WM93)}  & {\bf 7.55}  & 0.13 & 2 & 4.1   & 0.315  & -   \\
{\bf 40664 (1999 RF196)} & {\bf 9.61}  & 0.17 & 2 & 3.5   & 0.368  & -   \\
41510 (2000 QU171) & $>$ 24 & -   & 1 & 4.8   & 0.207  & -   \\
42704 (1998 MB32)  & $>$ 12 & -   & 1 & 5.3   & 0.333  & -   \\
{\bf 42835 (1999 NS56)}  & {\bf 2.49}  & 0.14 & 3 & 5.4   & 0.222  & -   \\
43174 (1999 XF180) & $>$ 16 & -   & 1 & 5.5   & 0.214  & -   \\
49653 (1999 JO85)  & $>$ 8 & -    & 1 & 5.5   & 0.209  & -   \\
{\bf 49923 (1999 XQ174)} & {\bf 5.21}  & 0.15 & 3 & 5.3   & 0.329  & L$^{2}$  \\
50511 (2000 DZ101) & $>$ 8 & -    & 1 & 3.7   & 0.205  & -   \\
{\bf 59748 (1999 LE14)}  & {\bf 6.65}  & 0.05 & 3 & 5.6   & 0.223  & -   \\
{\bf 66584 (1999 RM161)} & {\bf 9.06}  & 0.09 & 3 & 3.6   & 0.137  & -   \\
{\bf 68045 (2000 YN44)}  & {\bf 4.35}  & 0.12 & 3 & 3.3$^{M}$  & 0.254$^{M}$  & -   \\
{\bf 71001 Natspasoc}    & {\bf 3.15}  & 0.14 & 3 & 5.3   & 0.271  & LS$^{2}$ \\
{\bf 71145 (1999 XA183)} & {\bf 12}    & 0.09 & 2 & 6.1   & 0.246  & C$^{2}$  \\
{\bf 91533 (1999 RY199)} & {\bf 3.85}  & 0.53 & 3 & 5.0   & 0.283  & -   \\
109792 (2001 RN91) & $>$ 16 & -   & 1 & 2.5   & 0.101  & S$^{2}$  \\
114819 (2003 OU11) & $>$ 10 & -   & 1 & 3.3   & 0.211  & -  
\enddata
\tablecomments{Synodic rotational periods ($P_{rot}$) from our analysis, the amplitude of lightcurve $A(0)$ obtained using the equation of (3), and the reliability parameters (Q Notes) follow the definition by \citet{lag89} (see Sect. 4.1. for more details). For 34 objects marked with boldface, the rotational periods are presented for the first time. Diameter and albedo information were acquired mainly from the WISE IR data \citep{mas11} with the exception of those marked with A $\sbond$ AcuA \citep[Asteroid catalog using AKARI;][]{usu11} and M (assuming mean albedo value of 0.254 for the Maria family asteroids), while taxonomic types from \citep{nee10} are indicated with ``1'', types from the SDSS-based Asteroid Taxonomy \citep{has12} are marked with ``2''.}
\end{deluxetable}


\clearpage

\begin{deluxetable}{ccccccccccc}
\tabletypesize{\scriptsize}
\tablecaption{Summary of the Maria asteroids obtained the pole solutions}
\tablewidth{0pt}
\tablehead{
\colhead{Asteroid} & \colhead{$\lambda_1$} & \colhead{$\beta_1$} & \colhead{$\lambda_2$} & \colhead{$\beta_2$} &
\colhead{$\epsilon$} & \colhead{$P_{sid}$} & \colhead{$N_{LC}$} & \colhead{$N_{689}$} & \colhead{$N_{703}$} & \colhead{DAMIT} \\
\colhead{} & \colhead{(deg)} & \colhead{(deg)} & \colhead{(deg)} & \colhead{(deg)} &
\colhead{(deg)} & \colhead{(hr)} & \colhead{} & \colhead{} & \colhead{} & \colhead{} \\
}
\startdata
170 Maria      & {\bf 103} & {\bf -18} & 289 & -5  & {\bf 112} & 13.1313212 & 10 & 183 & 112 & -         \\
575 Renate     & 6   & +3  & {\bf 183} & {\bf -19} & {\bf 105} & 3.6603283  & 5  & 108 & 191 & -         \\
652 Jubilatrix & {\bf 143} & {\bf +34} & 321 & +27 & {\bf 69}  & 2.6626693  & 5  & 84  & 163 & -         \\
660 Crescentia & 58  & +1  & {\bf 241} & {\bf +15} & {\bf 90}  & 7.9140789  & 10 & 220 & 158 & -         \\
695 Bella      & {\bf 81}  & {\bf -58} & 308 & -49 & {\bf 149} & 14.2190116 & 15 & 203 & 144 & 87, -55   \\
               &     &     &     &     &     &            &    &     &     & 314, -56  \\
727 Nipponia   & {\bf 172} & {\bf +11} & 333 & +5  & {\bf 89}  & 5.0687459  & 10 & 239 & 185 & -         \\
787 Moskva     & 132 & +11 & {\bf 334} & {\bf +44} & {\bf 55}  & 6.0558066  & 30 & 164 & 132 & 126, +27  \\
               &     &     &     &     &     &            &    &     &     & 331, +59  \\
875 Nymphe     & {\bf 47}  & {\bf +32} & 200 & +46 & {\bf 52}  & 12.6212926 & 18 & 95  & 176 & 42, +29   \\
               &     &     &     &     &     &            &    &     &     & 196, +41  \\
897 Lysistrata & {\bf 109} & {\bf -55} & 299 & -29 & {\bf 136} & 11.2742816 & 7  & 186 & 134 & -         \\
1158 Luda      & 88  & +67 & {\bf 267} & {\bf +21} & {\bf 54}  & 6.8750690  & 12 & 106 & 177 & -         \\
1996 Adams     & {\bf 106} & {\bf +56} & 281 & +17 & {\bf 48}  & 3.3111390  & 3  & 82  & 172 & 107, +55  \\
3786 Yamada    & {\bf 77}  & {\bf +61} & 236 & +63 & {\bf 33}  & 4.0329422  & 9  & 18  & 131 & -         \\
6458 Nouda     & {\bf 63}  & {\bf -2}  & 240 & +21 & {\bf 81}  & 4.2030968  & 9  & -   & 163 & -           
\enddata
\tablecomments{The ecliptic longitude ($\lambda$) and latitude ($\beta$) of the asteroid pole orientation (usually the two solutions differ by 180 degrees), the spin vector obliquity ($\epsilon$) of the angle between the orbital plane and the spin pole (calculated from the lower chi-squared pole solution), the sidereal rotational period ($P_{sid}$), the number of dense lightcurves ($N_{LC}$), and the number of sparse points from USNO in Flagstaff (689), the Catalina Sky Survey (703). The information about spin axis ($\lambda, \beta$) of 4 objects was also found in DAMIT \citep[Database of Asteroid Models from Inversion Techniques;][]{dur10}. For the pole orientation, lower chi-squared value marked with boldface is preferred.}
\end{deluxetable}






\begin{thebibliography}{}
\bibitem[Alvared-Candal et al.(2004)]{alv04} Alvarez-Candal, A., Duffard, R., Angeli, C. A., et al. 2004, \icarus, 172, 388
\bibitem[Asphaug \& Scheeres(1999)]{asp99} Asphaug, E. \& Scheeres, D. J. 1999, \icarus, 139, 383
\bibitem[Binzel et al.(1989)]{bin89} Binzel, R. P., Farinella, P., Zappal\`a, V., \& Cellino, A. 1989, in Asteroid II, ed. R. P. Binzel, T. Gehrels, \& M. S. Matthews (Tucson, AZ: Univ. Arizona Press), 416
\bibitem[Bottke et al.(2001)]{bot01} Bottke, W. F., Vokrouhlick\'y, D., Bro\u{z}, M., et al. 2001, Science, 294, 1693
\bibitem[Bottke et al.(2002)]{bot02} Bottke, W. F., Morbidelli, A., Jedicke, R., et al. 2002, \icarus, 156, 399
\bibitem[Bottke et al.(2005)]{bot05} Bottke, W. F., Durda, D., Nesvorn\'y, D., et al. 2005, The origin and evolution of stony meteorites. in: Dynamics of Populations of Planetary Systems (Z. Knezevic \& A. Milani, Eds.), pp. 357-374. Cambridge University Press, Cambridge
\bibitem[Carbognani(2010)]{car10} Carbognani, A. 2010, \icarus, 205, 497
\bibitem[Cellino et al.(2002)]{cel02} Cellino, A., Bus, S. J., Doressoundiram, A., \& Lazzaro, D. 2002, in Asteroid III, ed. W. Bottke, A. Cellino, P. Paolicchi, \& R. Binzel (Tucson, AZ: Univ. Arizona Press), 633
\bibitem[Cellino et al.(2009)]{cel09} Cellino, A., Dell'Oro, A., \& Tedesco, E. F. 2009, \planss, 57, 173
\bibitem[Davis et al.(1994)]{dav94} Davis, D. R., Ryan, E. V., \& Farinella, P, 1994, \planss, 42, 599
\bibitem[\u{D}urech et al.(2009)]{dur09} \u{D}urech, J., Kaasalainen, M., Warner, B. D., et al. 2009, \aap, 493, 291
\bibitem[\u{D}urech et al.(2010)]{dur10} \u{D}urech, J., Sidorin, V., \& Kaasalainen, M. 2010, \aap, 513, 46
\bibitem[Dohnanyi(1969)]{doh69} Dohnanyi, J. S. 1969, \jgr, 74, 2531
\bibitem[Fieber-Beyer et al.(2011)]{fie11} Fieber-Beyer, S. K., Gaffey, M. J., Kelley, M. S., et al. 2011, \icarus, 213, 524
\bibitem[Giblin et al.(1998)]{gib98} Giblin, I., Martelli, G., Farinella, P., et al. 1998, \icarus, 134, 77
\bibitem[Gladman et al.(1997)]{gla97} Gladman, B. J., Migliorini, F., Morbidelli, A., et al. 1997, Science, 277, 197
\bibitem[Guillens et al.(2002)]{gui02} Guillens, S. A., Vieira Martins, R., \& Gomes, R. S. 2002, \aj, 124, 2322
\bibitem[Hanu\u{s} et al.(2011)]{han11} Hanu\u{s}, J., \u{D}urech, J., Bro\u{z}, M., et al. 2011, \aap, 530, 134
\bibitem[Hasselmann et al.(2012)]{has12} Hasselmann, P. H., Carvano, J. M., \& Lazzaro, D. 2012, SDSS-based Asteroid Taxonomy V1.1, EAR-A-I0035-5-SDSSTAX-V1.1, NASA Planetary Data System
\bibitem[Hirayama(1918)]{hir18} Hirayama, K. 1918, \aj, 31, 185
\bibitem[Hirayama(1922)]{hir22} Hirayama, K. 1922, JaJAG, 1, 55
\bibitem[Holsapple et al.(2002)]{hol02} Holsapple, K., Giblin, I., Housen, K., et al. 2002, in Asteroid III, ed. W. Bottke, A. Cellino, P. Paolicchi, \& R. Binzel (Tucson, AZ: Univ. Arizona Press), 443  
\bibitem[Housen \& Holsapple(1990)]{hou90} Housen, K. R. \& Holsapple, K. A. 1990, \icarus, 84, 226
\bibitem[Housen et al.(1991)]{hou91} Housen, K. R., Schmidt, R. M., \& Holsapple, K. A. 1991, \icarus, 94, 180
\bibitem[Howell(1989)]{how89} Howell, S. B. 1989, \pasp, 101, 616
\bibitem[Ito and Yoshida(2010)]{ito10} Ito, T. \& Yoshida, F. 2010, AdvGeo, 25, 161
\bibitem[Ivezi\'c et al.(2002)]{ive02} Ivezi\'c, \u{Z}., Lupton, R. H., Juri\'c, M., et al. 2002, \aj, 124, 2943
\bibitem[Kaasalainen \& Torppa(2001)]{kat01} Kaasalainen, M. \& Torppa, J. 2001, \icarus, 153, 24
\bibitem[Kaasalainen et al.(2001)]{kaa01} Kaasalainen, M., Torppa, J., \& Muinonen, K. 2001, \icarus, 153, 37
\bibitem[Kaasalainen et al.(2002)]{kaa02} Kaasalainen, M., Torppa, J., \& Piironen, J. 2002, \icarus, 159, 369
\bibitem[Kaasalainen(2004)]{kaa04} Kaasalainen, M. 2004, \aap, 422, L39
\bibitem[Kryszczy\'nska et al.(2012)]{kry12} Kryszczy\'nska, A., Colas, F., Poli\'n, M., et al. 2012, \aap, 546, 72
\bibitem[Kryszczy\'nska(2013)]{kry13} Kryszczy\'nska, A. 2013, \aap, 551, 102
\bibitem[La Spina et al.(2004)]{las04} La Spina, A., Paolicchi, P., Kryszczy\'nska, A., \& Pravec, P. 2004, \nat, 428, 400
\bibitem[Lagerkvist et al.(1989)]{lag89} Lagerkvist, C.-I., Harris, A. W., \& Zappal\`a, V. 1989, in Asteroid II, ed. R. P. Binzel, T. Gehrels, \& M. S. Matthews (Tucson, AZ: Univ. Arizona Press), 1162
\bibitem[Lagerkvist \& Magnusson(2011)]{lag11} Lagerkvist, C.-I. \& Magnusson, P. 2011, Asteroid Photometric Catalog V1.1, EAR-A-3-DDR-APC-LIGHTCURVE-V1.1, NASA Planetary Data System
\bibitem[Lenz \& Breger(2005)]{len05} Lenz, P. \& Breger, M. 2005, CoAst, 146, 53
\bibitem[Masiero et al.(2011)]{mas11} Masiero, J. R., Mainzer, A. K., Grav, T., et al. 2011, \apj, 741, 68
\bibitem[Masiero et al.(2013)]{mas13} Masiero, J. R., Mainzer, A. K., Bauer, J. M., et al. 2013, \apj, 770, 7
\bibitem[Michel et al.(2001)]{mic01} Michel, P., Benz, W., Tanga, P., et al. 2001, Science, 294, 1696
\bibitem[Milani et al.(2010)]{mil10} Milani, A., Knezevic, Z., Novakovic, B., \& Cellino, A. 2010, 207, 769  
\bibitem[Mizutani et al.(1990)]{miz90} Mizutani, H., Takagi, Y., \& Kawakami, S. 1990, \icarus, 87, 307
\bibitem[Morbidelli \& Moon(1995)]{mor95} Morbidelli, A. \& Moons, M. 1995, \icarus, 115, 60
\bibitem[Morbidelli et al.(1995)]{morb95} Morbidelli, A., Zappal\`a, V., Moons, M., et al. 1995, \icarus, 118, 132
\bibitem[Morbidelli and Vokrouhlick\'y(2003)]{mor03} Morbidelli, A. \& Vokrouhlick\'y, D. 2003, \icarus, 163, 120
\bibitem[Neese(2010)]{nee10} Neese, C. 2010, Asteroid Taxonomy V6.0, EAR-A-5-DDR-TAXONOMY-V6.0, NASA Planetary Data System
\bibitem[Nesvorn\'y et al.(2002)]{nes02} Nesvorn\'y, D., Morbidelli, A., Vokrouhlick\'y, D., et al. 2002, \icarus, 157, 155
\bibitem[Nesvorn\'y \& Bottke(2004)]{nes04} Nesvorn\'y, D. \& Bottke, W. F. 2004, \icarus, 170, 324
\bibitem[Nesvorn\'y et al.(2005)]{nes05} Nesvorn\'y, D., Jedicke, R., Whiteley, R., \& Ivezi\'c, \u{Z}. 2005, \icarus, 173, 132
\bibitem[Nesvorn\'y et al.(2010)]{nes10} Nesvorn\'y, D. 2010, Nesvorn\'y HCM Asteroid Families V1.0, EAR-A-VARGBDET-5-NESVORNYFAM-V1.0, NASA Planetary Data System
\bibitem[O'Brien \& Greenberg(2003)]{obr03} O'Brien, D. P. \& Greenberg, R. 2003, \icarus, 164, 334
\bibitem[Palmer(2009)]{pal09} Palmer, D. M. 2009, \apj, 695, 496
\bibitem[Paolicchi et al.(1996)]{pao96} Paolicchi, P., Verlicchi, A., \& Cellino, A., \icarus, 121, 126
\bibitem[Paolicchi(2005)]{pao05} Paolicchi, P. 2005, Mem. SAIt Suppl., 6, 110
\bibitem[Paolicchi and Kryszczy\'nska(2012)]{pao12} Paolicchi, P. \& Kryszczy\'nska, A. 2012, \planss, 73, 70
\bibitem[Parker et al.(2008)]{par08} Parker, A., Ivezi\'c, \u{Z}., Juri\'c, M., Lupton, R. H., et al. 2008, \icarus, 198, 138
\bibitem[Pravec \& Harris(2000)]{pra00} Pravec, P. \& Harris, A. W. 2000, \icarus, 148, 12
\bibitem[Pravec et al.(2002)]{pra02} Pravec, P., Harris, A. W., \& Michalowski, T. 2002, in Asteroid III, ed. W. Bottke, A. Cellino, P. Paolicchi, \& R. Binzel (Tucson, AZ: Univ. Arizona Press), 113
\bibitem[Robutel \& Laskar(2001)]{rob01} Robutel, P. \& Laskar, J. 2001, \icarus, 152, 4
\bibitem[Rubincam(2000)]{rub00} Rubincam, D. P. 2000, \icarus, 148, 2
\bibitem[Skogl\"{o}v and Erikson(2002)]{sko02} Skogl\"{o}v, E. \& Erikson, A. 2002, \icarus, 160, 24
\bibitem[Slivan(2002)]{sli02} Slivan, S. M. 2002, \nat, 419, 49
\bibitem[Slivan et al.(2003)]{sli03} Slivan, S. M., Binzel, R. P., Crespo da Silva, L. D., et al. 2003, \icarus, 162, 285
\bibitem[Slivan et al.(2008)]{sli08} Slivan, S. M., Binzel, R. P., Boroumand, S. C., et al. 2008, \icarus, 195, 226
\bibitem[Slivan et al.(2009)]{sli09} Slivan, S. M., Binzel, R. P., Kaasalainen, M., et al. 2009, \icarus, 200, 514
\bibitem[Tanga et al.(1999)]{tan99} Tanga, P., Cellino, A., Michel, P., et al. 1999, \icarus, 141, 65
\bibitem[Usui et al.(2011)]{usu11} Usui, F., Kuroda, D., Mueller, T. G., et al. 2011, \pasj, 63, 1117
\bibitem[Vokrouhlick\'y et al.(2003)]{vok03} Vokrouhlick\'y, D., Nesvorn\'y, D., \& Bottke, W. F. 2003, \nat, 425, 147
\bibitem[Vokrouhlick\'y et al.(2006)]{vok06} Vokrouhlick\'y, D., Bro\u{z}, M., Bottke, W. F., et al. 2006, \icarus, 182, 118
\bibitem[Warner et al.(2009)]{war09} Warner, B. D., Harris, A. W., \& Pravec, P. 2009, \icarus, 202, 134
\bibitem[Zappal\`a et al.(1990)]{zap90} Zappal\`a, V., Cellino, A., Barucci, A. M., et al. 1990, \aap, 231, 548
\bibitem[Zappal\`a et al.(1997)]{zap97} Zappal\`a, V., Cellino, A., di Martino, M., \& Migliorini, F. 1997, \icarus, 129, 1
\end{thebibliography}
\end{document}